\documentclass[3p,sort&compress]{elsarticle}
\newcommand{\singlePlotSize}{0.60\columnwidth}
\newcommand{\smallDoublePlotSize}{0.45\columnwidth}
\newcommand{\doublePlotSize}{0.49\columnwidth}
\newcommand{\doublePlusPlotSize}{0.51\columnwidth}

\journal{Physical Review D}

\usepackage{etoolbox}
\newtoggle{isPRD}
\togglefalse{isPRD}

\usepackage{atlasphysics} 

\usepackage{amssymb} 
\usepackage{amsmath} 

\usepackage{lineno}

\usepackage{paralist}
\usepackage{subfigure}
\usepackage{multirow} 
\usepackage{graphicx}
\usepackage{dcolumn}
\usepackage{bm}
\usepackage{wasysym}

\usepackage{dashrule}

\newcommand\htollllp{$H\to ZZ^{*}\to\ell^+\ell^-\ell^{'+}\ell^{'-}$}
\newcommand\htollllbrief{$H\to ZZ^{*}\to 4\ell$}
\newcommand\htoWWlvlv{$H\to WW^{*}\to \ell\nu\ell\nu$}

\newcommand\htogg{$H\to \gamma\gamma$}

\newcommand\zzstar{$ZZ^{*}$}
\newcommand\ZZbkg{\zzstar}

\newcolumntype{d}[1]{D{.}{.}{#1}}
\newcommand\progname[1]{{\sc #1}}
\newcommand\pval{\ensuremath{p_0}}

\def\tabref#1{Table~\ref{#1}} 
\def\Figref#1{Figure~\ref{#1}} 
\def\figref#1{Fig.\ \ref{#1}}   
\def\secref#1{Sec.\ \ref{#1}}
\newcommand{\tabscript}[3]{%
  \setlength{\fboxrule}{0pt}%
  \fbox{\ensuremath{#1^{#2}_{#3}}}%
}

\newcommand\lumia{\ensuremath{4.5~\ifb}}
\newcommand\lumib{\ensuremath{20.3~\ifb}}

\newcommand{\bdtzz}{{\ensuremath{{\mathrm{BDT}}_{ZZ^*}}}}
\newcommand{\bdtvbf}{{\ensuremath{\mathrm{BDT_{VBF}}}}}
\newcommand{\bdtvh}{{\ensuremath{\mathrm{BDT_{VH}}}}}
\newcommand{\ggfcat}{{\it ggF enriched}}
\newcommand{\vbfcat}{{\it VBF enriched}}
\newcommand{\vhlepcat}{{\it VH-leptonic enriched}}
\newcommand{\vhhadcat}{{\it VH-hadronic enriched}}

\newcommand{\hllllmassvalue}{124.51}

\newcommand{\hllllmassshort}{$m_{H}$ = \hllllmassvalue\ $\pm$ 0.52 \gev}
\newcommand{\atlasmassvalue}{125.36}

\newcommand{\atlasmassveryshort}{$m_{H}$ = \atlasmassvalue\ \gev}
\newcommand{\hllllmuvalue}{1.66}
\newcommand{\hllllmu}{$\mu$ = \hllllmuvalue\ $^{+0.39}_{-0.34}$ (stat) $^{+0.21}_{-0.14}$ (syst)}

\newcommand{\hllllmuCombmass}{$\mu$ = 1.50 $^{+0.35}_{-0.31}$ (stat) $^{+0.19}_{-0.13}$ (syst)}
\newcommand{\hllllmuCombmassCombCatValue}{1.44 $^{+0.34}_{-0.31}$ (stat) $^{+0.21}_{-0.11}$ (syst)}
\newcommand{\hllllmuCombmassCombCat}{$\mu$ = \hllllmuCombmassCombCatValue}
\newcommand{\hllllmuCombmassCombCatForAbstract}{1.44 $^{+0.40}_{-0.33}$}
\newcommand{\hllllmuggF}{1.66 $^{+0.45}_{-0.41}$ (stat) $^{+0.25}_{-0.15}$ (syst)}
\newcommand{\hllllmuggFForAbstract}{1.7 $^{+0.5}_{-0.4}$}
\newcommand{\hllllmuVBF}{0.26 $^{+1.60}_{-0.91}$ (stat) $^{+0.36}_{-0.23}$ (syst)}
\newcommand{\hllllmuVBFForAbstract}{0.3 $^{+1.6}_{-0.9}$}
\newcommand{\hllllmuVBFOvermuggF}{$0.2^{+1.2}_{-0.5}$}
\newcommand{\vecth}{{\ensuremath{\bm{\theta}}}}
\newcommand{\vecalpha}{{\ensuremath{\bm{\alpha}}}}
\newcommand{\hllllmuggFform}{1.66\ ^{+0.45}_{-0.41} {\rm{~(stat)~}} ^{+0.25}_{-0.15} {\rm{~(syst)}}}
\newcommand{\hllllmuVBFform}{0.26\ ^{+1.60}_{-0.91} {\rm{~(stat)~}} ^{+0.36}_{-0.23} {\rm{~(syst)}}}

\usepackage{hypernat}
\usepackage{hyperref}

\newcommand{\papertitle}{Measurements of Higgs boson production and couplings in the four-lepton channel in $pp$ collisions at center-of-mass energies of 7 and 8\,\tev\ with the ATLAS detector}

\newcommand{\papertitleWithBreaks}{Measurements of Higgs boson production and couplings \\ 
                                   in the four-lepton channel in $pp$ collisions \\ 
                                   at center-of-mass energies of 7 and 8\,\tev\ with the ATLAS detector}

\newcommand{\paperabstract}{The final ATLAS Run 1 measurements of  Higgs boson production and couplings
in the decay channel \htollllp{}, where $\ell,\ell^{'}=e\text{ or }\mu$, are presented. These
measurements were performed using $pp$ collision data corresponding to integrated luminosities
of \lumia~and \lumib\ at center-of-mass energies of 7 \tev\ and 8 \tev, respectively, recorded with
the ATLAS detector at the LHC.  The \htollllbrief{} signal is observed with a significance of 8.1
standard deviations, with an expectation of 6.2 standard deviations, at \atlasmassveryshort, the
combined ATLAS measurement of the Higgs boson mass from the \htogg\ and \htollllbrief\ channels.
The production rate relative to the Standard Model expectation, the signal strength, is measured in
four different production categories in the \htollllbrief\ channel. The measured signal strength, at
this mass, and with all categories combined, is \hllllmuCombmassCombCatForAbstract.  The signal
strength for Higgs boson production in gluon fusion or in association with $t\bar{t}$ or $b\bar{b}$
pairs is found to be \hllllmuggFForAbstract, while the signal strength for vector-boson fusion
combined with $WH/ZH$ associated production is found to be \hllllmuVBFForAbstract.}

\usepackage{preprintcover}  
\PreprintCoverPaperTitle{\papertitle}  
\PreprintIdNumber{CERN-PH-EP-2014-170}  
\PreprintCoverAbstract{\paperabstract}  
\PreprintJournalName{Physical Review D}  

\clearpage

\newcommand{\Cc}{\ensuremath{\kappa}}

\begin{document}

\title{\papertitleWithBreaks}
\author{G. Aad {\it et.\ al}\footnote{Full author list given at the end of the article.} \\ (ATLAS Collaboration)}
\address{}


\date{\today}

\begin{abstract}

\paperabstract

\end{abstract}


\maketitle

\section{Introduction}
\label{sec:intro}

In the Standard Model (SM) the Brout-Englert-Higgs (BEH) mechanism is the source of electroweak
symmetry breaking and results in the appearance of a fundamental scalar particle, the Higgs
boson~\cite{Englert:1964et,Higgs:1964pj,Guralnik:1964eu}. The ATLAS and CMS experiments have
reported the observation of a particle in the search for the SM Higgs
boson~\cite{ATLAS:2012af,Chatrchyan:2012ia}, where the most sensitive channels
are \htollllbrief{}, \htoWWlvlv{} and \htogg{}.
An important step in the confirmation of the new particle as the SM Higgs boson is the measurement
of its properties, which are completely defined in the SM once its mass is known. Previous ATLAS
studies~\cite{Aad:2013wqa, Aad:2013xqa} have shown that this particle is consistent with the SM Higgs boson.

The Higgs boson decay to four leptons, \htollllbrief{}, where $\ell=e\text{ or }\mu$, provides good
sensitivity for the measurement of the Higgs boson properties due to its high signal-to-background
ratio, which is about 2 for each of the four final states: $\mu^+\mu^-\mu^+\mu^-$~($4\mu$),
$e^+e^-\mu^+\mu^-$ ($2e2\mu$), $\mu^+\mu^-e^+e^-$ ($2\mu2e$), and $e^+e^-e^+e^-$~($4e$), where the
first lepton pair is defined to be the one with the dilepton invariant mass closest to the $Z$ boson
mass.  The contribution to these final states from $H\to ZZ^{*}, Z^{(*)} \to \tau^+ \tau^-$ decays
is below the per mille level in the current analysis.  The largest background in this search comes
from continuum $(Z^{(*)}/\gamma^*)(Z^{(*)}/\gamma^*)$ production, referred to as \zzstar\
hereafter. For the four-lepton events with an invariant mass, $m_{4\ell}$, below about $160 $\,\gev,
there are also important background contributions from $Z+\rm jets$ and $t\bar{t}$ production with
two prompt leptons, where the additional charged lepton candidates arise from decays of hadrons with
$b$- or $c$-quark content, from photon conversions or from misidentification of jets.

Interference effects are expected between the Higgs boson signal and SM background processes. For
the \htollllbrief{} channel, the impact of this interference on the mass spectrum near the resonance
is negligible~\cite{Kauer:2012hd}.  This analysis does not account for interference effects in the
mass spectra.

In the SM, the inclusive production of the \htollllbrief{} final state is dominated by the gluon
fusion (ggF) Higgs boson production mode, which represents 86\% of the total production cross
section for $m_H=125$ \gev\ at $\sqrt{s}=8$ TeV. Searching for Higgs boson production in the
vector-boson fusion (VBF) and the vector-boson associated production (VH) modes allows further
exploration of the coupling structure of the new particle.  The corresponding fractions of the
production cross section for VBF and VH are predicted to be 7\% and 5\%, respectively.

This paper presents the final ATLAS Run 1 results of the measurement of the SM Higgs boson
production in the \htollllbrief\ decay mode, where the production is studied both inclusively and
with events categorized according to the characteristics of the different production modes.  The
categorized analysis allows constraints to be placed on possible deviations from the expected
couplings of the SM Higgs boson.  The data sample used corresponds to an integrated luminosity of
\lumia\ at a center-of-mass energy of 7 TeV and \lumib\ at a center-of-mass energy of 8 TeV,
collected in the years 2011 and 2012, respectively.  The method adopted to extract the production
rates simultaneously provides a measurement of the Higgs boson mass.  The measurement of the Higgs
boson mass for this channel, performed in combination with the \htogg\ decay mode, is discussed in
Ref.~\cite{combmasspaper} and is only covered briefly here.  This paper contains a full description
of the signal and background simulation, the object reconstruction and identification, the event
selection and the background estimations of the \htollllbrief\ decay mode, providing the details for
other Run 1 final results, including the combined mass measurement, reported elsewhere.  The
corresponding final Run 1 CMS results for the \htollllbrief\ decay mode have been reported in
Ref.~\cite{CMS:2013mxa}.

The present analysis improves on the earlier result~\cite{Aad:2013wqa} with the following
changes:
\begin{inparaenum}[\itshape a\upshape)]
\item the electron identification uses a multivariate likelihood instead of a cut-based method,
  improving the background rejection at a fixed efficiency;
\item the electron transverse energy (\et ) measurement has been improved by a refined cluster energy
  reconstruction in the calorimeter and by combining the electron cluster energy with the track
  momentum for low-\et\ electrons;
\item the energy scale for electrons and momentum scale for muons have both been improved;
\item the inclusion of final-state radiation (FSR) off charged leptons has been extended to
  noncollinear photons;
\item a multivariate discriminant against the \zzstar\ background has been introduced to improve the
  signal-to-background ratio for the ggF production mode;
\item the estimates of the reducible $\ell\ell+\rm{jets}$ and $t\bar{t}$ background processes have
  been improved;
\item the sensitivity for different production modes has been improved, both by introducing a new VH
  category with two jets in the final state and by using multivariate techniques for this category
  and the VBF category.
\end{inparaenum}

The ATLAS detector is briefly described in \secref{sec:ATLASDetector}, and the signal and background
simulation is presented in \secref{sec:SandBSim}.  The object reconstruction and identification, the
event selection and categorization, and the background estimation are presented in Secs.\
\ref{sec:objRecId}, \ref{sec:eventSelection} and \ref{sec:Background}, respectively.  The
multivariate discriminants and the signal and background modeling are discussed in Secs.\ \ref{sec:MVA} and
\ref{sec:SignalModel}.  Finally, the systematic uncertainties and the results are presented in
Secs.\ \ref{sec:systematics} and \ref{sec:results}.

\vspace{10 mm}
\section{The ATLAS Detector}
\label{sec:ATLASDetector}
The ATLAS detector~\cite{Aad:2008zzm} is a multipurpose particle detector with approximately
forward-backward symmetric cylindrical geometry.\footnote{The ATLAS experiment uses a right-handed
coordinate system with its origin at the nominal interaction point (IP) in the center of the
detector and the $z$-axis along the beam pipe. The $x$-axis points from the IP to the center of the
LHC ring, and the $y$-axis points upward. Cylindrical coordinates $(r,\phi)$ are used in the
transverse plane, $\phi$ being the azimuthal angle around the beam pipe. The pseudorapidity is
defined in terms of the polar angle $\theta$ as $\eta=-\ln\tan(\theta/2)$.} The inner tracking
detector (ID) consists of a silicon pixel detector, which is closest to the interaction point, and a
silicon microstrip detector surrounding the pixel detector, both covering $|\eta|<2.5$, followed by
a transition radiation straw-tube tracker (TRT) covering $|\eta|< 2$.  The ID is surrounded by a
thin superconducting solenoid providing a $2\:\rm{T}$ axial magnetic field.  A highly segmented
lead/liquid-argon (LAr) sampling electromagnetic calorimeter measures the energy and the position of
electromagnetic showers with $\left| \eta \right|<3.2$.  The LAr calorimeter includes a presampler
(for $|\eta|<1.8$) and three sampling layers, longitudinal in shower depth, for $|\eta|<2.5$.  LAr
sampling calorimeters are also used to measure hadronic showers in the endcaps
($1.5<\left| \eta \right|<3.2$) and electromagnetic and hadronic showers in the forward
($3.1<\left|\eta \right|<4.9$) regions, while an iron/scintillator tile calorimeter measures
hadronic showers in the central region ($\left| \eta \right|<1.7$).


The muon spectrometer (MS) surrounds the calorimeters and is designed to detect muons in the
pseudorapidity range up to $\left| \eta \right|= 2.7$. The MS consists of one barrel
($\left| \eta \right|<1.05$) and two endcap regions. A system of three large superconducting
air-core toroid magnets, each with eight coils, provides a magnetic field with a bending integral of
about 2.5 T$\cdot$m in the barrel and up to 6 T$\cdot$m in the endcaps.  Monitored drift-tube
chambers in both the barrel and endcap regions and cathode strip chambers covering $|\eta|>2$ are
used as precision chambers, whereas resistive plate chambers in the barrel and thin gap chambers in
the endcaps are used as trigger chambers, covering up to $|\eta|=2.4$. The chambers are arranged in
three layers, so high-\pt\ particles traverse at least three stations with a lever arm of several
meters.

A three-level trigger system selects events to be recorded for offline analysis.

\section{Signal and Background Simulation}
\label{sec:SandBSim}

  
The \htollllbrief{} signal is modeled using the \progname{Powheg-Box} Monte Carlo (MC) event
generator~\cite{powheg1,powheg2,powheg3,powheg4,powheg5}, which provides separate calculations for
the ggF and VBF production mechanisms with matrix elements up to next-to-leading order (NLO) in the
QCD coupling constant. The description of the Higgs boson transverse momentum (\pt) spectrum in the
ggF process is reweighted to follow the calculation of Refs.~\cite{deFlorian:2012mx,
  Grazzini:2013mca}, which includes QCD corrections up to next-to-next-to-leading order (NNLO) and
QCD soft-gluon resummations up to next-to-next-to-leading logarithm (NNLL). The effects of nonzero
quark masses are also taken into account~\cite{Bagnaschi:2011tu}.  \progname{Powheg-Box} is
interfaced to \progname{Pythia8.1}~\cite{pythia,pythia81} for showering and hadronization, which in
turn is interfaced to \progname{Photos}~\cite{Golonka:2005pn,Davidson:2010ew} for QED radiative
corrections in the final state. \progname{Pythia8.1} is used to simulate the production of a Higgs
boson in association with a $W$ or a $Z$~boson (VH) or with a $t\bar{t}$ pair ($t\bar{t}H$).  The
production of a Higgs boson in association with a $b\bar{b}$ pair ($b\bar{b}H$) is included in the
signal yield assuming the same $m_{H}$ dependence as for the $t\bar{t}H$ process, while the signal
efficiency is assumed to be equal to that for ggF production.

The Higgs boson production cross sections and decay branching ratios, as well as their
uncertainties, are taken from
Refs.~\cite{LHCHiggsCrossSectionWorkingGroup:2011ti,LHCHiggsCrossSectionWorkingGroup:2012vm}. The
cross sections for the ggF process have been calculated to
NLO~\cite{Djouadi:1991tka,Dawson:1990zj,Spira:1995rr} and
NNLO~\cite{Harlander:2002wh,Anastasiou:2002yz,Ravindran:2003um} in QCD.  In addition, QCD soft-gluon
resummations calculated in the NNLL approximation are applied for the ggF
process~\cite{Catani:2003zt}. NLO electroweak (EW) radiative corrections are also
applied~\cite{Aglietti:2004nj,Actis:2008ug}.  These results are compiled in
Refs.~\cite{deFlorian:2012yg,Anastasiou:2012hx,Baglio:2010ae} assuming factorization between QCD and
EW corrections. For the VBF process, full QCD and EW corrections up to
NLO~\cite{Ciccolini:2007jr,Ciccolini:2007ec,Arnold:2008rz} and approximate NNLO
QCD~\cite{Bolzoni:2010xr} corrections are used to calculate the cross section.
The cross sections for the associated $WH/ZH$ production processes are calculated at
NLO~\cite{Han:1991ia} and at NNLO~\cite{Brein:2003wg} in QCD, and NLO EW radiative corrections are
applied~\cite{Ciccolini:2003jy}.  The cross section for associated Higgs boson production with a
$t\bar{t}$ pair is calculated at NLO in
QCD~\cite{Beenakker:2001rj,Beenakker:2002nc,Dawson:2002tg,Dawson:2003zu}.  The cross section for the
$b\bar{b}H$ process is calculated in the four-flavor scheme at NLO in
QCD~\cite{Dawson:2003kb,Dittmaier:2003ej,Dawson:2005vi} and in the five-flavor scheme at
NNLO in QCD~\cite{Harlander:2003ai} and combined via the Santander matching
scheme~\cite{Harlander:2011aa,LHCHiggsCrossSectionWorkingGroup:2012vm}.

The Higgs boson decay widths for the $WW$ and $ZZ$ four-lepton final states are provided by
\progname{Prophecy4f}~\cite{Bredenstein:2006rh,Bredenstein:2006ha}, which includes the complete NLO
QCD+EW corrections and interference effects between identical final-state fermions.  The other Higgs
boson decay widths, e.g.\ $\gamma\gamma$, $\tau\tau$, $b\bar{b}$, etc., are obtained with
\progname{Hdecay}~\cite{Djouadi:1997yw} and combined with the \progname{Prophecy4f} results to
obtain the \htollllbrief{} branching ratios.  Table~\ref{tab:MCSignal} gives the production cross
sections and branching ratios for \htollllbrief{}, which are used to normalize the signal
simulation, for several values of $m_H$.
     
The QCD scale uncertainties for $m_H = 125$ \gev~\cite{LHCHiggsCrossSectionWorkingGroup:2011ti}
amount to +7\% and $-$8\% for the ggF process, from $\pm1$\%\ to $\pm2$\%\ for the VBF and
associated $WH/ZH$ production processes and +4\% and $-$9\% for the associated $t\bar{t}H$
production process. The uncertainties on the production cross section due to uncertainties on the
parton distribution functions (PDF) and the strong coupling constant, $\alpha_{s}$, is $\pm 8\%$ for
gluon-initiated processes and $\pm 4\%$ for quark-initiated processes, estimated by following the
prescription in Ref.~\cite{Botje:2011sn} and by using the PDF sets of CTEQ~\cite{Lai:2010vv},
MSTW~\cite{Martin:2009iq} and NNPDF~\cite{Ball:2011mu}.  The PDF uncertainties are assumed to be
100\%\ correlated among processes with identical initial states, regardless of whether they are
signal or background~\cite{atlasCmsCombinationProc}.

\begin{table*}[!hb]
 \centering
 \footnotesize
 \caption{Calculated SM Higgs boson production cross sections for gluon fusion, vector-boson fusion
   and associated production with a $W$ or $Z$ boson or with a $b\bar{b}$ or $t\bar{t}$ pair in $pp$
   collisions at $\sqrt{s}$ of 7 \tev\ and 8 \tev~\cite{LHCHiggsCrossSectionWorkingGroup:2011ti}.
   The quoted uncertainties correspond to the total theoretical systematic uncertainties calculated
   by adding in quadrature the QCD scale and PDF$+\alpha_s$ uncertainties.  The decay branching
   ratio ($B$) for $H \rightarrow 4\ell$ with $\ell = e, \mu$, is reported in the last
   column~\cite{LHCHiggsCrossSectionWorkingGroup:2011ti}. \label{tab:MCSignal}}
 \vspace{0.1cm}
 \begin{tabular}{rcccccc}
   \hline \hline
   \noalign{\vspace{0.05cm}}
   $m_{H}$ & $\sigma\left(gg\to H\right)$ & $\sigma\left(qq'\to Hqq'\right)$ & $\sigma\left(q\bar{q}\to WH\right)$ &
   $\sigma\left(q\bar{q}\to ZH\right)$   & $\sigma\left(q\bar{q}/gg\to b\bar{b}H/t\bar{t}H\right)$ & $B$ (\htollllbrief)   \\ 
   $[{\rm GeV}]$ & $[{\rm pb}]$ & $[{\rm pb}]$ & $[{\rm pb}]$ & $[{\rm pb}]$ & $[{\rm pb}]$ & $[10^{-3}]$\\
   \noalign{\vspace{0.05cm}}
   \hline\hline
   \multicolumn{7}{c}{\tabscript{\sqrt{s}=7\,\tev}{}{}}\\
   \hline\hline
   \noalign{\vspace{0.05cm}}
   123\rule[-1mm]{0mm}{4.7mm} & $15.6 \pm 1.6$   & $1.25 \pm 0.03$   & $0.61 \pm 0.02$      & $0.35 \pm 0.01 $ & $0.26 \pm 0.04$ & $0.103\pm 0.005$  \\
   125\rule[-1mm]{0mm}{4.7mm} & $15.1 \pm 1.6$                       & $1.22 \pm 0.03$      & $0.58 \pm 0.02 $                      & $0.34  \pm 0.01 $     & $0.24 \pm 0.04$        & $0.125\pm 0.005$  \\

   127\rule[-1mm]{0mm}{4.7mm} & $14.7 \pm 1.5$                       & $1.20 \pm 0.03$      & $0.55 \pm 0.02$                       & $0.32  \pm 0.01 $     & $0.23 \pm 0.03$       & $0.148\pm 0.006$  \\
   \noalign{\vspace{0.05cm}}
   \hline\hline
   \multicolumn{7}{c}{\tabscript{\sqrt{s}=8\,\tev}{}{}}\\
   \hline\hline
   \noalign{\vspace{0.05cm}}
   123\rule[-1mm]{0mm}{4.7mm} & $19.9 \pm 2.1$                               & $1.61 \pm 0.05$                       & $0.74  \pm  0.02$                     & $0.44 \pm 0.02$               & $0.35 \pm 0.05$  & $0.103\pm 0.005$   \\             
   125\rule[-1mm]{0mm}{4.7mm} & $19.3 \pm 2.0$                               & $1.58 \pm 0.04$        & $0.70  \pm  0.02$                     & $0.42 \pm 0.02$               & $0.33 \pm 0.05$                       &  $0.125\pm 0.005$  \\
   127\rule[-1mm]{0mm}{4.7mm} & $18.7 \pm 1.9$  & $1.55 \pm 0.04$                       & $0.67 \pm 0.02$        & $0.40 \pm 0.02$               & $0.32 \pm 0.05$        &  $0.148\pm 0.006$        \\             
   \noalign{\vspace{0.05cm}}
   \hline\hline
 \end{tabular}
\end{table*}

The \zzstar\ continuum background is modeled using \progname{Powheg-Box}~\cite{Melia:2011tj} for
quark--antiquark annihilation and \progname{gg2ZZ}~\cite{Binoth:2008pr} for gluon fusion. The
PDF+$\alpha_s$ and QCD scale uncertainties are parametrized as functions of $m_{4\ell}$ as
recommended in Ref.~\cite{LHCHiggsCrossSectionWorkingGroup:2012vm}. For the \zzstar\ background at
$m_{4\ell} = 125$ \gev, the quark-initiated (gluon-initiated) processes have a QCD scale uncertainty
of $\pm 5\%$ ($\pm 25\%)$, and $\pm 4\%$ ($\pm 8\%$) for the PDF and $\alpha_{s}$ uncertainties,
respectively.

The $Z+$jets production is modeled using \progname{Alpgen}~\cite{Mangano:2002ea} and is divided into
two sources: $Z+$light-jets, which includes $Zc\bar{c}$ in the massless $c$-quark approximation and
$Zb\bar{b}$ with $b\bar{b}$ from parton showers, and $Zb\bar{b}$ using matrix-element calculations
that take into account the $b$-quark mass.  The MLM~\cite{Mangano:2006rw} matching scheme is used to
remove any double counting of identical jets produced via the matrix-element calculation and the
parton shower, but this scheme is not implemented for $b$-jets.  Therefore, $b\bar{b}$ pairs with
separation $\Delta R\equiv\sqrt{\left(\Delta\phi\right)^2+\left(\Delta\eta\right)^2} > 0.4$ between
the $b$-quarks are taken from the matrix-element calculation, whereas for $\Delta R<0.4$ the
parton-shower $b\bar{b}$~pairs are used.  In this search the $Z+\rm jets$ background is normalized
using control samples from data.  For comparison between data and simulation, the NNLO QCD
\progname{FEWZ}~\cite{Melnikov:2006kv,Anastasiou:2003ds} and NLO QCD
\progname{MCFM}~\cite{Campbell:2010ff,Campbell:2000bg} cross-section calculations are used to
normalize the simulations for inclusive $Z$ boson and $Zb\bar{b}$ production, respectively. The
$t\bar{t}$ background is modeled using \progname{Powheg-Box} interfaced to \progname{Pythia8.1} for
parton shower and hadronization, \progname{photos} for QED radiative corrections and
\progname{tauola}~\cite{Jadach:1993hs,Golonka:2003xt} for the simulation of $\tau$ lepton
decays. \progname{Sherpa}~\cite{Gleisberg:2008ta} is used for the simulation of $WZ$ production.

Generated events are processed through the ATLAS detector simulation~\cite{simuAtlas} within the
\progname{Geant4} framework~\cite{GEANT4}.  Additional $pp$ interactions in the same and nearby
bunch crossings (pileup) are included in the simulation.  The simulation samples are weighted to
reproduce the observed distribution of the mean number of interactions per bunch crossing in the
data.

\section{Object Reconstruction and Identification} 
\label{sec:objRecId}

The \htollllbrief\ channel has a small rate but is a relatively clean final state where the
signal-to-background ratio {\it vis-\`{a}-vis} the reducible backgrounds alone, i.e. ignoring the
\zzstar\ background, is above 6 for the present analysis.  Significant effort was made to obtain a
high efficiency for the reconstruction and identification of electrons and muons, while keeping the
loss due to background rejection as small as possible.  In particular, this becomes increasingly
difficult for electrons as \et\ decreases.

Electrons are reconstructed using information from the ID and the electromagnetic calorimeter.  For
electrons, background discrimination relies on the shower shape information available from the
highly segmented LAr EM calorimeter, high-threshold TRT hits, as well as compatibility of the
tracking and calorimeter information.  Muons are reconstructed as tracks in the ID and MS, and their
identification is primarily based on the presence of a matching track or tag in the MS.  Finally,
jets are reconstructed from clusters of calorimeter cells and calibrated using a dedicated scheme
designed to adjust the energy measured in the calorimeter to that of the true jet energy on average.

\subsection{Electron reconstruction and identification}
\label{sec:ele}

Electron candidates are clusters of energy deposited in the electromagnetic calorimeter associated
with ID tracks~\cite{Aad:2011mk,ref:run1-egamma-calib}.  All candidate electron tracks are fitted
using a Gaussian-sum filter~\cite{GSFConf} (GSF) to account for bremsstrahlung energy losses.  The
GSF fit brings the candidate electron $E/p$ distribution closer to unity and improves the measured
electron direction, resulting in better impact parameter resolution.  For the 2012 (8\,\tev) data
set, the electron reconstruction was modified to allow for large bremsstrahlung energy losses.  A
second pass was added to the ATLAS track pattern recognition that allows for an electron hypothesis
with larger energy loss to be tried after a first pass with a pion hypothesis.  Furthermore, the
track-to-cluster matching algorithm was improved, for example by incorporating an additional test
that extrapolates tracks to the calorimeter using the measured cluster energy rather than the track
momentum.  These improvements increased the electron reconstruction efficiency on average by 5\% for
electrons with \et\ above 15 \gev, with a 7\% improvement for \et\ at 15 \gev, as measured with
data~\cite{ElectronEff2012}.

The electron identification is based on criteria that require the longitudinal and transverse shower
profiles to be consistent with those expected for electromagnetic showers, the track and cluster
positions to match in \eta\ and $\phi$, and the presence of high-threshold TRT hits.  To maintain
both large acceptance and good discrimination, the selection is kept ``loose'' for a large number of
discriminating variables; for comparison, the most stringent electron identification would induce an
additional 15\% reduction in electron efficiency. Compared to the previous
measurement~\cite{Aad:2013wqa}, the electron identification was improved for the 2012 data set by
moving from a cut-based method to a likelihood method. The likelihood allows the inclusion of
discriminating variables that are difficult to use with explicit cuts without incurring significant
efficiency losses.  For example, the GSF fit measures a significant difference between the momenta
at the start and end of the electron trajectory for only a fraction of true electrons so that
requiring a large difference for all electrons would not be an efficient selection cut.  The
likelihood improves the rejection of light-flavor jets and photon conversions by a factor of two for
the same signal efficiency.  For the 2011 (7\,\tev) data set, the electron reconstruction proceeds
as described above, but without the improved pattern recognition and cluster-to-track matching. The
electron identification used for the 2011 data set is the same cut-based identification as in the
previous measurement~\cite{Aad:2013wqa}.  Detailed descriptions of the likelihood identification
used for the 2012 data set, the cut-based identification used for the 2011 data set and the
corresponding efficiency measurements can be found in Refs.~\cite{ElectronEff2012, ElectronEff2011}.

Finally, the electron transverse energy is computed from the cluster energy and the track direction
at the interaction point.  The cluster energy is the sum of the calibrated energy deposited in the
cells in a fixed-size window in $\eta \times \phi$, different for the barrel and end-cap.  The
cluster energy is corrected for energy lost before the calorimeter, deposited in neighboring cells
and beyond the calorimeter. Further corrections for the response dependence are applied as a
function of the impact point within the central cluster cell.  The cluster energy measurement was
improved compared to the previous analysis~\cite{Aad:2013wqa} and is described
elsewhere~\cite{ref:run1-egamma-calib}.  Several of the steps in the energy calibration were
significantly improved including:
\begin{inparaenum}[\itshape a\upshape)]
\item the addition of a multivariate technique to extract the cluster energy from the energy deposit
  in simulation,
\item additional corrections for response details not included in the simulation, and
\item equalization of the energy scales of the longitudinal calorimeter layers.
\end{inparaenum}
These improvements resulted in a significant reduction in the overall energy scale uncertainty (for
example for $|\eta| < 1.37$ the uncertainty is reduced from 0.4\% to 0.04\% for electrons of \et\ =
40 \gev~\cite{ref:run1-egamma-calib}) and have an important impact on the systematic uncertainty of
the Higgs boson mass measurement~\cite{combmasspaper}.  In addition, a combined fit of the cluster
energy and track momentum is applied to electrons with \et\ below 30 \gev\ when the cluster \et\ and
the track \pt\ agree within their uncertainties.  The combined fit improves the resolution of
$m_{4\ell}$ for the $4e$ and $ 2\mu 2e$ final states by about 4\%.

\subsection{Muon reconstruction and identification}
\label{sec:muon}

Four types of muon candidates are distinguished, depending on how they are reconstructed.  Most muon
candidates are identified by matching a reconstructed ID track with either a complete or partial
track reconstructed in the MS~\cite{ATLAS-CONF-2013-088,MCPpaper2014}.  If a complete MS track is
present, the two independent momentum measurements are combined (combined muons), otherwise the
momentum is measured using the ID, and the partial MS track serves as identification (segment-tagged
muons).  The muon reconstruction and identification coverage is extended by using tracks
reconstructed in the forward region ($2.5 < |\eta| < 2.7$) of the MS, which is outside the ID
coverage (standalone muons). In the center of the barrel region ($|\eta|< 0.1$), which lacks MS
geometrical coverage, ID tracks with \pt $ > 15$ \gev\ are identified as muons if their calorimetric
energy deposition is consistent with a minimum ionizing particle (calorimeter-tagged muons).  The
inner detector tracks associated with muons that are identified inside the ID acceptance are
required to have a minimum number of associated hits in each of the ID subdetectors to ensure good
track reconstruction.  The muon candidates outside the ID acceptance that are reconstructed only in
the MS are required to have hits in each of the three stations they traverse.  At most one
standalone or calorimeter-tagged muon is used per event.

\subsection{Final-state radiation  recovery}
\label{sec:fsr}

The QED process of radiative photon production in $Z$ decays is well modeled by simulation.  Some of
the FSR photons can be identified in the calorimeter and incorporated into the four-lepton
measurement.  A dedicated method to include the FSR photons in the reconstruction of $Z$ bosons was
developed. This method includes a search for collinear and noncollinear FSR photons, with the
collinear search described in Ref.~\cite{ATLAS-CONF-2012-143}.  Collinear photons are only
associated with muons\footnote{Photons collinear to electrons are included in the calorimeter
  shower.}  ($\Delta R_{\mathrm{cluster},\mu}\leq0.15$), and noncollinear photons can be associated
with either muons or electrons ($\Delta R_{\mathrm{cluster},\ell}>0.15$).

At most one FSR photon is used per event, with priority given to collinear photons.  The probability
of having more than one FSR per event with significant energy is negligible.  The collinear photons
are required to have a transverse energy of \et\ $>$ 1.5 \gev\ and a fraction of the total energy
deposited in the front sampling layer of the calorimeter greater than 0.1.  If more than one
collinear photon is found, only the one with the highest \et\ is kept.  noncollinear photons must
have \et\ $>10$ \gev, be isolated (\et\ below 4 \gev\ within a cone of size $\Delta R = 0.4$,
excluding the photon itself), and satisfy strict (``tight'') identification
criteria~\cite{ATLAS:2012ana}.  Again, only the highest-\et\ noncollinear photon is retained, and
only if no collinear photon is found.

The inclusion of an FSR photon in a four-lepton event is discussed below
in~\secref{sec:inclusiveAnalysis}.  The collinear FSR selection recovers 70\% of the FSR photons
within the selected fiducial region with a purity of about 85\%, where misidentified photons come
from pileup and muon ionization. The noncollinear FSR selection has an efficiency of approximately
60\% and a purity greater than 95\% within the fiducial region.

\begin{figure}[!htb]
  \centering 
    \subfigure[\label{fig:CollinearFSR}]{\includegraphics[width=\doublePlotSize]{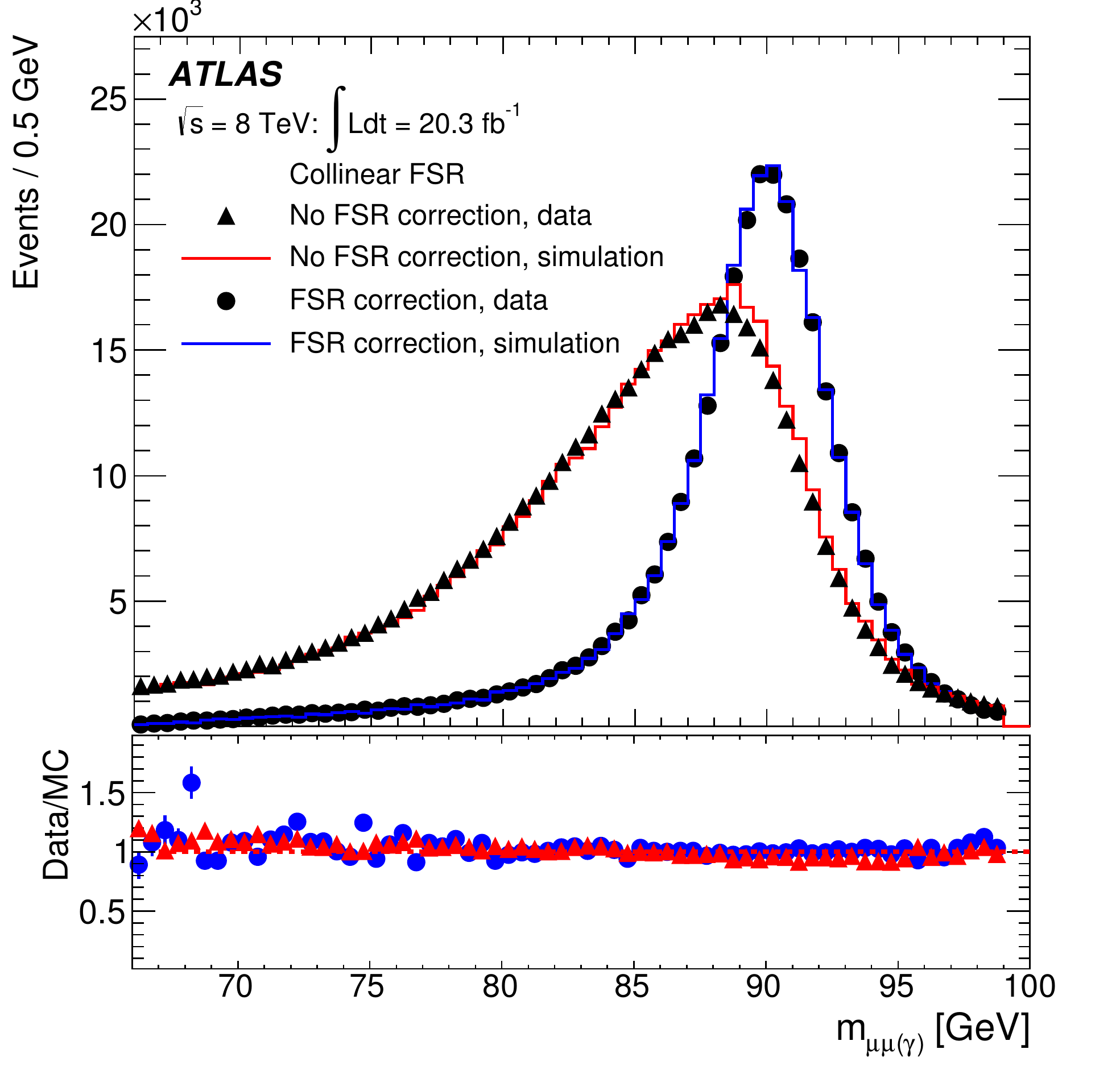}}
    \subfigure[\label{fig:NonCollinearFSR}]{ \includegraphics[width=\doublePlotSize]{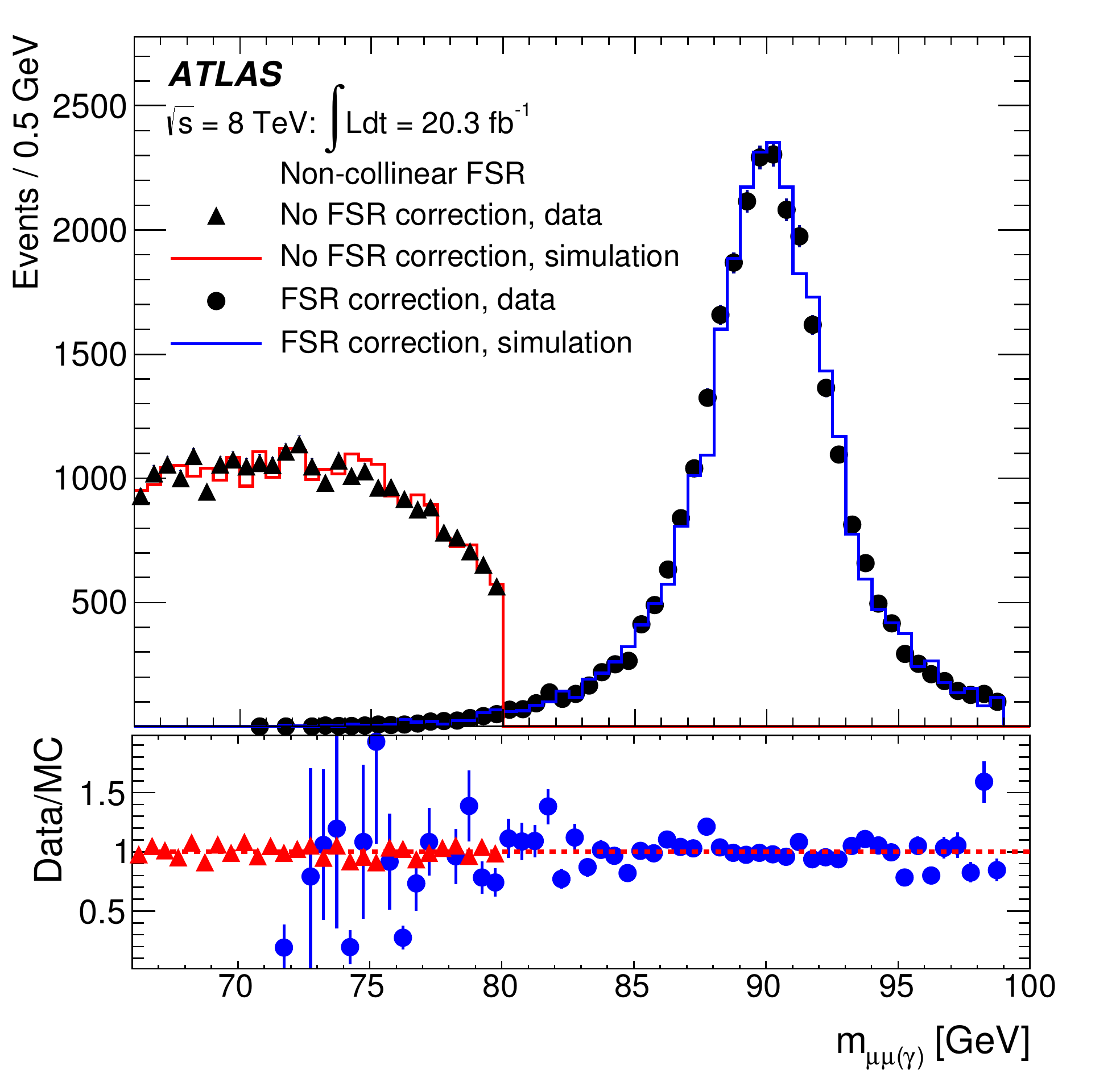}}
    \caption{\subref{fig:CollinearFSR} The invariant mass distributions of $Z\rightarrow
      \mu^+\mu^-(\gamma)$ events in data before collinear FSR correction (filled triangles) and
      after collinear FSR correction (filled circles), for events with a collinear FSR photon
      satisfying the selection criteria as described in Sec.~\ref{sec:fsr}.  The prediction of the
      simulation is shown before correction (red histogram) and after correction (blue histogram).
      \subref{fig:NonCollinearFSR} The invariant mass distributions of $Z\rightarrow
      \mu^+\mu^-(\gamma)$ events with a noncollinear FSR photon satisfying the selection criteria
      as described in Sec.~\ref{sec:fsr}.  The prediction of the simulation is shown before
      correction (red histogram) and after correction (blue histogram).
      \label{fig:FSRZMassCorrection}}
\end{figure}

In Fig.~\ref{fig:FSRZMassCorrection}, the invariant mass distributions are shown for
$Z\rightarrow\mumu$ candidate events where either a collinear Fig.~\ref{fig:CollinearFSR} or
noncollinear Fig.~\ref{fig:NonCollinearFSR} FSR photon is found.  The invariant mass distributions
are shown both before and after the addition of the FSR photons, for both data and simulation.  Good
agreement between data and simulation is observed.

\subsection{Jet reconstruction}
\label{sec:jet}
Jets are reconstructed using the anti-$k_t$ algorithm~\cite{Cacciari:2005hq,Cacciari:2008gp} with a
distance parameter $R=0.4$. The inputs to the reconstruction are three-dimensional clusters of
energy~\cite{Lampl:1099735,Aad:2011he} in the calorimeter, calibrated to the electromagnetic energy
scale and corrected for contributions from in-time and out-of-time pileup~\cite{ATLAS:2012lla}, and
the position of the primary interaction vertex (see \secref{sec:eventSelection}). The algorithm for
this clustering suppresses noise by keeping only cells with a significant energy deposit and their
neighboring cells.  Subsequently, the jets are calibrated to the hadronic energy scale using $\pt$-
and $\eta$-dependent correction factors determined from simulation (2011 data set) and from data
(2012 data set)~\cite{Aad:2011he,Aad:2014bia}. The uncertainty on these correction factors is
determined from control samples in data. To reduce the number of jet candidates originating from
pileup vertices, jets with $\pt<50$ \gev\ within the ID acceptance ($|\eta|<2.4$) are required to
have more than 50\% (75\% for 2011 data) of the summed scalar \pt\ of the tracks associated with the
jet (within $\Delta R =0.4$ around the jet axis) come from tracks of the primary
vertex~\cite{TheATLAScollaboration:2013pia}.

\section{Event Selection}
\label{sec:eventSelection}

The data are subjected to quality requirements: if any relevant detector component is not operating
correctly during a period when an event is recorded, the event is rejected.  Events are required to
have at least one vertex with three associated tracks with \pt\ $>$ 400 \mev, and the primary vertex
is chosen to be the reconstructed vertex with the largest track $\sum \pt^2$.  Identical
requirements are applied to all four-lepton final states.  For the inclusive analysis, four-lepton
events are selected and classified according to their channel: $4\mu$, $2e2\mu$, $2\mu2e$,
$4e$. These events are subsequently categorized according to their production mechanism to provide
measurements of each corresponding signal strength.

\subsection{Inclusive analysis}
\label{sec:inclusiveAnalysis}

Four-lepton events were selected with single-lepton and dilepton triggers.  The \pt\ (\et)
thresholds for single-muon (single-electron) triggers increased from 18 \gev\ to 24 \gev\ (20
\gev\ to 24 \gev) between the 7 \tev\ and 8 \tev\ data, in order to cope with the increasing
instantaneous luminosity. The dilepton trigger thresholds for 7 \tev\ data are set at 10
\gev\ \pt\ for muons, 12 \gev\ \et\ for electrons and (6,10) \gev\ for (muon,electron) mixed-flavor
pairs.  For the 8 \tev\ data, the thresholds were raised to 13 \gev\ for the dimuon trigger, to 12
\gev\ for the dielectron trigger and (8,12) \gev\ for the (muon,electron) trigger; furthermore, a
dimuon trigger with different thresholds on the muon \pt, 8 and 18 \gev, was added.  The trigger
efficiency for events passing the final selection is above 97\% in the $4\mu$, $2\mu2e$ and $2e2\mu$
channels and close to 100\% in the $4e$ channel for both 7 \tev\ and 8 \tev\ data.

Higgs boson candidates are formed by selecting two same-flavor, opposite-sign lepton pairs (a lepton
quadruplet) in an event.  Each lepton is required to have a longitudinal impact parameter less than
10 mm with respect to the primary vertex, and muons are required to have a transverse impact
parameter of less than 1 mm to reject cosmic-ray muons.  These selections are not applied to
standalone muons, that have no ID track.  Each electron (muon) must satisfy $\et>7$ \gev\ ($\pt>6$
\gev) and be measured in the pseudorapidity range $\left|\eta \right|<2.47$
($\left|\eta\right|<2.7$). The highest-\pt\ lepton in the quadruplet must satisfy $\pt >20$ \gev,
and the second (third) lepton in $\pt$ order must satisfy $\pt>15$ \gev\ ($\pt>10$ \gev).  Each
event is required to have the triggering lepton(s) matched to one or two of the selected leptons.

Multiple quadruplets within a single event are possible: for four muons or four electrons there are
two ways to pair the masses, and for five or more leptons there are multiple ways to choose the
leptons.  Quadruplet selection is done separately in each subchannel: $4\mu$, $2e2\mu$, $2\mu2e$,
$4e$, keeping only a single quadruplet per channel.  For each channel, the lepton pair with the mass
closest to the $Z$ boson mass is referred to as the leading dilepton and its invariant mass,
$m_{12}$, is required to be between 50 \gev\ and 106 \gev. The second, subleading, pair of each
channel is chosen from the remaining leptons as the pair closest in mass to the $Z$ boson and in the
range $m_{\rm{min}}< m_{34} < 115$ \gev, where $m_{\rm{min}}$ is 12 \gev\ for $m_{4\ell}<$ 140 \gev,
rises linearly to 50 \gev\ at $m_{4\ell}=$ 190 \gev\, and then remains at 50 \gev\ for $m_{4\ell} >
$ 190 \gev.  Finally, if more than one channel has a quadruplet passing the selection, the channel
with the highest expected signal rate is kept, i.e. in the order: $4\mu$, $2e2\mu$, $2\mu2e$, $4e$.
The rate of two quadruplets in one event is below the per mille level. 

Events with a selected quadruplet are required to have their leptons a distance $\Delta R>0.1$ from
each other if they are of the same flavor and $\Delta R>0.2$ otherwise.  For $4\mu$ and $4e$ events,
if an opposite-charge same-flavor dilepton pair is found with $m_{\ell\ell}$ below 5 \gev\ the event
is removed.

The $Z+\rm{jets}$ and $t\bar{t}$ background contributions are further reduced by applying impact
parameter requirements as well as track- and calorimeter-based isolation requirements to the
leptons. The transverse impact parameter significance, defined as the impact parameter in the transverse plane
divided by its uncertainty, $|d_0|/\sigma_{d_0}$, for all muons (electrons) is required to be lower
than 3.5 (6.5).  The normalized track isolation discriminant, defined as the sum of the transverse
momenta of tracks, inside a cone of size $\Delta R=0.2$ around the lepton, excluding the lepton
track, divided by the lepton \pt, is required to be smaller than 0.15.

The relative calorimetric isolation for electrons in the 2012 data set is computed as the sum of the
cluster transverse energies $\et$, in the electromagnetic and hadronic calorimeters, with a
reconstructed barycenter inside a cone of size $\Delta R=0.2$ around the candidate electron cluster,
divided by the electron $\et$.  The electron relative calorimetric isolation is required to be
smaller than 0.2. The cells within $0.125 \times 0.175$ in $\eta \times \phi$ around the electron
barycenter are excluded.  The pileup and underlying event contribution to the calorimeter isolation
is subtracted event by event~\cite{Cacciari:2007fd}.  The calorimetric isolation of electrons in the
2011 data set is cell-based (electromagnetic and hadronic calorimeters) rather than cluster-based,
and the calorimeter isolation relative to the electron \et\ requirement is 0.3 instead of 0.2.  In
the case of muons, the relative calorimetric isolation discriminant is defined as the sum, $\Sigma
\et$, of the calorimeter cells above $3.4\sigma$, where $\sigma$ is the quadrature sum of the
expected electronic and pileup noise, inside a cone of size $\Delta R<0.2$ around the muon
direction, divided by the muon $\pt$.  Muons are required to have a relative calorimetric isolation
less than 0.3 (0.15 in case of standalone muons).  For both the track- and calorimeter-based
isolations any contributions arising from other leptons of the quadruplet are subtracted.

As discussed in \secref{sec:fsr}, a search is performed for FSR photons arising from any of the
lepton candidates in the final quadruplet, and at most one FSR photon candidate is added to the
4$\ell$ system. The FSR correction is applied only to the leading dilepton, and priority is given to
collinear photons.  The correction is applied if $ 66 <m_{\mu\mu} < 89$ \gev\ and $m_{\mu\mu\gamma}
< 100$ \gev. If the collinear-photon search fails then the noncollinear FSR photon with the
highest \et\ is added, provided it satisfies the following requirements: $m_{\ell\ell} < 81$ \gev\
and $m_{\ell\ell\gamma}<100$ \gev.  The expected fraction of collinear (noncollinear) corrected
events is 4\% (1\%).

For the 7 \tev\ data, the combined signal reconstruction and selection efficiency for $\mH=125$ \gev\ is
$39$\% for the $4\mu$ channel, $25$\% for the $2e2\mu$/$2\mu2e$ channels and $17$\% for the $4e$
channel. The improvements in the electron reconstruction and identification for the 8 \tev\ data lead
to increases in these efficiencies by 10\%--15\% for the channels with electrons, bringing their
efficiencies to $27$\% for the $2e2\mu$/$2\mu2e$ channels and $20$\% for the $4e$ channel.

After the FSR correction, the lepton four-momenta of the leading dilepton are recomputed by means of
a $Z$-mass-constrained kinematic fit.  The fit uses a Breit-Wigner $Z$ line shape and a single
Gaussian to model the lepton momentum response function with the Gaussian $\sigma$ set to the
expected resolution for each lepton.  The $Z$-mass constraint improves the $m_{4\ell}$ resolution by
about 15\%.  More complex momentum response functions were compared to the single Gaussian and found
to have only minimal improvement for the $m_{4\ell}$ resolution.

Events satisfying the above criteria are considered candidate signal events for the inclusive
analysis, defining a signal region independent of the value of $m_{4\ell}$.

\subsection{Event categorization}
\label{sec:eventCategorisation}

\begin{figure}[h]
  \centering
  \includegraphics[width=\doublePlotSize]{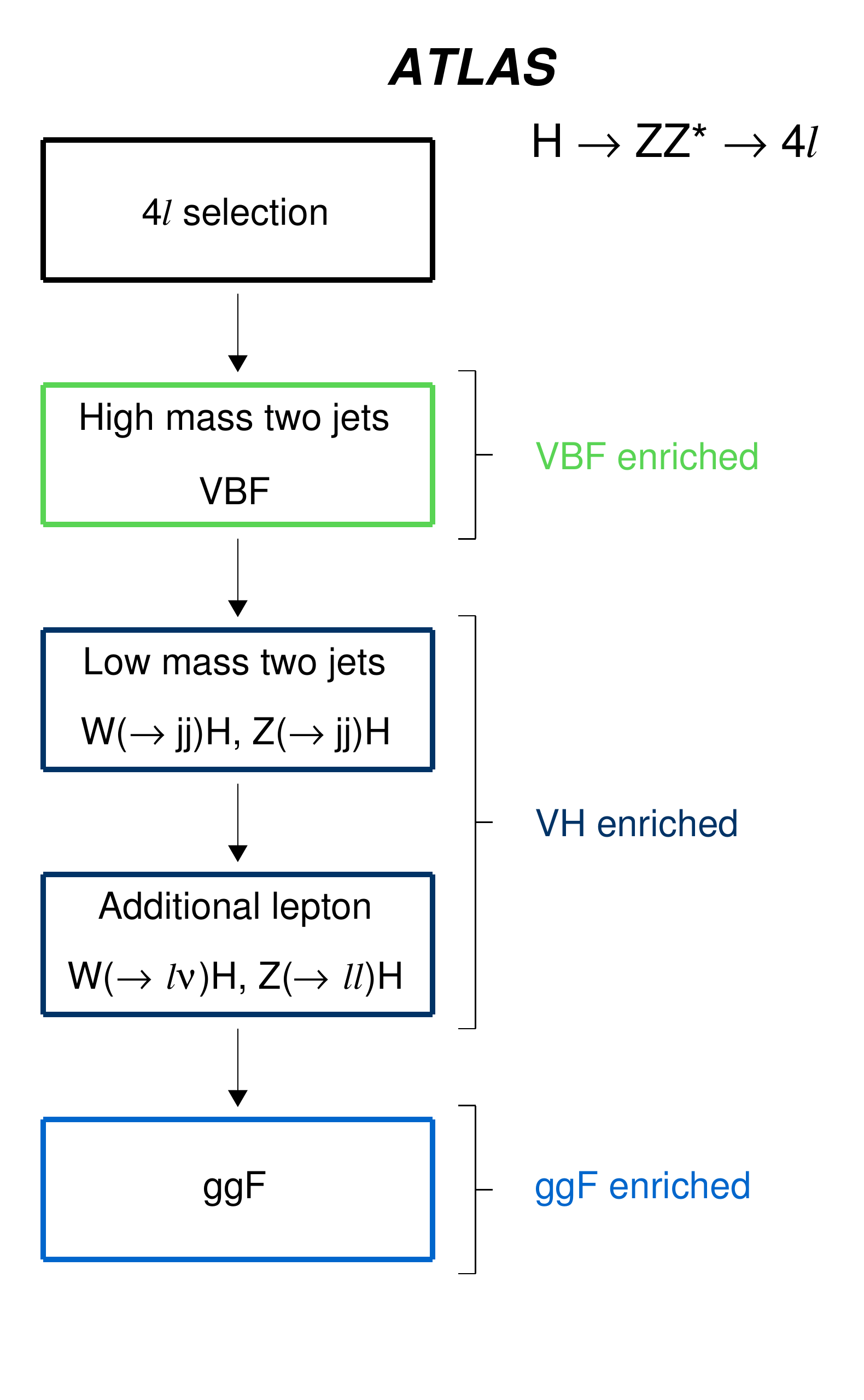}
  \caption{Schematic view of the event categorization.  Events are required to pass the four-lepton
    selection, and then they are assigned to one of four categories which are tested sequentially:
    \vbfcat, \vhhadcat, \vhlepcat, or \ggfcat.  \label{fig:cat_scheme}}
\end{figure}

To measure the rates for the ggF, VBF, and VH production mechanisms, discussed in
\secref{sec:SandBSim}, each $H\rightarrow 4\ell$ candidate selected by the criteria described above
is assigned to one of four categories (\vbfcat, \vhhadcat, \vhlepcat, or \ggfcat), depending on
other event characteristics. A schematic view of the event categorization is shown in
\figref{fig:cat_scheme}.

The \vbfcat\ category is defined by events with two high-\pt\ jets. The kinematic requirements for
jets are $\pt >$ 25 (30)\,\gev\ for $|\eta| < 2.5$ ($2.5<|\eta| < 4.5$). If more than two jets
fulfill these requirements, the two highest-\pt\ jets are selected as VBF jets.  The event is
assigned to the \vbfcat\ category if the invariant mass of the dijet system, $m_{jj}$, is greater
than 130 \gev, leading to a signal efficiency of approximately 55\%.  This category has a
considerable contamination from ggF events, with 54\% of the expected events in this category
arising from production via gluon fusion.

\begin{figure}[h]
  \centering
  \includegraphics[width=\singlePlotSize]{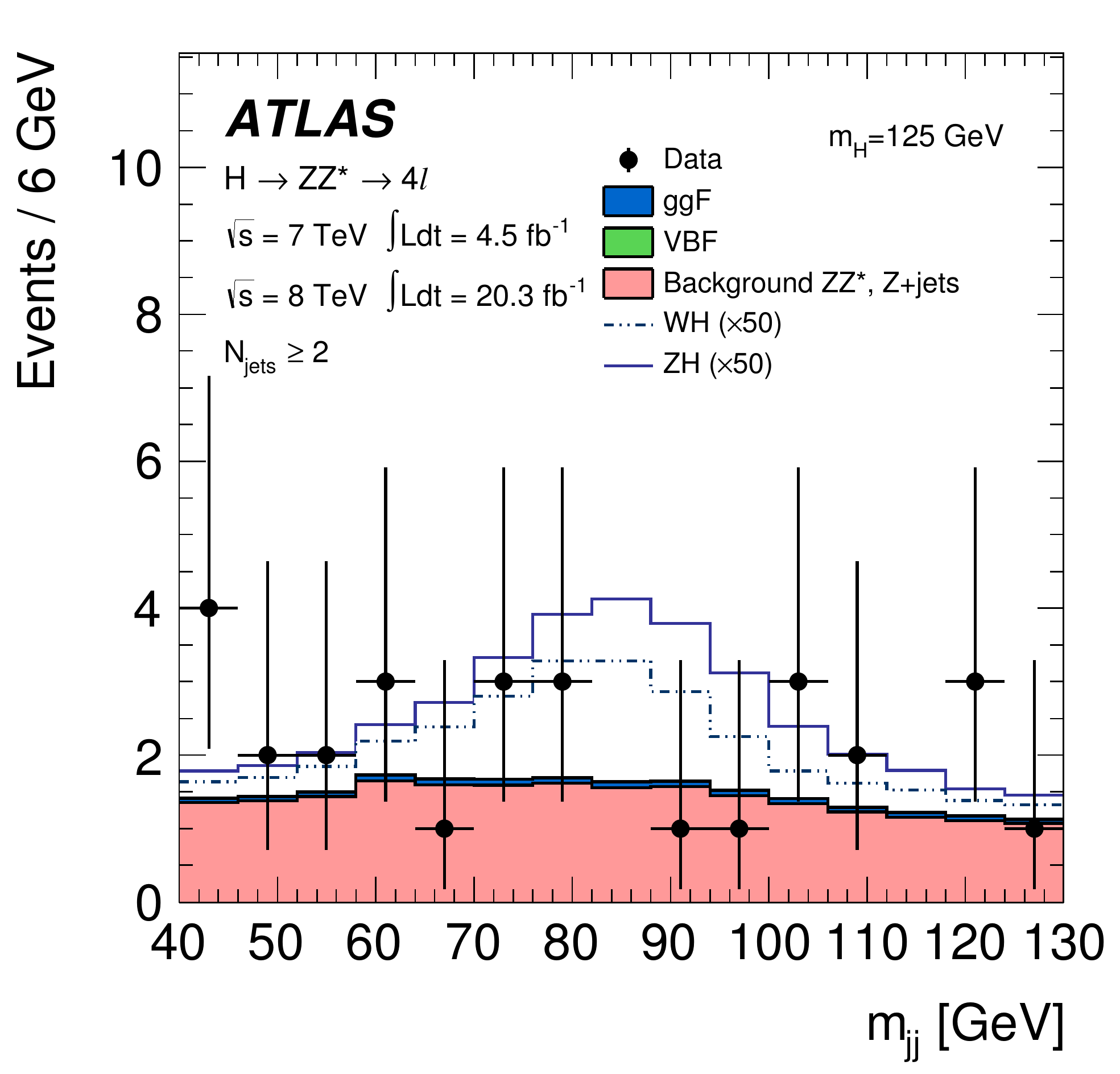}
  \caption{Distributions of the dijet invariant mass for the events with at least two jets for the
    data (filled circles), the expected signal (solid and dot-dot-dashed histograms) and the
    backgrounds (filled histograms).  The $WH$ and $ZH$ hadronic signals are scaled by a factor 50
    and the $ZH$ distribution is added on top of the $WH$ distribution.\label{fig:mjj_noBDT}}
\end{figure}

Events that do not satisfy the \vbfcat\ criteria are considered for the \vhhadcat\ category. The
same jet-related requirements are applied but with 40 $<m_{jj}<$ 130 \gev, as presented in
\figref{fig:mjj_noBDT}. Moreover, the candidate has to fulfill a requirement on the output weight of
a specific multivariate discriminant, presented in \secref{sec:BDTCat}. The signal efficiency for
requiring two jets is 48\% for VH and applying the multivariate discriminant brings the overall
signal efficiency to 25\%.

Events failing to satisfy the above criteria are next considered for the \vhlepcat\ category.  Events are
assigned to this category if there is an extra lepton ($e$ or $\mu$), in addition to the four leptons
forming the Higgs boson candidate, with $\pt >$ 8\,\gev\ and satisfying the same lepton
requirements.  The signal efficiency for the extra vector boson for the \vhlepcat\ category is
around 90\% (100\%) for the $W$ ($Z$), where the $Z$ has two leptons which can pass the extra lepton
selection.

Finally, events that are not assigned to any of the above categories categories are associated with
the \ggfcat\ category. Table~\ref{tab:categoryexpected} shows the expected yields for Higgs boson
production and \zzstar\ background events in each category from each of the production mechanisms,
for $\mH$ = 125 \gev\ and \lumia\ at $\sqrt{s}=7$ \tev\ and \lumib\ at $\sqrt{s}=8$ \tev.

\begin{table*}[hbt!]
  \centering
  \caption{The expected number of events in each category (\ggfcat, \vbfcat, \vhhadcat\ and
    \vhlepcat), after all analysis criteria are applied, for each signal production mechanism
    (ggF/$b\bar{b}H/t\bar{t}H$, VBF, VH) at $\mH =$125 \gev, for \lumia\ at $\sqrt{s}=7$ \tev\ and
    \lumib\ at $\sqrt{s}=8$ \tev. The requirement $m_{4\ell}>110$ \gev\ is
    applied.  \label{tab:categoryexpected}}
  \begin{tabular}{lccc}
    \hline\hline
    Category         & $gg \to H, q\bar{q}/gg \to b\bar{b}H/t\bar{t}H$ &  $qq'\to Hqq'$ & $q\bar{q}\to W/ZH$ \\\hline
    \multicolumn{4}{c}{$\sqrt{s}=7$ \tev}  \\
    \hline		  
    \ggfcat          & $2.06 \ \pm\ 0.25$  & $0.114  \ \pm\ 0.005$  & $0.067 \ \pm\ 0.003$   \\
    \vbfcat          & $0.13 \ \pm\ 0.04$  & $0.137  \ \pm\ 0.009$  & $0.015 \ \pm\ 0.001$   \\
    \vhhadcat        & $0.053\ \pm\ 0.018$ & $0.007  \ \pm\ 0.001$  & $0.038 \ \pm\ 0.002$   \\
    \vhlepcat        & $0.005\ \pm\ 0.001$ & $0.0007 \ \pm\ 0.0001$ & $0.023 \ \pm\ 0.002$   \\
    \hline
    \multicolumn{4}{c}{$\sqrt{s}=8$ \tev}  \\
    \hline		  
    \ggfcat          & $12.0  \ \pm\ 1.4$   & $0.52 \ \pm\ 0.02$    & $0.37 \ \pm\ 0.02$     \\
    \vbfcat          & $ 1.2  \ \pm\ 0.4$   & $0.69 \ \pm\ 0.05$    & $0.10 \ \pm\ 0.01$     \\
    \vhhadcat        & $ 0.41 \ \pm\ 0.14$  & $0.030\ \pm\ 0.004$   & $0.21 \ \pm\ 0.01$     \\
    \vhlepcat        & $0.021 \ \pm\ 0.003$ & $0.0009\ \pm\ 0.0002$ & $0.13 \ \pm\ 0.01$     \\
    \hline
  \end{tabular}
\end{table*}


\section{Background Estimation}  
\label{sec:Background}

The rate of the \zzstar\ background is estimated using simulation normalized to the SM cross section
as described in \secref{sec:SandBSim}, while the rate and composition of the reducible
$\ell\ell+\rm{jets}$ and $t\bar{t}$ background processes are evaluated with data-driven methods.
The composition of the reducible backgrounds depends on the flavor of the subleading dilepton pair,
and different approaches are taken for the $\ell\ell+\mu\mu$ and the $\ell\ell+ee$ final states.
These two cases are discussed in Secs.\ \ref{sec:llmumu} and \ref{sec:llee}, respectively, and the
yields for all reducible backgrounds in the signal region are summarized in Tables~\ref{tab:fitSR}
and~\ref{tab:bkg_overviewllee}. Finally, the small contribution from the $WZ$ reducible background
is estimated from simulation. The background estimation follows the methods previously described in
Refs.~\cite{ATLAS:2012af, ATLAS-CONF-2013-013} with several improvements and additional
cross-checks.


\subsection{$\ell\ell+\mu\mu$ background} 
\label{sec:llmumu}

The $\ell\ell+\mu\mu$ reducible background arises from $Z+{\rm jets}$ and $t\bar{t}$ processes,
where the $Z+{\rm jets}$ contribution has a $Zb\bar{b}$ heavy-flavor quark component in which the
heavy-flavor quarks decay semileptonically, and a component arising from $Z$ + light-flavor jets
with subsequent $\pi$/\kaon~in-flight decays.  The number of background events from $Z+{\rm jets}$
and $t\bar{t}$ production is estimated from an unbinned maximum likelihood fit, performed
simultaneously to four orthogonal control regions, each of them providing information on one or more
of the background components. The fit results are expressed in terms of yields in a reference
control region, defined by applying the analysis event selection except for the isolation and impact
parameter requirements to the subleading dilepton pair. The reference control region is also used
for the validation of the estimates. Finally, the background estimates in the reference control
region are extrapolated to the signal region.

The control regions used in the maximum likelihood fit are designed to minimize contamination from
the Higgs boson signal and the \zzstar\ background.  The four control regions are:

\begin{inparaenum}[\itshape a\upshape)]
\item {\em Inverted requirement on impact parameter significance.} Candidates are selected following
  the analysis event selection, but \begin{inparaenum}[\itshape 1\upshape)] \item without applying
    the isolation requirement to the muons of the subleading dilepton and \item requiring that at
    least one of the two muons fails the impact parameter significance requirement. \end{inparaenum}
  As a result, this control region is enriched in $Zb\bar{b}$ and $t\bar{t}$ events.

\item {\em Inverted requirement on isolation.}  Candidates are selected following the analysis event
  selection, but requiring that at least one of the muons of the subleading dilepton fails the
  isolation requirement. As a result, this control region is enriched in $Z$ + light-flavor-jet
  events ($\pi/K$ in-flight decays) and $t\bar{t}$ events.

\item {\em $e\mu$ leading dilepton ($e\mu+\mu\mu$).} Candidates are selected following the analysis
  event selection, but requiring the leading dilepton to be an electron-muon pair. Moreover, the
  isolation and impact parameter requirements are not applied to the muons of the subleading
  dilepton, which are also allowed to have the same or opposite charge sign. Events containing a
  $Z$-boson candidate decaying into $e^+e^-$ or $\mu^+\mu^-$ pairs are removed with a requirement on
  the mass. This control region is dominated by $t\bar{t}$ events.

\item {\em Same-sign subleading dilepton.} The analysis event selection is applied, but for the
  subleading dilepton neither isolation nor impact parameter significance requirements are applied
  and the leptons are required to have the same charge sign (SS).  This same-sign control region is
  not dominated by a specific background; all the reducible backgrounds have a significant
  contribution.
\end{inparaenum}

The expected composition for each control region is shown in \tabref{tab:muCRcompFitSim}. The
uncertainties on the relative yields between the control regions and the reference control region
are introduced in the maximum likelihood fit as nuisance parameters.  The residual contribution from
\zzstar\ and the contribution from $WZ$ production, where--contrary to the $Z+{\rm jets}$ and
$t\bar{t}$ backgrounds--only one of the leptons in the subleading dilepton is expected to be a
nonisolated backgroundlike muon, are estimated for each control region from simulation.

\begin{table*}[h]
  \centering
  \caption{Expected contribution of the $\ell\ell+\mu\mu$ background sources in each of the control
    regions.\label{tab:muCRcompFitSim}}
  \vspace{0.2cm}

  \setlength{\tabcolsep}{12pt}

  \begin{tabular}{lcccc}
    \hline\hline
    &\multicolumn{4}{c}{Control region}\\
    Background              & Inverted $d_0$        & Inverted isolation          & $e\mu+\mu\mu$     & Same-sign      \\
    \noalign{\vspace{0.05cm}}
    \hline                                                                                   
    \noalign{\vspace{0.05cm}}
    $Zb\bar{b}$             &  $32.8 \pm 0.5$\%  &  $26.5 \pm 1.2$\%  &  $0.3 \pm 1.2$\%    & $30.6 \pm 0.7$\% \\
    $Z$ + light-flavor jets &  $9.2 \pm 1.3$\%   &  $39.3 \pm 2.6$\%  &  $0.0 \pm 0.8$\%    & $16.9 \pm 1.6$\% \\
    $t\bar{t}$              &  $58.0 \pm 0.9$\%  &  $34.2 \pm 1.6$\%  &  $99.7 \pm 1.0$\%   & $52.5 \pm 1.1$\% \\
    \noalign{\vspace{0.05cm}}
    \hline\hline
  \end{tabular}
\end{table*}

\begin{figure*}
  \centering
  \subfigure[\label{fig:OSad0smuSimFit}]{\includegraphics[width=\doublePlotSize]{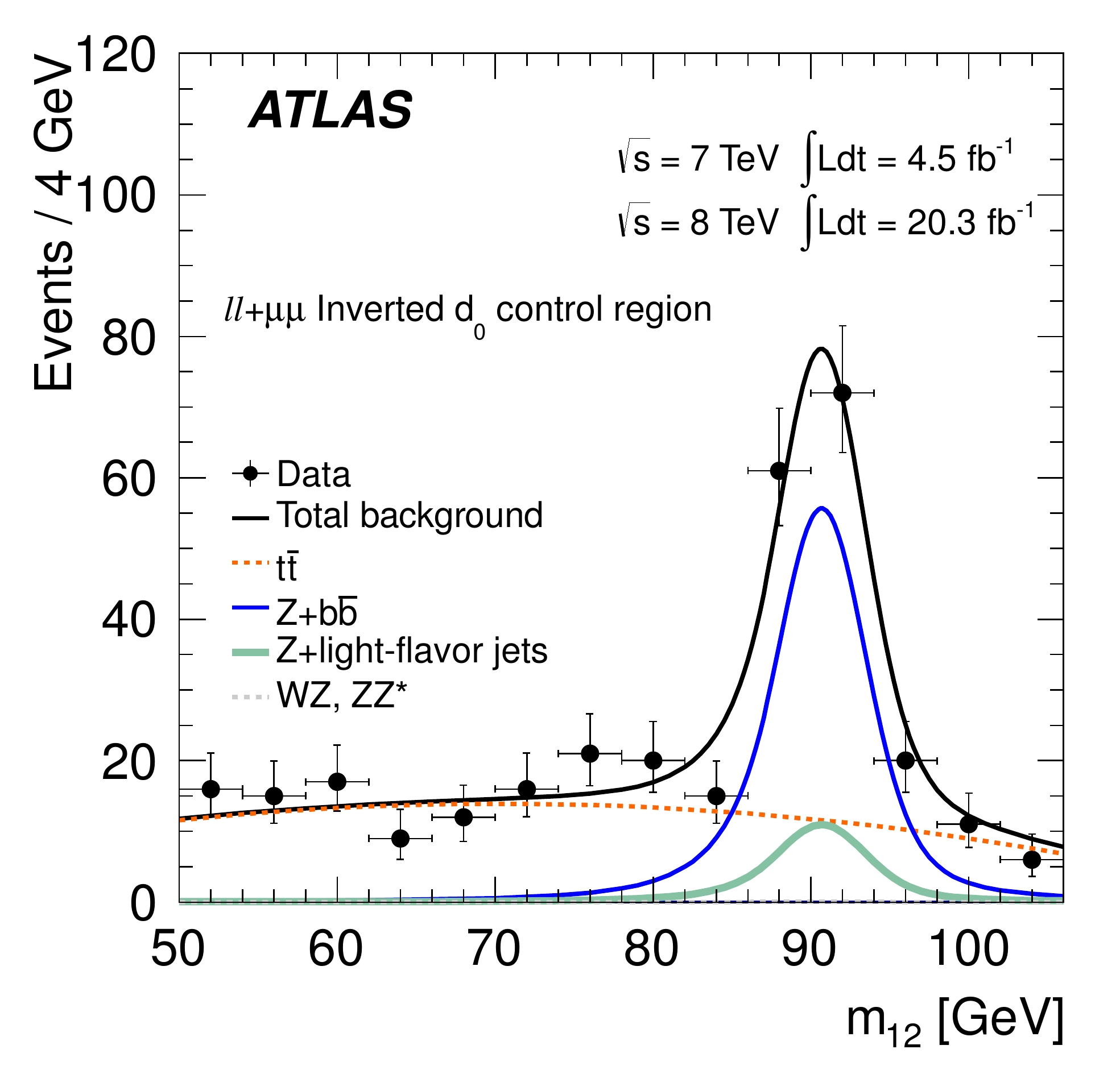}}
  \subfigure[\label{fig:OSd0aisoSimFit}]{\includegraphics[width=\doublePlotSize]{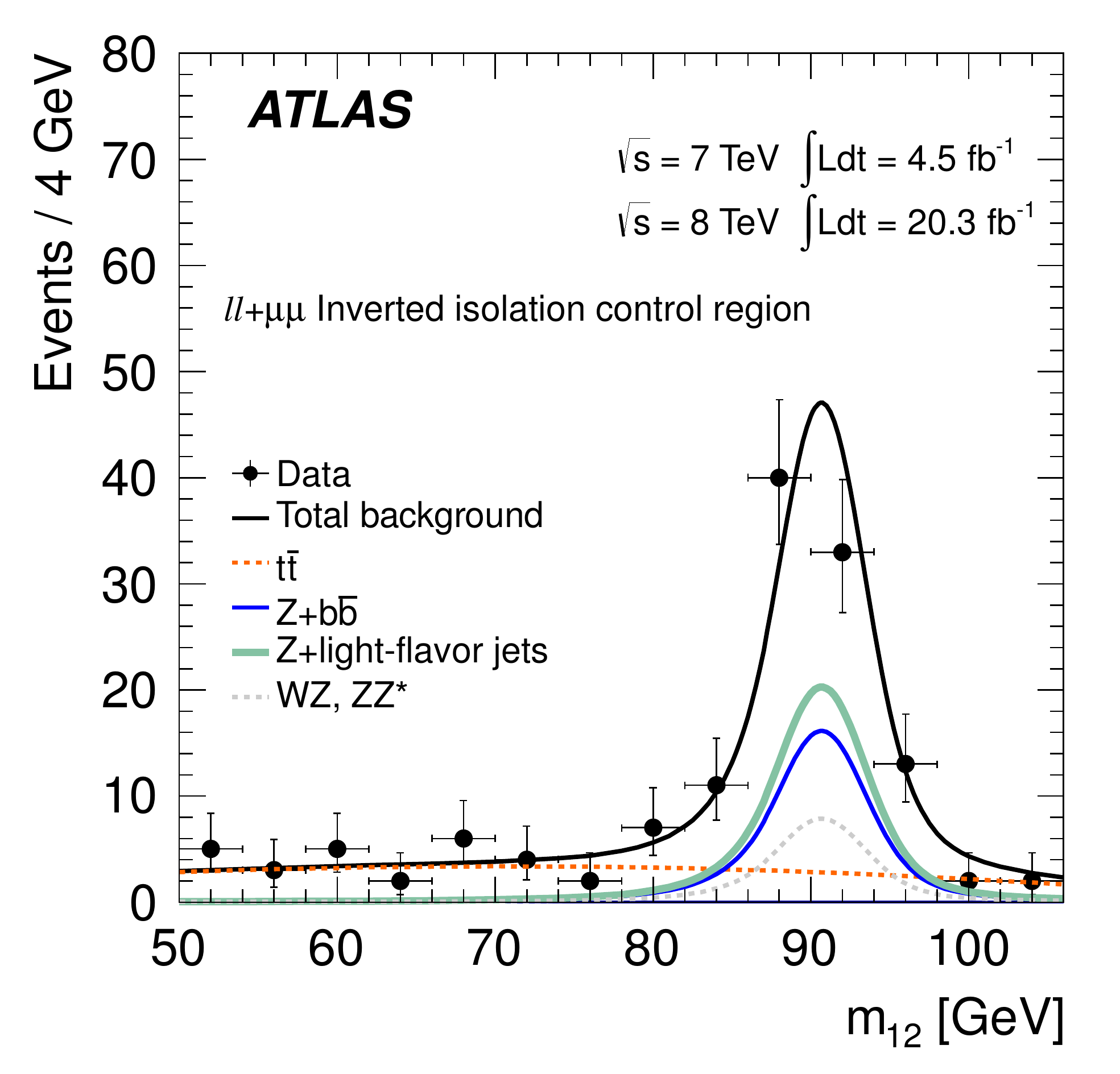}}
  \subfigure[\label{fig:emuSimFit}]{\includegraphics[width=\doublePlotSize]{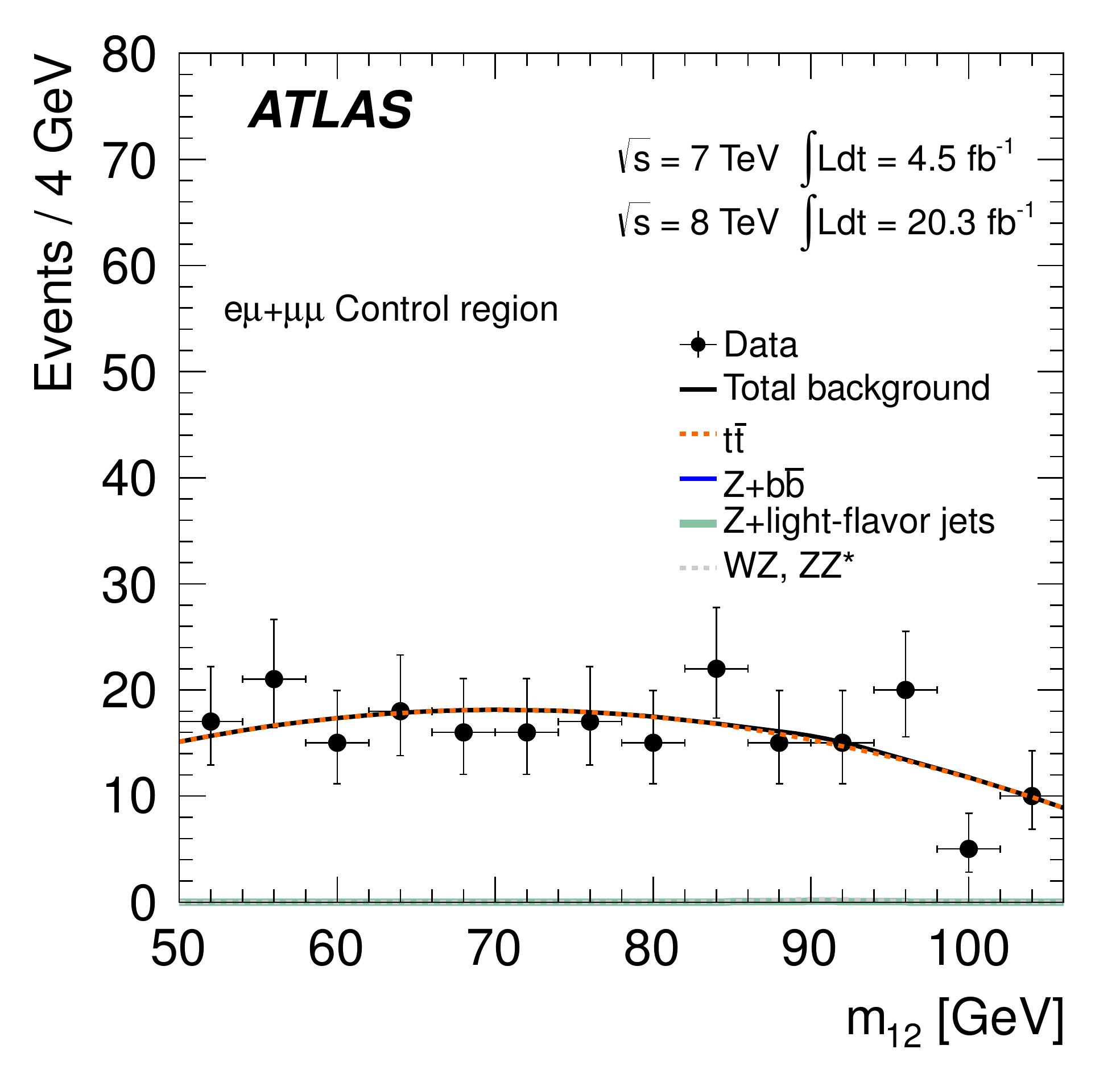}}
  \subfigure[\label{fig:SSSimFit}]{\includegraphics[width=\doublePlotSize]{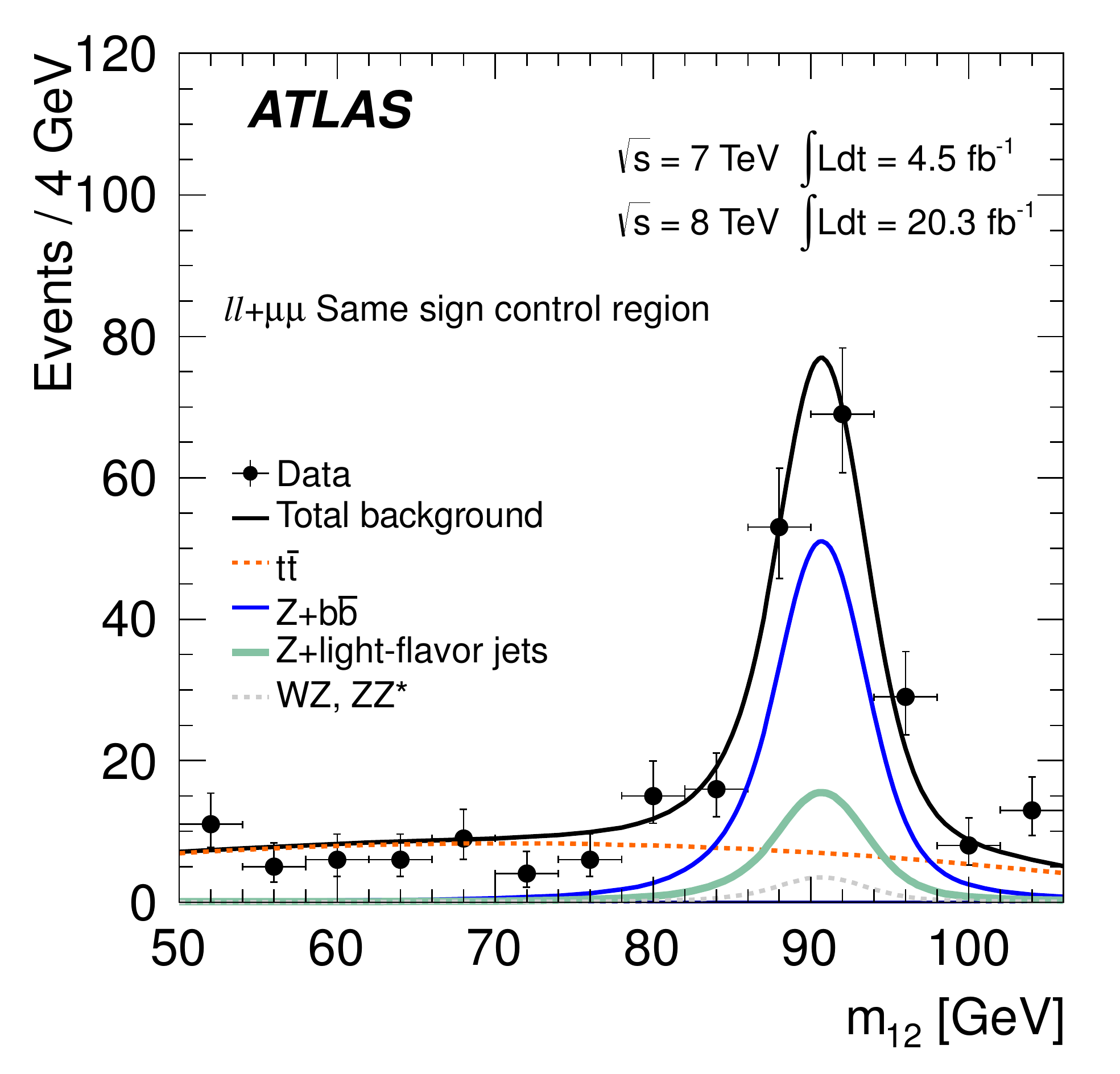}}
  \caption{The observed $m_{12}$ distributions (filled circles) and the results of the maximum
    likelihood fit are presented for the four control regions: \subref{fig:OSad0smuSimFit} inverted
    requirement on impact parameter significance, \subref{fig:OSd0aisoSimFit} inverted requirement
    on isolation, \subref{fig:emuSimFit} $e\mu$ leading dilepton, where the backgrounds besides
    $t\bar{t}$ are small and not visible, and \subref{fig:SSSimFit} same-sign subleading
    dilepton. The fit results are shown for the total background (black line) as well as the
    individual components: $Z+{\rm jets}$ decomposed into $Z+b\bar{b}$ (blue line) and
    $Z+$light-flavor jets (green line), $t\bar{t}$ (dashed red line), and the combined $WZ$ and $ZZ$
    (dashed gray line), where the $WZ$ and $ZZ$ contributions are estimated from
    simulation.\label{fig:muSimFit}}
\end{figure*}

In all the control regions, the observable is the mass of the leading dilepton, $m_{12}$, which
peaks at the $Z$ mass for the resonant ($Z+{\rm jets}$) component and has a broad distribution for
the nonresonant ($t\bar{t}$) component. For the $t\bar{t}$ component the $m_{12}$ distribution is
modeled by a second-order Chebyshev polynomial, while for the $Z+{\rm jets}$ component it is modeled
using a convolution of a Breit-Wigner distribution with a Crystal Ball function.  The shape
parameters are derived from simulation. In the combined fit, the shape parameters are constrained to
be the same in each of the control regions, and are allowed to fluctuate within the uncertainties
obtained from simulation.  The results of the combined fit in the four control regions are shown in
\figref{fig:muSimFit}, along with the individual background components, while the event yields in
the reference control region are summarized in \tabref{tab:fractocr12}. As a validation of the fit
method, the maximum likelihood fit is applied to the individual control regions yielding estimates
compatible to those of the combined fit; these are also summarized in \tabref{tab:fractocr12}.

\begin{table*}[h]
  \centering
  \caption{Data-driven $\ell\ell+\mu\mu$ background estimates for the $\sqrt{s}=7$ \tev\ and
    $\sqrt{s}=8$ \tev\ data, expressed as yields in the reference control region, for the combined
    fit and fits to the individual control regions. In the individual control regions only the total
    $Z$+jets contribution can be determined, while the $e\mu+\mu\mu$ control region is only
    sensitive to the $t\bar{t}$ background. The statistical uncertainties are also
    shown.\label{tab:fractocr12}}
  \begin{tabular}{lcccc}
    \noalign{\vspace{0.05cm}}
    \hline
    \hline
    \noalign{\vspace{0.05cm}}
    \multicolumn{5}{c}{Reducible background yields for $4\mu$ and $2e2\mu$ in reference control region}\\
    \noalign{\vspace{0.15cm}}
    Control region              & $Zb\bar{b} $   & $Z$ + light-flavor jets & Total $Z$ + jets & $t\bar{t}$ \\ 
    \noalign{\vspace{0.15cm}}
    \hline    
    \noalign{\vspace{0.15cm}}
    Combined fit                & $159 \pm 20$   & $49 \pm 10$             & $208 \pm 22$     & $210 \pm 12$ \\ 

    \multicolumn{5}{c}{\hdashrule[0.5ex]{15cm}{0.5pt}{2mm}} \\

    Inverted impact parameter   &                &                         & $206 \pm 18$     & $208 \pm 23$  \\ 
    Inverted isolation          &                &                         & $210 \pm 21$     & $201 \pm 24$  \\ 
    $e\mu+\mu\mu$               &                &                         & --               & $201 \pm 12$  \\
    Same-sign dilepton          &                &                         & $198 \pm 20$     & $196 \pm 22$  \\
    \noalign{\vspace{0.15cm}}
    \hline
    \hline
  \end{tabular}
\end{table*}

\begin{figure}[h]
  \centering \includegraphics[width=\singlePlotSize]{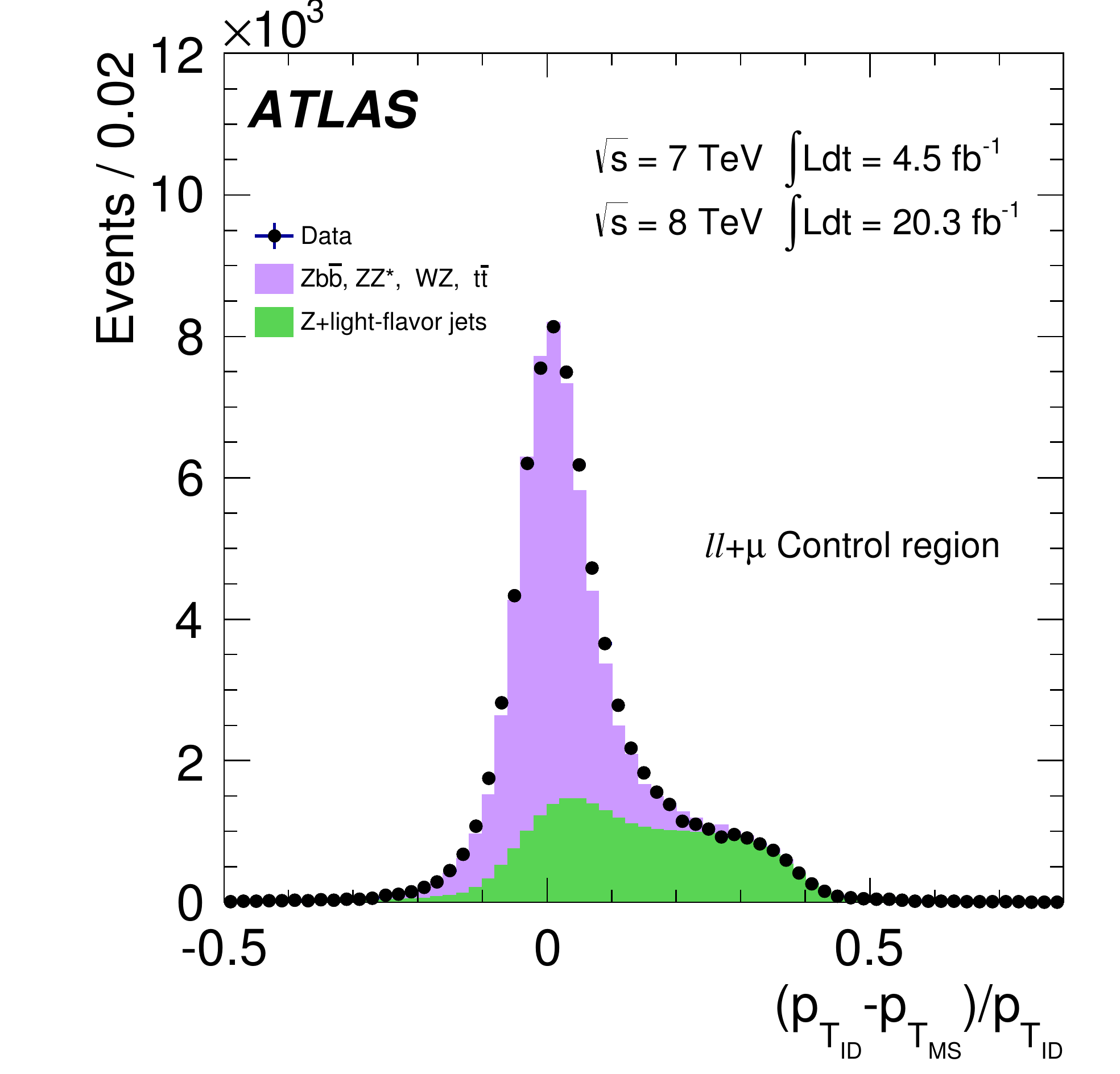} \caption{The
    distribution of the difference between the transverse momentum measured in the ID and in the MS
    normalized to the ID measurement, $(p_{\rm T_{ID}}-p_{\rm T_{MS}})/p_{\rm T_{ID}}$, for combined
    muons accompanying a $Z\to\ell\ell$ candidate.  The data (filled circles) are compared to the
    background simulation (filled histograms) which has the $Z+$ light-flavor background shown
    separately to distinguish the contribution from $\pi/K$ in-flight decays.  The additional muon
    is selected to be a combined muon with \pt$ > 6$ \gev, which fulfills the $\Delta R$ requirement
    for the lepton separation of the analysis and in the case of $Z (\to \mu^+ \mu^-) + \mu$ final
    state, the opposite sign pairs are required to have $m_{\mu^+\mu^-} > 5$ \gev\ to remove
    $J/\psi$ decays.  \label{fig:e123lprop_msid}}
\end{figure}

The estimated yields in the reference control region are extrapolated to the signal region by
multiplying each background component by the probability of satisfying the isolation and impact
parameter significance requirements, estimated from the relevant simulated sample. The systematic
uncertainty in these transfer factors, stemming mostly from the size of the simulated sample, is 6\%
for $Zb\bar{b}$, 60\% for $Z$ + light-flavor jets and 16\% for $t\bar{t}$. Furthermore, these
simulation-based efficiencies are validated with data using muons accompanying $Z\to\ell\ell$
candidates, where the leptons composing the $Z$ boson candidate are required to satisfy isolation
and impact parameter criteria. Events with four leptons, or with an opposite-sign dimuon with mass
less than 5 \gev, are excluded. Based on the data/simulation agreement of the efficiencies in this
control region an additional systematic uncertainty of $1.6\%$ is added.
\Figref{fig:e123lprop_msid} shows the relative difference between the ID and MS \pt\ measurements
for combined muons for a subset of the $Z+X$ control region where the $X$ represents a single
combined muon.  The contribution from $\pi/K$ in-flight decays is clearly visible and well described
by the simulation.

The reducible background estimates in the signal region are given in Table~\ref{tab:fitSR},
separately for the $\sqrt{s}=7\:\TeV$ and $8\:\TeV$ data. The uncertainties are separated into
statistical and systematic contributions, where in the latter the transfer factor uncertainty and
the fit systematic uncertainties are included.

\begin{table*}[h] 
  \centering
  \caption{Estimates for the $\ell\ell+\mu\mu$ background in the signal region for the full
    $m_{4\ell}$ mass range for the $\sqrt{s}=7$ \tev\ and $\sqrt{s}=8$ \tev\ data.  The
    $Z+\rm{jets}$ and $t\bar{t}$ background estimates are data-driven and the $WZ$ contribution is
    from simulation.  The decomposition of the $Z+\rm{jets}$ background in terms of the $Zb\bar{b}$
    and the $Z$ + light-flavor-jets contributions is also provided. \label{tab:fitSR}}
  \begin{tabular}{ccc} \noalign{\vspace{0.05cm}} \hline
    \hline \noalign{\vspace{0.05cm}}
    Background & $4\mu$ & $2e2\mu$ \\
    \hline
    \noalign{\vspace{0.05cm}}
    \multicolumn{3}{c}{$\sqrt{s}=7\:\TeV$}\\
    \noalign{\vspace{0.05cm}}
    \hline
    \noalign{\vspace{0.05cm}}
    $Z+\rm{jets}$                & $0.42 \pm 0.21\rm{ (stat)} \pm 0.08\rm{ (syst)} $ & $0.29 \pm 0.14\rm{ (stat)} \pm 0.05\rm{ (syst)}$ \\
    $t\bar{t}$                   & $0.081 \pm 0.016\rm{(stat)} \pm 0.021\rm{(syst)}$ & $0.056 \pm 0.011\rm{(stat)} \pm 0.015\rm{(syst)}$\\  
    $WZ$ expectation             & $0.08 \pm 0.05$                                   & $0.19 \pm 0.10$ \\
    \noalign{\vspace{0.05cm}}
    \multicolumn{3}{c}{\hdashrule[0.5ex]{10cm}{0.5pt}{2mm}} \\
    \noalign{\vspace{0.05cm}}
    \multicolumn{3}{c}{$Z+\rm{jets}$ decomposition} \\
    $\tabscript{Zb\bar{b}}{}{} $ & $0.36 \pm 0.19\rm{ (stat)} \pm 0.07\rm{ (syst)}$  & $0.25 \pm 0.13\rm{ (stat)} \pm 0.05\rm{ (syst)}$ \\
    $Z$ + light-flavor jets      & $0.06 \pm 0.08\rm{ (stat)} \pm 0.04\rm{ (syst)}$  & $0.04 \pm 0.06\rm{ (stat)} \pm 0.02\rm{ (syst)}$ \\
    \noalign{\vspace{0.05cm}}
    \hline
    \noalign{\vspace{0.05cm}}
    \multicolumn{3}{c}{$\sqrt{s}=8\:\TeV$}\\
    \noalign{\vspace{0.05cm}}
    \hline
    \noalign{\vspace{0.05cm}}
    $Z+\rm{jets}$                & $3.11 \pm 0.46\rm{ (stat)} \pm 0.43\rm{ (syst)} $ & $2.58 \pm 0.39\rm{ (stat)} \pm 0.43\rm{ (syst)}$ \\
    $t\bar{t}$                   & $0.51 \pm 0.03\rm{(stat)} \pm 0.09\rm{(syst)}$ & $0.48 \pm 0.03\rm{(stat)} \pm 0.08\rm{(syst)}$\\  
    $WZ$ expectation             & $0.42 \pm 0.07$                                   & $0.44 \pm 0.06$          \\
    \noalign{\vspace{0.05cm}}
    \multicolumn{3}{c}{\hdashrule[0.5ex]{10cm}{0.5pt}{2mm}} \\
    \noalign{\vspace{0.05cm}}
    \multicolumn{3}{c}{$Z+\rm{jets}$ decomposition} \\
    $\tabscript{Zb\bar{b}}{}{} $ & $2.30 \pm 0.26\rm{ (stat)} \pm 0.14\rm{ (syst)}$  & $2.01 \pm 0.23\rm{ (stat)} \pm 0.13\rm{ (syst)}$ \\
    $Z$ + light-flavor jets      & $0.81 \pm 0.38\rm{ (stat)} \pm 0.41\rm{ (syst)}$  & $0.57 \pm 0.31\rm{ (stat)} \pm 0.41\rm{ (syst)}$ \\
    \noalign{\vspace{0.05cm}}
    \hline
    \hline
  \end{tabular}
\end{table*}

\subsection{$\ell\ell+ee$ background} 
\label{sec:llee}

The background for subleading electron pairs arises from jets misidentified as electrons. The
background is classified into three distinct sources: light-flavor jets ($f$), photon conversions
($\gamma$) and heavy-flavor semileptonic decays ($q$). These sources are identified exactly in
simulated background events. In addition, corresponding data control regions are defined which are
enriched in events associated with each of these sources, thus allowing data-driven classification
of reconstructed events into matching categories. For the background estimation, two types of
control regions are defined:
\begin{itemize}
\item the first, denoted as $3\ell+X$, in which the identification requirements for the
  lower-\pt\ electron of the subleading pair are relaxed;
\item the second, denoted as $\ell\ell+XX$, which comes in two variants: one in which the
  identification requirements for both electrons of the subleading pair are relaxed, and another in
  which an inverted selection is applied to the subleading pair.
\end{itemize}
In both cases, the leading pair satisfies the complete event selection.  The final background
estimate is obtained from the $3\ell+X$ region, while the estimates from the $\ell\ell+XX$ region
are used as cross-checks.

The efficiencies needed to extrapolate the different background sources from the control regions
into the signal region are obtained separately for each of the $f$, $\gamma$, $q$ background
sources, in \pt\ and \eta\ bins, from simulation. These simulation-based efficiencies are corrected
to correspond to the efficiency measured in data using a third type of control region, denoted as
$Z+X$, enhanced for each $X$ component. The $Z+X$ control region has a leading lepton pair,
compatible with the decay of a $Z$ boson, passing the full event selection and an additional object
($X$) that satisfies the relaxed identification for the specific control region to be extrapolated.
The $Z+X$ data sample is significantly larger than the background control data samples.  For all of
the methods, the extrapolation from the background control region, $3\ell+X$ or $\ell\ell+XX$, to
the signal region cannot be done directly with the efficiencies from the $Z+X$ data control region
due to differences in the fractions of $f$, $\gamma$, $q$ for the $X$ of the two control regions.
In the following, the $q$ contribution in the simulation is increased by a factor of 1.4 to match
the data.

\subsubsection{Background estimation from $3\ell+X$} 
\label{sec:3L1}

This method uses the $3\ell+X$ data control region with one loosely identified lepton for
normalization. The control region is then fit using templates derived from simulation to determine
the composition in terms of the three background sources $f$, $\gamma$, $q$, and these components
are extrapolated individually to the signal region using the efficiency from the $Z+X$ control
region.

The background estimation from the $3\ell+X$ region uses data that has quadruplets built as for the
full analysis, with the exception that the full selection is applied to only the three
highest-\pt\ leptons.  Relaxed requirements are applied to the lowest-\pt\ electron: only a track
with a minimum number of silicon hits which matches a cluster is required and the electron
identification and isolation/impact parameter significance selection criteria are not applied.  In
addition, the subleading electron pair is required to have the same sign for both charges (SS) to
minimize the contribution from the \zzstar\ background. A residual \zzstar\ component with a
magnitude of 5\% of the background estimate survives the SS selection, and is subtracted to get the
final estimate.

By requiring only a single electron with relaxed selection, the composition of the control region is
simplified when compared with the other $\ell\ell+XX$ control regions, and the yields of the
different background components can be extracted with a two-dimensional fit.  Two variables, the
number of hits in the innermost layer of the pixel detector ($n^{\rm B\mbox{-}layer}_{\rm hits}$)
and the ratio of the number of high-threshold to low-threshold TRT hits ($r_{\rm TRT}$),\footnote{A
  large number of hits above a high signal pulse-height threshold is an indication of the presence
  of transition radiation, which is more probable for electrons than for pions.} allow the
separation of the $f$, $\gamma$ and $q$ components, since most photons convert after the innermost
pixel layer, and hadrons faking electrons have a lower $r_{\rm TRT}$ compared to conversions and
heavy-flavor electrons.  Templates for the fit are taken from the $Z+X$ simulation after applying
corrections from data.

\begin{figure*}
  \centering
  \subfigure[\label{fig:splot_blay}]{\includegraphics[width=\doublePlotSize]{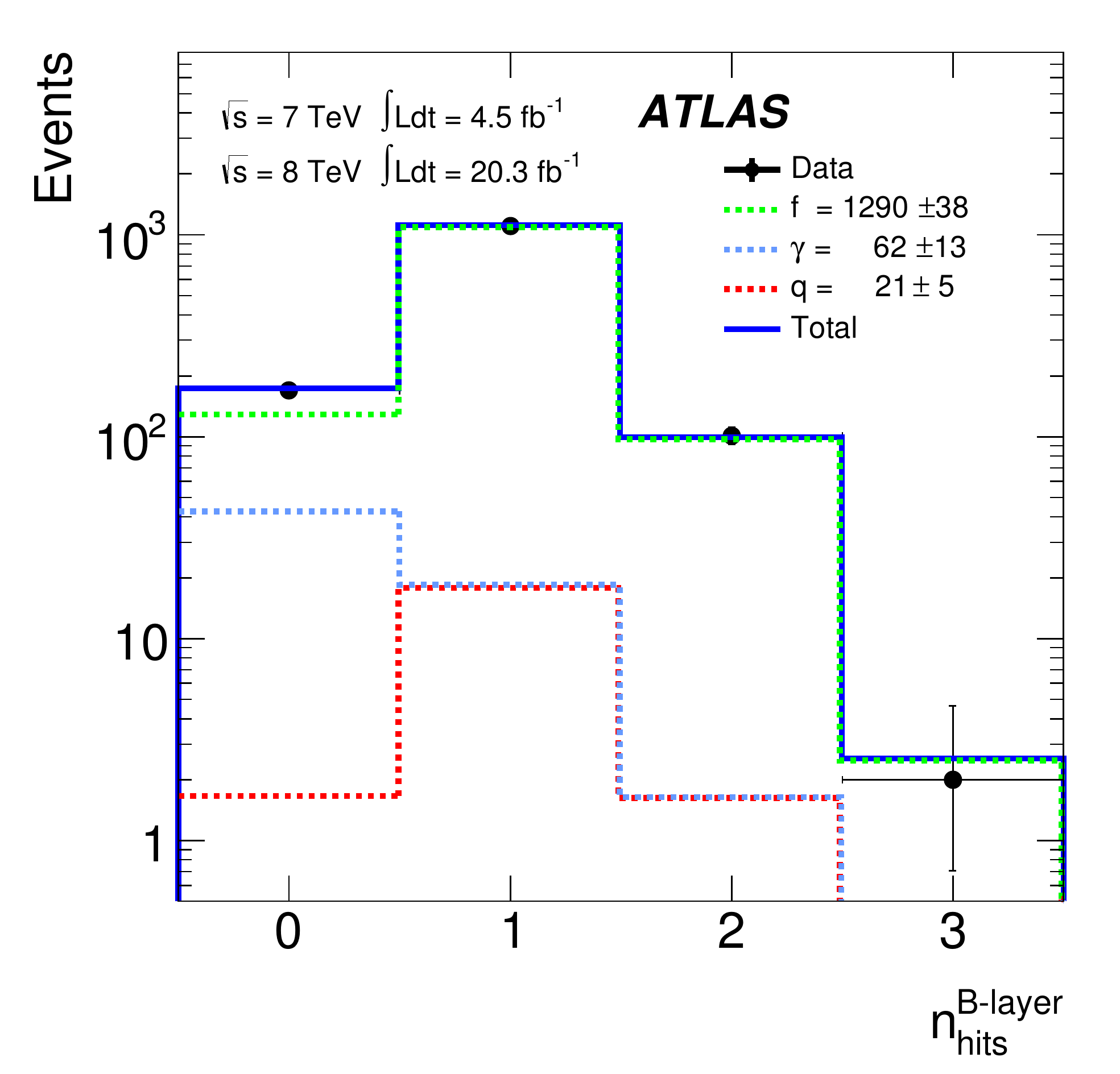}}
  \subfigure[\label{fig:splot_trt}]{\includegraphics[width=\doublePlotSize]{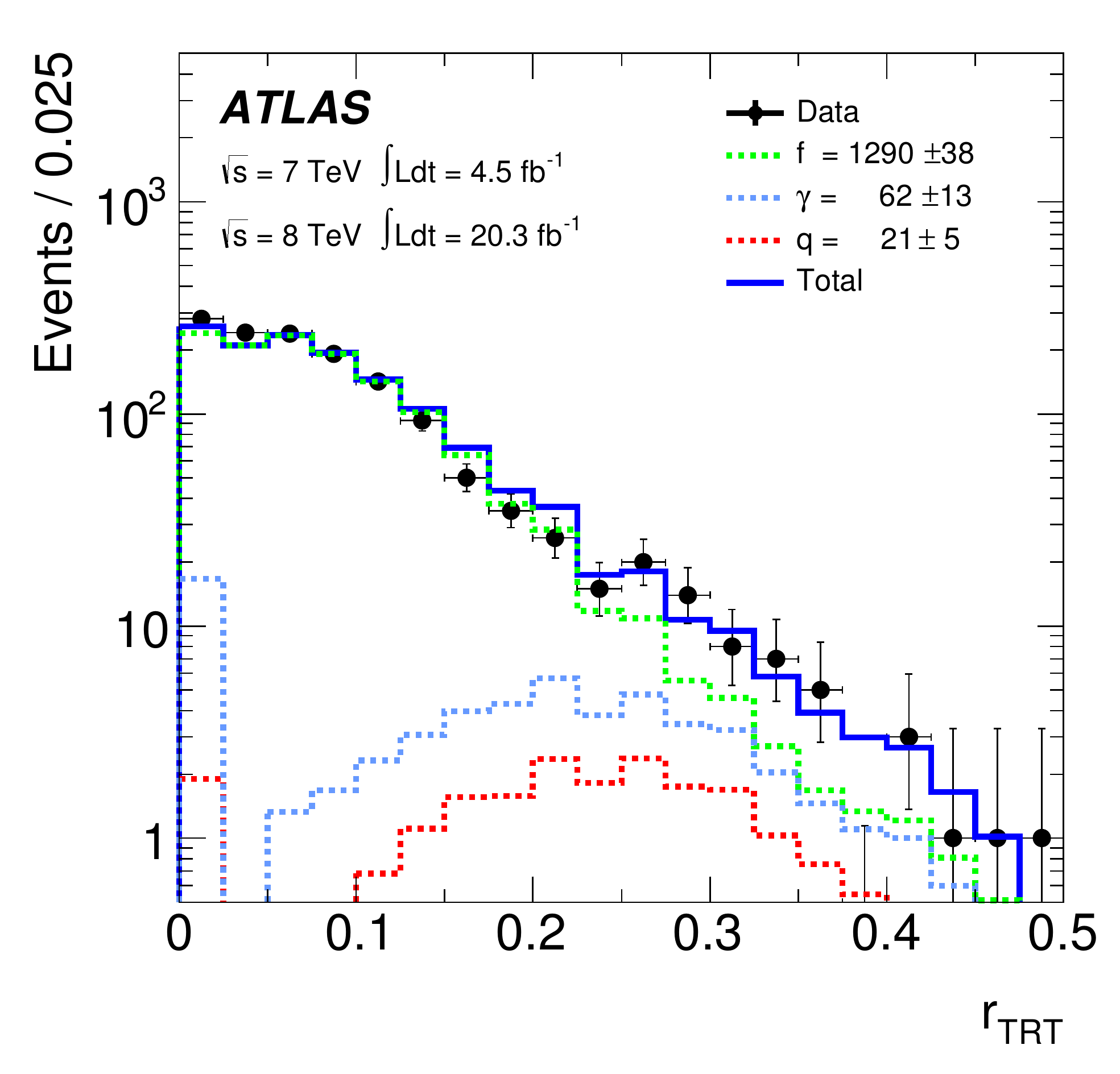}} 
  \caption{The results of a simultaneous fit to \subref{fig:splot_blay} $n^{\rm B\mbox{-}layer}_{\rm
      hits}$, the number of hits in the innermost pixel layer, and \subref{fig:splot_trt} $r_{\rm
      TRT}$, the ratio of the number of high-threshold to low-threshold TRT hits, for the background
    components in the $3\ell+X$ control region. The fit is performed separately for the $2\mu 2e$
    and $4e$ channels and summed together in the present plots.  The data are represented by the
    filled circles.  The sources of background electrons are denoted as: light-flavor jets faking an
    electron ($f$, green dashed histogram), photon conversions ($\gamma$, blue dashed histogram) and
    electrons from heavy-flavor quark semileptonic decays ($q$, red dashed histogram). The total
    background is given by the solid blue histogram. \label{fig:ThreePlusOneFit}}
\end{figure*}

The results of the fit are shown in Fig.~\ref{fig:ThreePlusOneFit}, for the $2\mu 2e$ and $4e$
channels combined.  The \emph{sPlot} method~\cite{Nima_splot} is used to unfold the contributions
from the different background sources as a function of electron \pt. The background estimates for
the $f$, $\gamma$ and $q$ components in the control region, averaged over the $2\mu2e$ and $4e$
channels, are summarized in Table~\ref{tab:ThreePlusOne_data_results}.

To extrapolate the $f$, $\gamma$ and $q$ components from the $3\ell+X$ control region to the signal
region, the efficiency for the different components to satisfy all selection criteria is obtained
from the $Z+X$ simulation. As previously mentioned, the simulation efficiency for each component is
corrected by comparing with data using the $Z+X$ control region with an adjusted selection to enrich
it for each specific component. For the $f$ component, the simulation efficiency is corrected by a
factor between 1.6 and 2.5, rising with increasing \pt.  The simulation is found to model well the
efficiency of the $\gamma$ component, to within approximately 10\%. For the $q$ component, the
efficiency is found to be modeled well by simulation, but there is an additional correction,
obtained from simulation, to estimate the number of background opposite-sign (OS) events from the
number of SS events, which is OS/SS $\approx1.7$. The systematic uncertainty is dominated by these
simulation efficiency corrections, corresponding to 30\%, 20\%, 25\% uncertainties for $f, \gamma,
q$, respectively.  The extrapolation efficiency and signal yields are also given in
Table~\ref{tab:ThreePlusOne_data_results}.  After removing the residual \zzstar\ background
($\approx5\%$), the final results for the $2\mu2e$ and $4e$ reducible backgrounds are given in
Table~\ref{tab:bkg_overviewllee}.

\begin{table*}[h]
  \centering
  \setlength{\tabcolsep}{12pt}
  \caption{The fit results for the $3\ell+X$ control region, the extrapolation factors and the
    signal region yields for the reducible $\ell\ell+ee$ background. The second column gives the fit
    yield of each component in the $3\ell+X$ control region.  The corresponding extrapolation
    efficiency and signal region yield are in the next two columns. The background values represent
    the sum of the $2\mu2e$ and $4e$ channels.  The uncertainties are the combination of the
    statistical and systematic uncertainties.  \label{tab:ThreePlusOne_data_results}}
  \vspace{0.2cm}
  \begin{tabular}{cccc} 
    \hline\hline 
    \noalign{\vspace{0.05cm}}
      2$\mu$2$e$ and 4$e$ Type & Fit yield in control region  & Extrapolation factor & Yield in signal region \\ 
      \noalign{\vspace{0.05cm}}
      \hline
      \noalign{\vspace{0.05cm}}
       & \multicolumn{3}{c}{$\sqrt{s}=7$ \tev\ data}\\ 
      \noalign{\vspace{0.05cm}}
      $f$      & 391   $\pm$ 29   & 0.010 $\pm$ 0.001    & 3.9  $\pm$ 0.9 \\ 
      $\gamma$ & 19    $\pm$ 9    & 0.10  $\pm$ 0.02     & 2.0  $\pm$ 1.0 \\
      $q$      & 5.1   $\pm$ 1.0  & 0.10  $\pm$ 0.03     & 0.51 $\pm$ 0.15 \\
      \noalign{\vspace{0.05cm}}
       & \multicolumn{3}{c}{$\sqrt{s}=8$ \tev\ data}\\ 
      \noalign{\vspace{0.05cm}}
      $f$      & 894   $\pm$ 44   & 0.0034 $\pm$ 0.0004  & 3.1 $\pm$ 1.0 \\
      $\gamma$ & 48    $\pm$ 15   & 0.024  $\pm$ 0.004   & 1.1 $\pm$ 0.6 \\
      $q$      & 18.3  $\pm$ 3.6  & 0.10   $\pm$ 0.02    & 1.8 $\pm$ 0.5 \\
      \noalign{\vspace{0.05cm}}
      \hline \hline
  \end{tabular} 
\end{table*}

\subsubsection{Background estimation from the $\ell\ell+XX$ region using the transfer-factor method} 
\label{sec:TF}

The transfer-factor method starts from the $\ell\ell+XX$ control region in data with two leptons
with inverted selection requirements.  Using the predicted sample composition from simulation, two
approaches are taken to obtain transfer factors: one using the $Z+X$ simulation corrected by data,
and the other using the $Z+X$ data control region that is enriched to obtain a $q$ component
matching that of the $\ell\ell+XX$ control region.

The $\ell\ell+XX$ data control region has relaxed electron likelihood identification on the $X$ pair
and requires each $X$ to fail one selection among the full electron identification, isolation and
impact parameter significance selections, leading to a sample of around 700 events for each of the
$2\mu+XX$ and $2e+XX$ channels.  The inverted selection removes most of the \zzstar\ background from
the control region as well as the Higgs signal.  The main challenge is to correctly estimate the
extrapolation efficiency, or transfer factor, from the $\ell\ell+XX$ control region to the signal
region using the $Z+X$ sample, since the background composition of $f$, $\gamma$ and $q$ is
different for the $Z+X$ and $\ell\ell+XX$ control regions and each of their extrapolation
efficiencies is significantly different.

In order to aid in the understanding of the control region composition and to improve the
uncertainty on the estimate of the extrapolation to the signal region, each $X$ is assigned to one
of two reconstruction categories: electron-like (E) or fake-like (F), for both data and
simulation. For the E category, a selection is applied to enhance the electron content, and the
remaining $X$ fall into the F category. For the $\ell\ell+XX$ control region, the composition of $X$
in terms of the background source is balanced between fakes ($f$) and electrons ($\gamma,q$) for the
E category corresponding to component fractions of 50\% $f$, 20\% $\gamma$, and 30\% $q$, and is
dominated by fakes for the F category with 92\% $f$, 5\% $\gamma$ and 3\% $q$.

The two approaches taken to estimate the background from the $\ell\ell+XX$ data control region
differ in the way they estimate the extrapolation to the signal region with $Z+X$ events.  Both
approaches separate $XX$ into the four reconstruction categories: EE, EF, FE and FF.  The first
approach uses the $Z+X$ simulation to determine the transfer factors for $X$ in bins of \pt\ and
\eta, where the extrapolation efficiency of each background component of the $\ell\ell+XX$
simulation is combined according to the composition seen in the $\ell\ell+XX$ simulation.  In
addition, the simulation extrapolation efficiency is corrected to agree with data as previously
described in \secref{sec:3L1}.  For the background estimate, the transfer factors are applied to the
$\ell\ell+XX$ data control region, accounting for the inverted selection.  The result is corrected
by subtracting a small residual \zzstar\ contribution, and including a $WZ$ contribution that is
removed by the inverted selection on the $XX$; both are estimated with simulation. The background
estimate with the transfer-factor method is given in Table~\ref{tab:bkg_overviewllee}.

The second approach differs in the manner in which the background composition of the $Z+X$ control
region is brought into agreement with the $\ell\ell+XX$ control region.  The most important
difference lies in the heavy-flavor component fraction, which is three times larger in the
$\ell\ell+XX$ control region and has a significantly larger transfer factor than either the $f$ or
$\gamma$ backgrounds.  This approach modifies the composition of the $Z+X$ data control region by
requiring a $b$-jet in each event.  By tuning the selection of a multivariate
$b$-tagger~\cite{ATLAS-CONF-2014-004}, the $q$ and $f$ composition of the $Z+X$ control region can
be brought into agreement with that of the $\ell\ell+XX$ control region to the level of 5\%--10\%, as
seen with simulation.  The transfer factors are extracted from the $Z+X$ data control region and
applied in bins of \pt\ and \eta\ as for the other approach, and the systematic uncertainty is
estimated in part by varying the operating point used for the multivariate $b$-tagger.  Finally, the
$WZ$ contribution is accounted for with simulation, as previously.  The background estimate from the
transfer factors based on $b$-enriched samples is given in Table~\ref{tab:bkg_overviewllee}.

\subsubsection{Reco-truth unfolding method} 
\label{sec:RecoTruth}

A third method uses the $\ell\ell+XX$ data control region; however, the two subleading electrons
have only the electron identification relaxed and do not have an inverted selection applied as for
the transfer-factor method.  This control region thus contains all backgrounds, including the
\zzstar\ background, and the \htollllbrief\ signal.  The extrapolation to the signal region is
performed with the $Z+X$ simulation. This method was used as the baseline for previous
publications~\cite{ATLAS:2012af, Aad:2013wqa}, but is now superseded by the $3\ell+X$ method, which
provides the smallest uncertainties of the data-driven methods.  Using the simulation, each of the
paired reconstruction categories (EE, EF, FE and FF) of the $\ell\ell+XX$ sample is decomposed into
its background origin components ($ee$, $ff$, $\gamma\gamma$, $qq$ and the 12 cross combinations),
where the $e$ background category is introduced to contain the isolated electrons from \zzstar\ and
\htollllbrief.  This $4\times16$ composition table is summed with efficiency weights, in bins of
\pt\ and \eta, obtained from the $Z+X$ simulation, which is corrected from comparison with data as
previously mentioned.  To remove the \zzstar\ and \htollllbrief\ contributions from this estimate,
the background origin category $ee$ is removed from the sum, and an estimated residual of
$1.2\pm0.4$ \zzstar\ events is subtracted to obtain the final result, which is also given in
Table~\ref{tab:bkg_overviewllee}.

\subsubsection{Summary of reducible background estimates for $\ell\ell+ee$}
\label{sec:bkgSummaryllee}

The summary of the reducible backgrounds for the $\ell\ell+ee$ final states is given for the full
mass region in Table~\ref{tab:bkg_overviewllee}.  In addition to the previously discussed methods,
the results are presented for the full analysis applied to $\ell\ell+ee$ events in data where the
subleading $ee$ pair is required to have the same-sign charge, and $m_{4\ell}$ is required to be
below 160 \gev\ to avoid a $ZZ$ contribution; the region with $m_{4\ell} < 160$ \gev\ contains 70\%
of the expected reducible backgrounds.  Although limited in statistical precision, this agrees well
with the other estimates.

\begin{table*}[h]
  \centering \caption{Summary of the $\ell\ell+ee$ data-driven background estimates for the
    $\sqrt{s}=7$ \tev\ and $\sqrt{s}=8$ \tev\ data for the full $m_{4\ell}$ mass range.  OS (SS)
    stands for opposite-sign (same-sign) lepton pairs.  The ``$\dagger$'' symbol indicates the estimates
    used for the background normalization; the other estimates are used as cross-checks. The first
    uncertainty is statistical, while the second is systematic.  The SS data full analysis is
    limited to the region with $m_{4\ell}$ below 160 \gev\ to avoid a $ZZ$ contribution; this region
    contains 70\% of the expected background.  \label{tab:bkg_overviewllee}}

  \setlength{\tabcolsep}{12pt}

  \vspace{0.2cm} 
  \begin{tabular}{lcc} 
    \hline\hline
    \noalign{\vspace{0.05cm}}
    Method & $\sqrt{s}=7$ \tev\ data & $\sqrt{s}=8$ \tev\ data \\ \hline
    \noalign{\vspace{0.05cm}}
    \multicolumn{3}{c}{$2\mu2e$} \\ 
    \noalign{\vspace{0.05cm}}
    \hline
    $3\ell+X^\dagger$                            & $2.9  \pm 0.5  \pm 0.5 $             & $2.9  \pm 0.3  \pm 0.6 $\\ 
    $\ell\ell+ XX$ transfer factor              & $2.2  \pm 0.3  \pm 1.1 $             & $2.5  \pm 0.1  \pm 0.9 $\\  
    $\ell\ell+ XX$ transfer factor $b$-enriched & $2.8  \pm 0.5  \pm 0.8 $             & $3.2  \pm 0.2  \pm 0.9 $\\ 
    $\ell\ell+ XX$ reco-truth                   & $2.8  \pm 0.4  \pm 1.0 $             & $2.9  \pm 0.3  \pm 0.3  $\\
    $2\mu2e$ SS data full analysis              & 1                                    & 2 \\
    \hline
    \noalign{\vspace{0.05cm}}
    \multicolumn{3}{c}{$4e$}\\ 
    \noalign{\vspace{0.05cm}}
    \hline
    $3\ell+X^\dagger$                            & $3.3  \pm 0.5  \pm 0.5 $            & $2.9  \pm 0.3  \pm 0.5 $\\ 
    $\ell\ell+ XX$ transfer factor              & $2.0  \pm 0.3  \pm 0.9 $            & $2.4  \pm 0.1  \pm 0.9 $\\ 
    $\ell\ell+ XX$ transfer factor $b$-enriched & $3.4  \pm 0.9  \pm 0.8 $            & $2.9  \pm 0.2  \pm 0.8 $\\ 
    $\ell\ell+ XX$ reco-truth                   & $2.6  \pm 0.4  \pm 0.9 $            & $2.8  \pm 0.3  \pm 0.3  $\\
    $4e$ SS data full analysis                  & 2                                   & 2 \\

    \hline \hline
  \end{tabular}
\end{table*}


\subsection{Shape of the reducible background contributions}
\label{sec:bkgSummary}

\begin{figure*}
  \centering
  \subfigure[\label{fig:sub_2crmu}]{\includegraphics[width=\doublePlotSize]{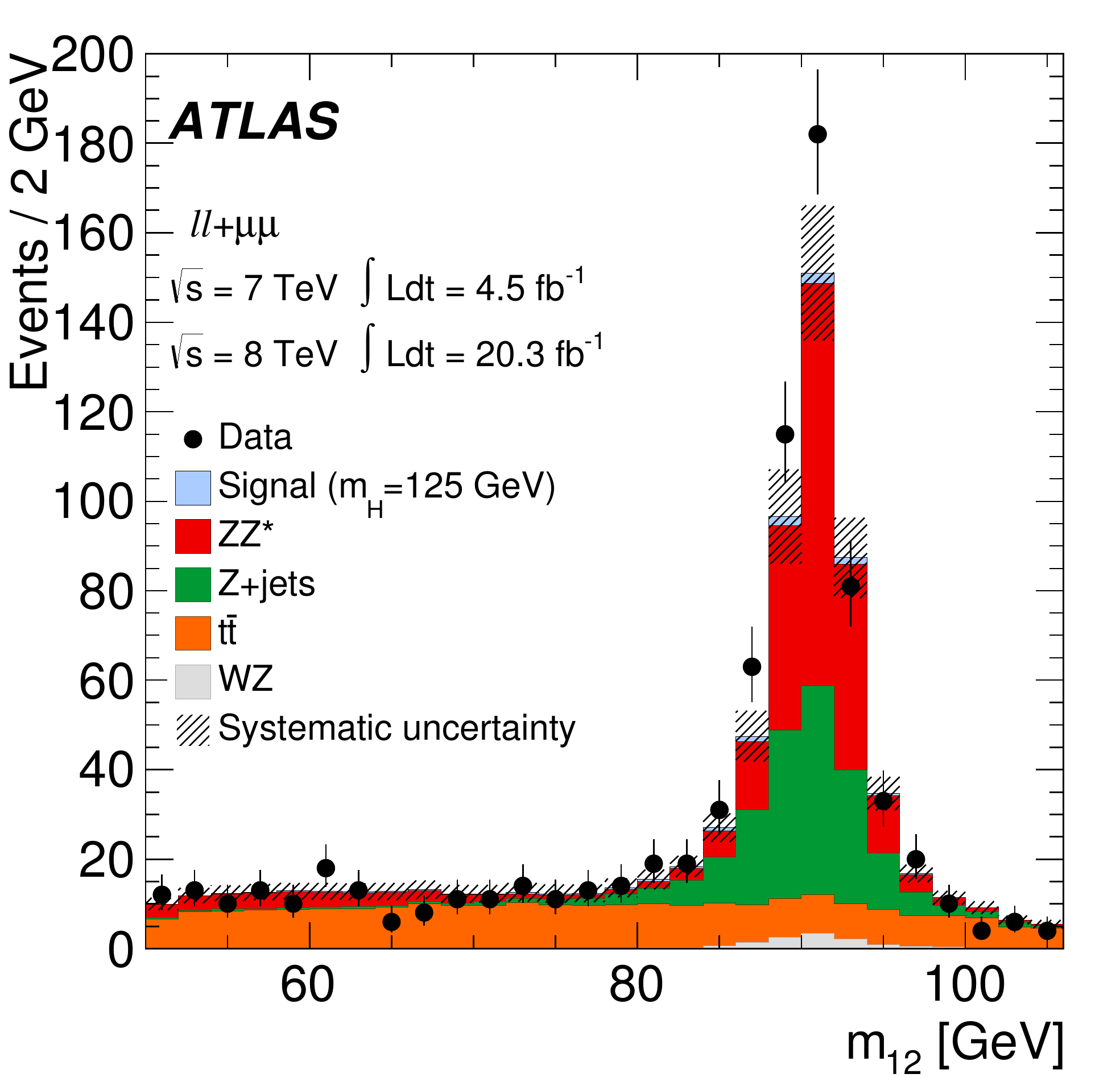}}
  \subfigure[\label{fig:sub_2cre}]{\includegraphics[width=\doublePlotSize]{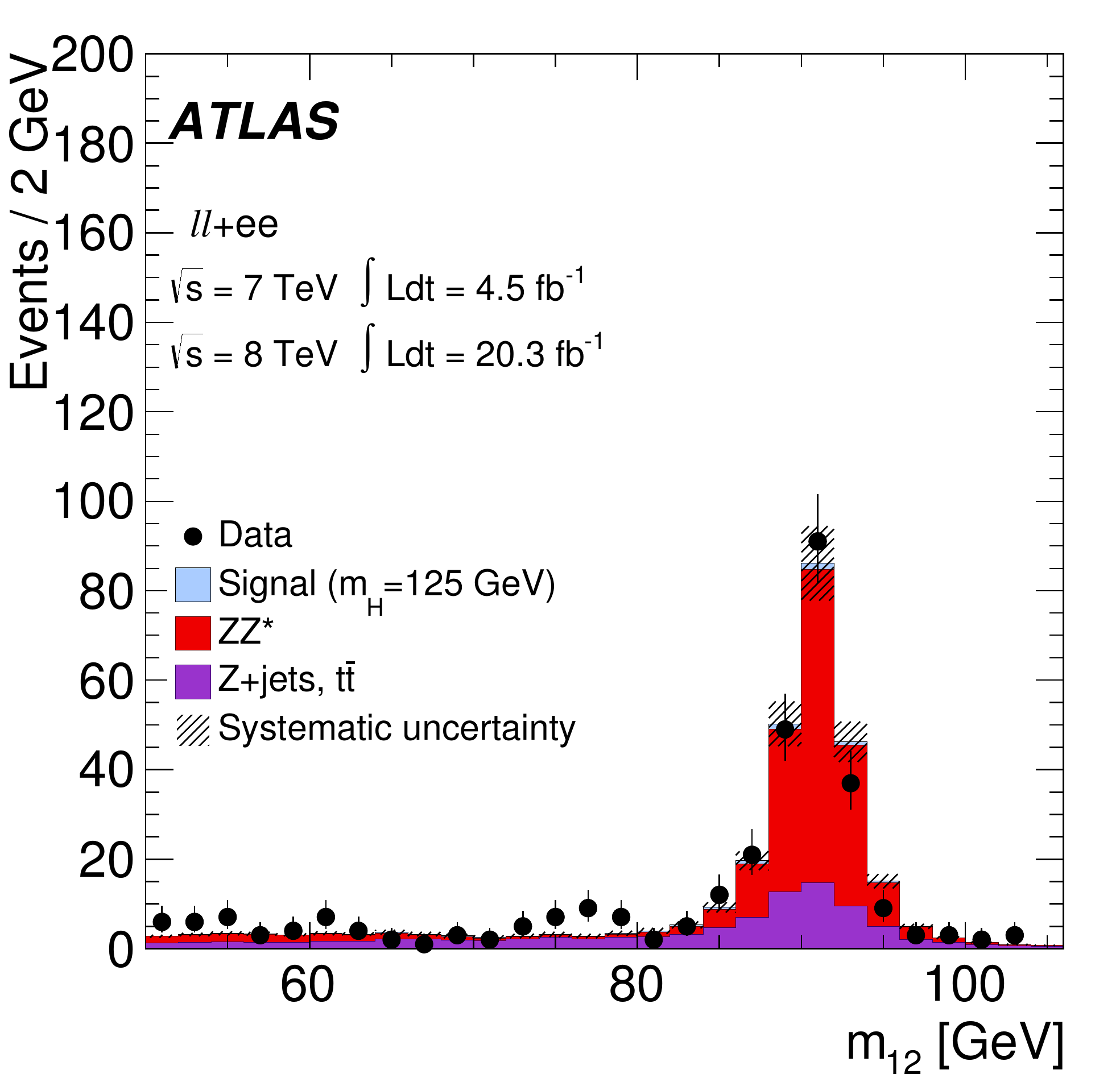}}
  \subfigure[\label{fig:sub_crmu}]{\includegraphics[width=\doublePlotSize]{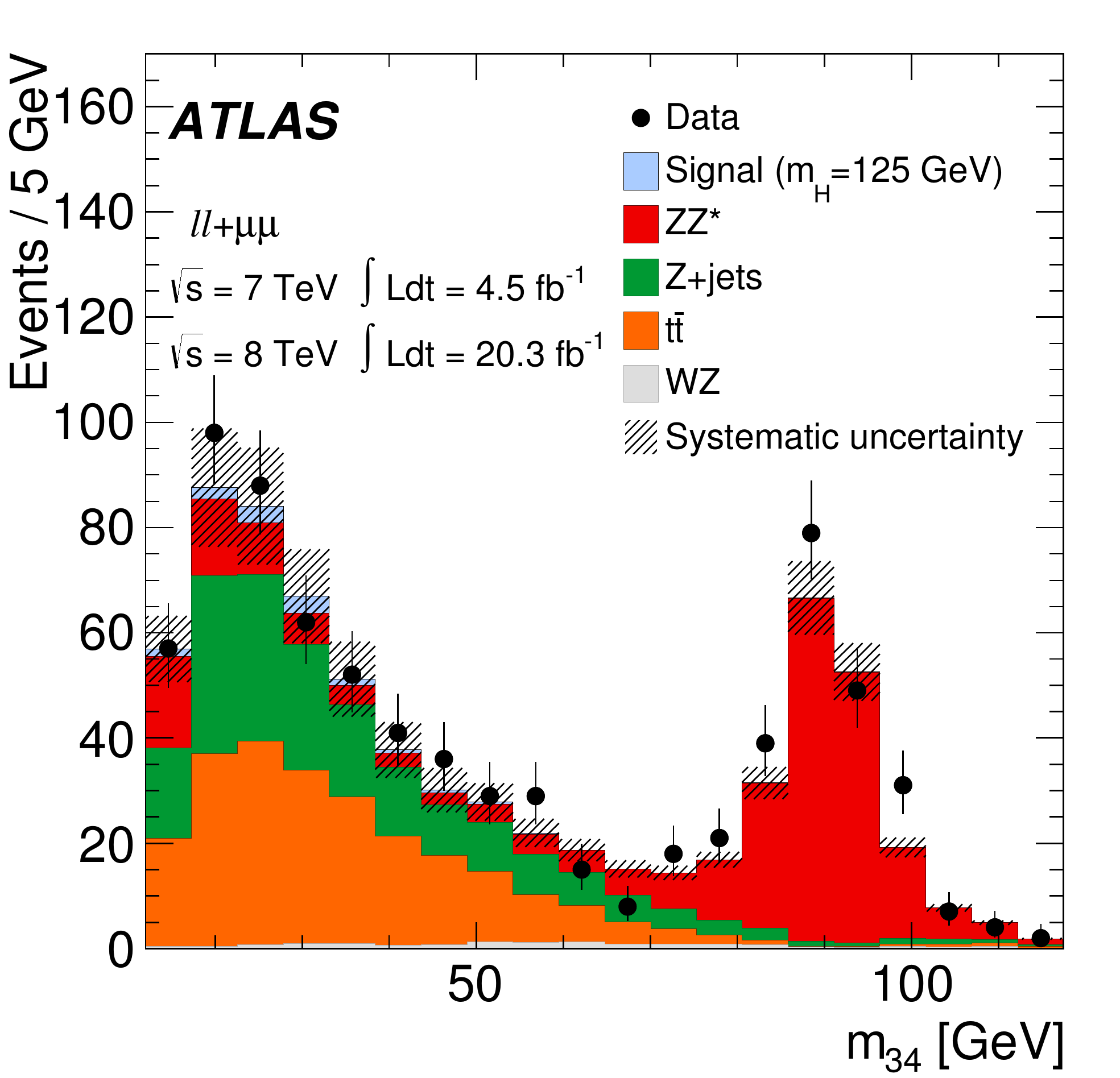}}
  \subfigure[\label{fig:sub_cre}]{\includegraphics[width=\doublePlotSize]{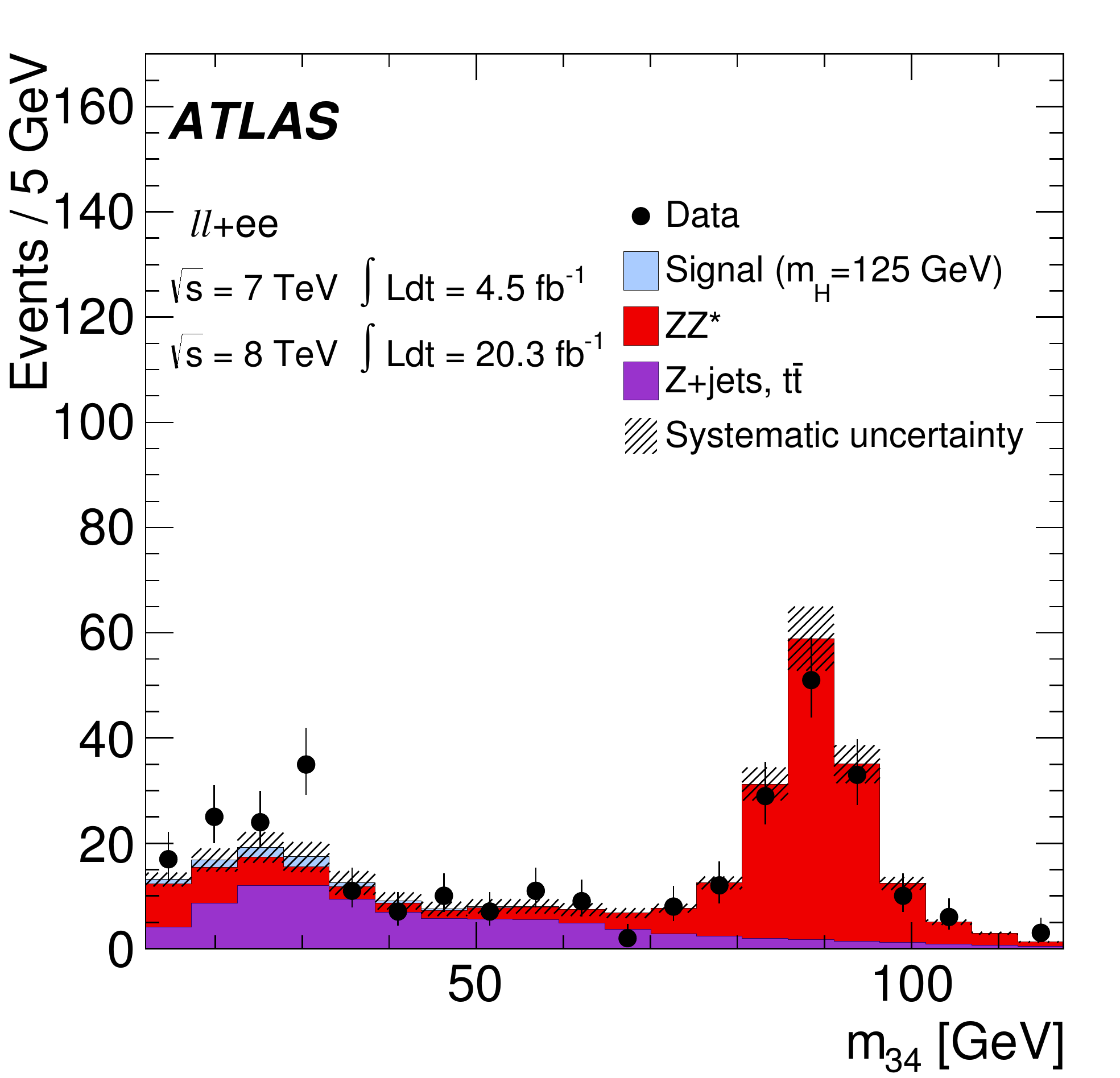}}
  \caption{Invariant mass distributions of the lepton pairs in the control sample defined by a $Z$
    boson candidate and an additional same-flavor lepton pair, including all signal and background
    contributions, for the $\sqrt{s}=7\:\tev$ and $\sqrt{s}=8\:\tev$ data sets. The sample is divided
    according to the flavor of the additional lepton pair.  In \subref{fig:sub_2crmu} and
    \subref{fig:sub_crmu} the $m_{12}$ and $m_{34}$ distributions are presented for $\ell\ell
    +\mu^+\mu^-$ events, where $\ell\ell$ is $\mu^+\mu^-$ or $e^+e^-$.  In \subref{fig:sub_2cre} and
    \subref{fig:sub_cre} the $m_{12}$ and $m_{34}$ distributions are presented for $\ell\ell
    +e^+e^-$ events.  The data are shown as filled circles and the different backgrounds as filled
    histograms with the total background systematic uncertainty represented by the hatched areas.
    The kinematic selection of the analysis is applied. Isolation and impact parameter significance
    requirements are applied to the first lepton pair only.  The simulation is normalized to the
    data-driven background estimates. \label{fig:sub_cr}}
\end{figure*}

The $m_{4\ell}$ distributions of the reducible backgrounds are required for the normalization and
shape of these backgrounds in the mass fit region, discussed below.  The shape of the distribution
for the $\ell\ell+\mu\mu$ background is taken from simulation and the uncertainty comes from varying
the track isolation and impact parameter significance selections.  The corresponding distribution
for the $\ell\ell+ee$ background comes from the $3\ell+X$ sample, after reweighting with the
transfer factor to match the kinematics of the signal region.  The uncertainty in the $\ell\ell+ee$
background shape is taken as the difference between the shapes obtained from the control regions of
the two other methods: transfer factor and reco--truth.  The estimates in the $120 < m_{4\ell} <
130$ \gev\ mass window are provided in Table~\ref{tab:yields}.  \Figref{fig:sub_cr} presents the
$m_{12}$ and $m_{34}$ distributions for the $\ell\ell+\mu\mu$ and $\ell\ell+ee$ control regions
where the full selection has been applied except for subleading lepton impact parameter significance
and isolation requirements, which are not applied.  Good agreement is seen between the data and the
sum of the various background estimates.  The shape of the background in the $m_{4\ell}$
distribution extrapolated to the signal region can be seen in \figref{fig:finalMasses}.

\subsection{Background for categories}
\label{sec:bckcat}

For the reducible background, the fraction of background in each category is evaluated using
simulation.  Applying these fractions to the background estimates from Tables~\ref{tab:fitSR}
and~\ref{tab:bkg_overviewllee} gives the reducible background estimates per category shown in
Table~\ref{tab:cat_bkg_overview}.  The systematic uncertainties include the differences observed
between the fractions obtained from simulation and those from the reducible background data control
regions.  The expected \zzstar\ background evaluated from simulation for each category is given in
Table~\ref{tab:couplingResultsYields}.  To obtain the reducible background in the signal region, the
shapes of the $m_{4\ell}$ distributions for the reducible backgrounds discussed in
\secref{sec:bkgSummary} are used.

\begin{table*}[htbb!]
  \centering
  \caption{Summary of the background estimates for the data recorded at $\sqrt{s}=7$ TeV and
    $\sqrt{s}=8$ TeV for the full $m_{4\ell}$ mass range.  The quoted uncertainties include the
    combined statistical and systematic components.\label{tab:cat_bkg_overview}}

  \setlength{\tabcolsep}{8pt}

  \vspace{0.2cm}
  \begin{tabular}{ccccc}
    \hline\hline
    \noalign{\vspace{0.05cm}}
    Channel         & \ggfcat         & \vbfcat        & \vhhadcat & \vhlepcat\\ 
    \hline
    \noalign{\vspace{0.05cm}}
    & \multicolumn{4}{c}{$\sqrt{s}=7$ TeV}\\ 
    \noalign{\vspace{0.05cm}}
    \hline
    $\ell\ell+\mu\mu$ & $0.98 \pm 0.32$  & $ 0.12\pm 0.08$ & $0.04 \pm 0.02$  & $0.004 \pm 0.004$\\
    $\ell\ell+ee$     & $5.5  \pm 1.2 $  & $0.51 \pm 0.6$  & $0.20 \pm 0.16$  & $0.06 \pm 0.11$ \\
    \hline
    \noalign{\vspace{0.05cm}}
    & \multicolumn{4}{c}{$\sqrt{s}=8$ TeV}\\ 
    \noalign{\vspace{0.05cm}}
    \hline
    $\ell\ell+\mu\mu$ & $6.7  \pm 1.4$   & $0.6 \pm 0.6$   & $0.21 \pm 0.13$  & $0.003 \pm 0.003$\\
    $\ell\ell+ee$     & $5.1  \pm 1.4$   & $0.5 \pm 0.6$   & $0.19 \pm 0.15$  & $0.06 \pm 0.11$ \\
    \hline\hline
  \end{tabular}
\end{table*}

\vspace{10 mm}

\section{Multivariate Discriminants} 
\label{sec:MVA}
The analysis sensitivity is improved by employing three multivariate discriminants to distinguish
between the different classes of four-lepton events: one to separate the Higgs boson signal from the
\zzstar\ background in the inclusive analysis, and two to separate the VBF- and VH-produced Higgs
boson signal from the ggF-produced Higgs boson signal in the \vbfcat\ and \vhhadcat\
categories. These discriminants are based on boosted decision trees (BDT) \cite{TMVA}.

\subsection{BDT for \zzstar\ background rejection}
\label{sec:zz_bdt}

\begin{figure*}
  \centering
  \subfigure[\label{fig:variables_KD}]{\includegraphics[width=\doublePlotSize]{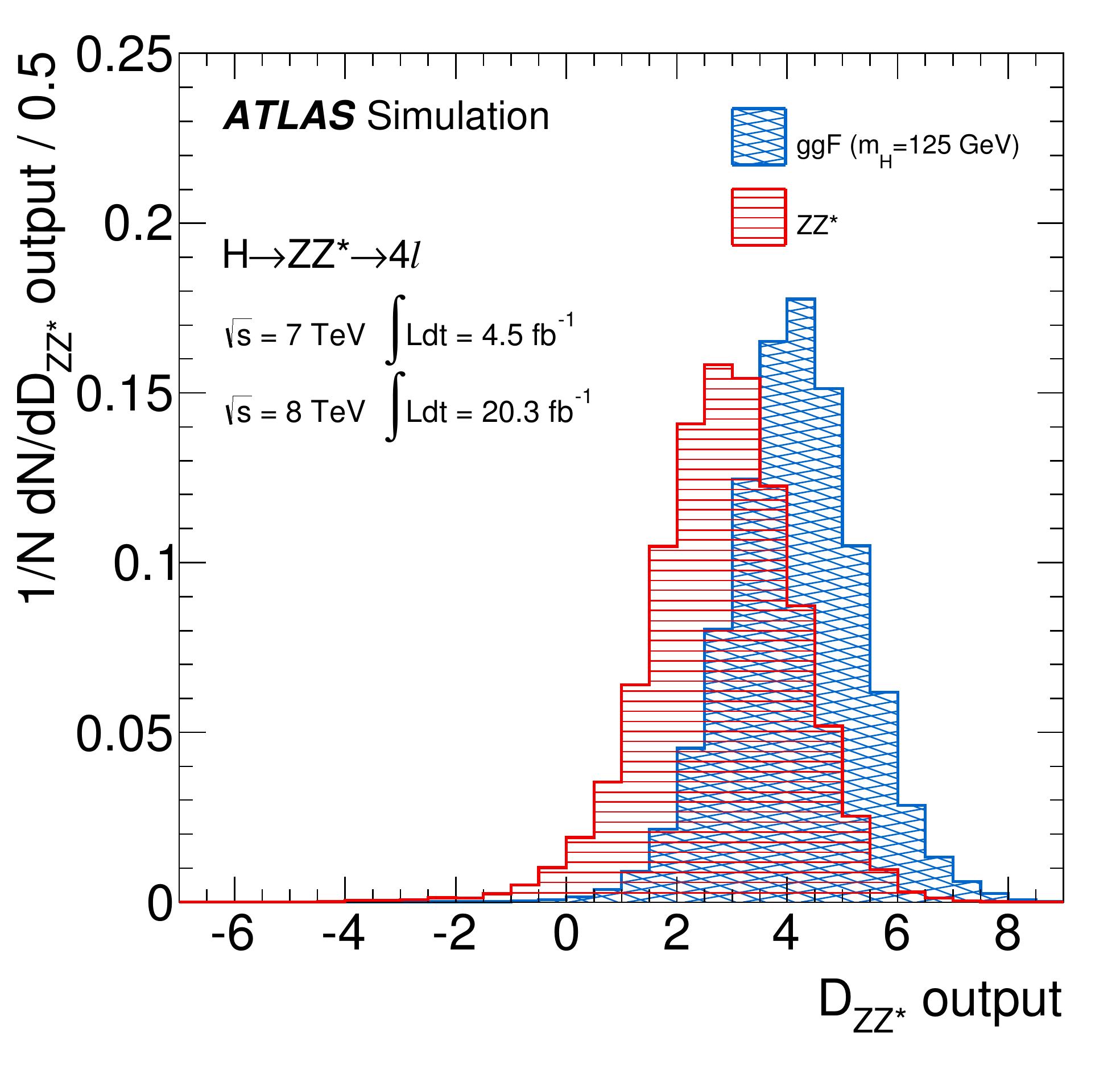}}
  \subfigure[\label{fig:variables_pt}]{\includegraphics[width=\doublePlotSize]{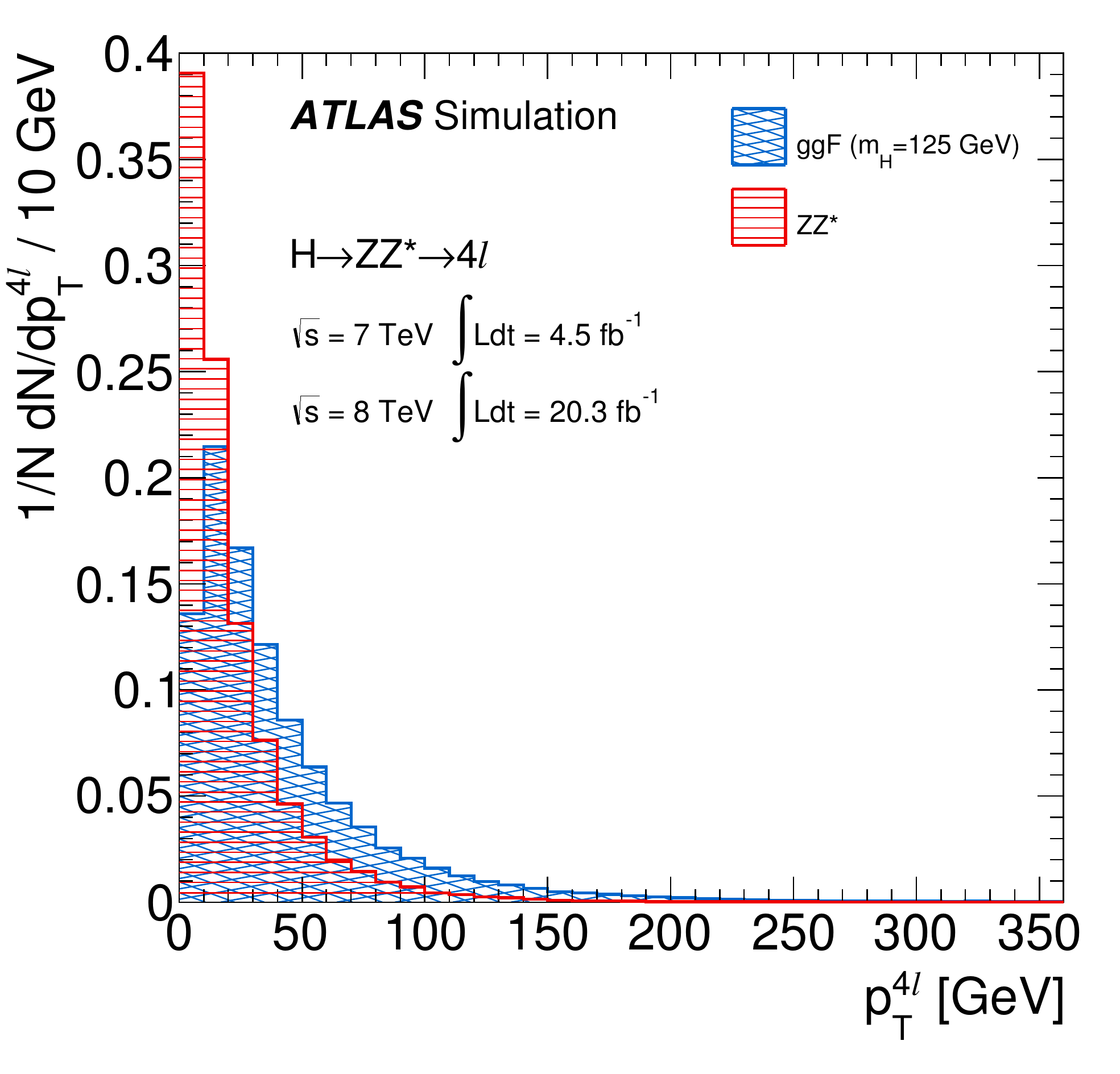}}
  \subfigure[\label{fig:variables_eta}]{\includegraphics[width=\doublePlotSize]{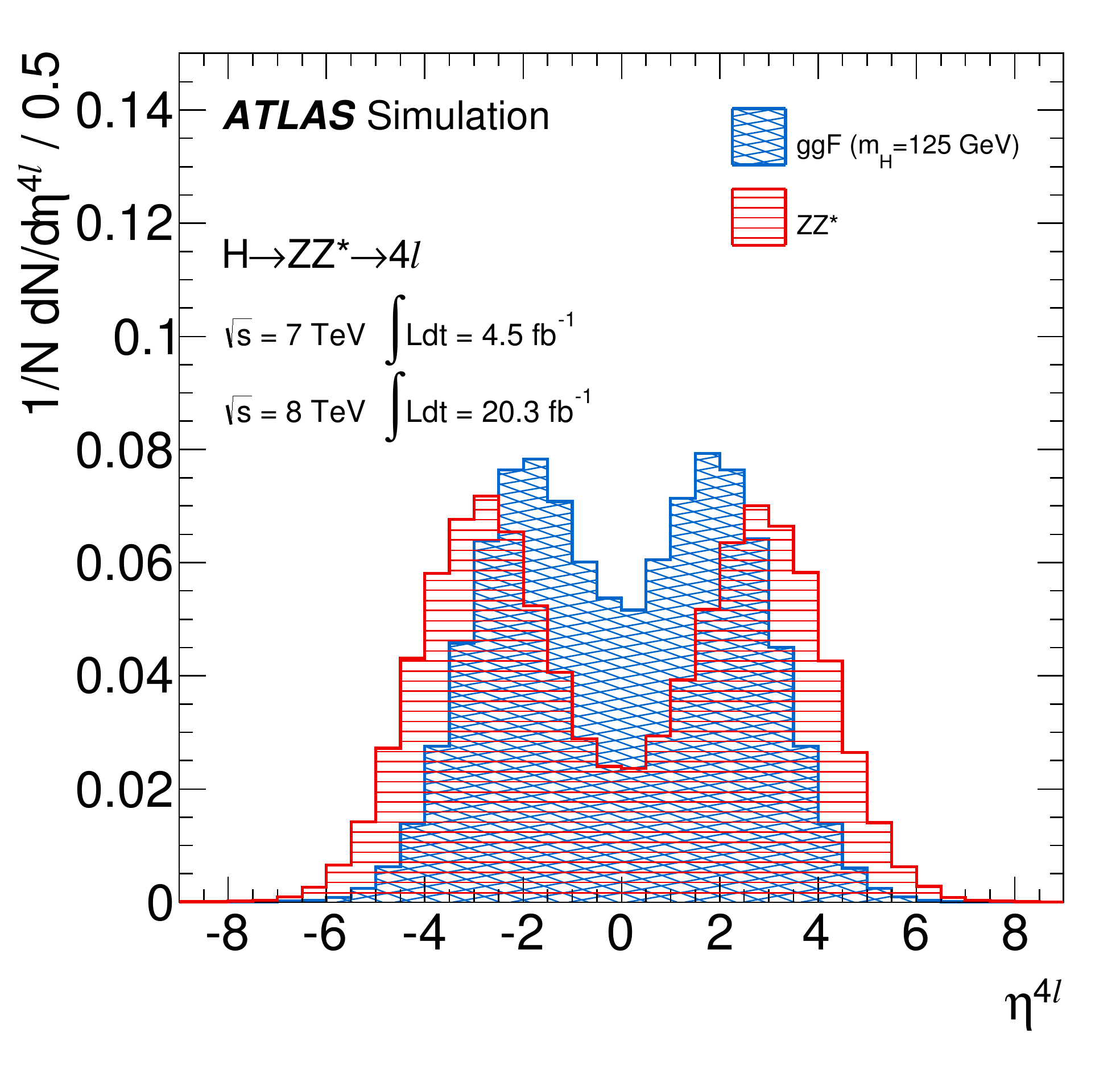}}
  \subfigure[\label{fig:variables_BDT}]{\includegraphics[width=\doublePlotSize]{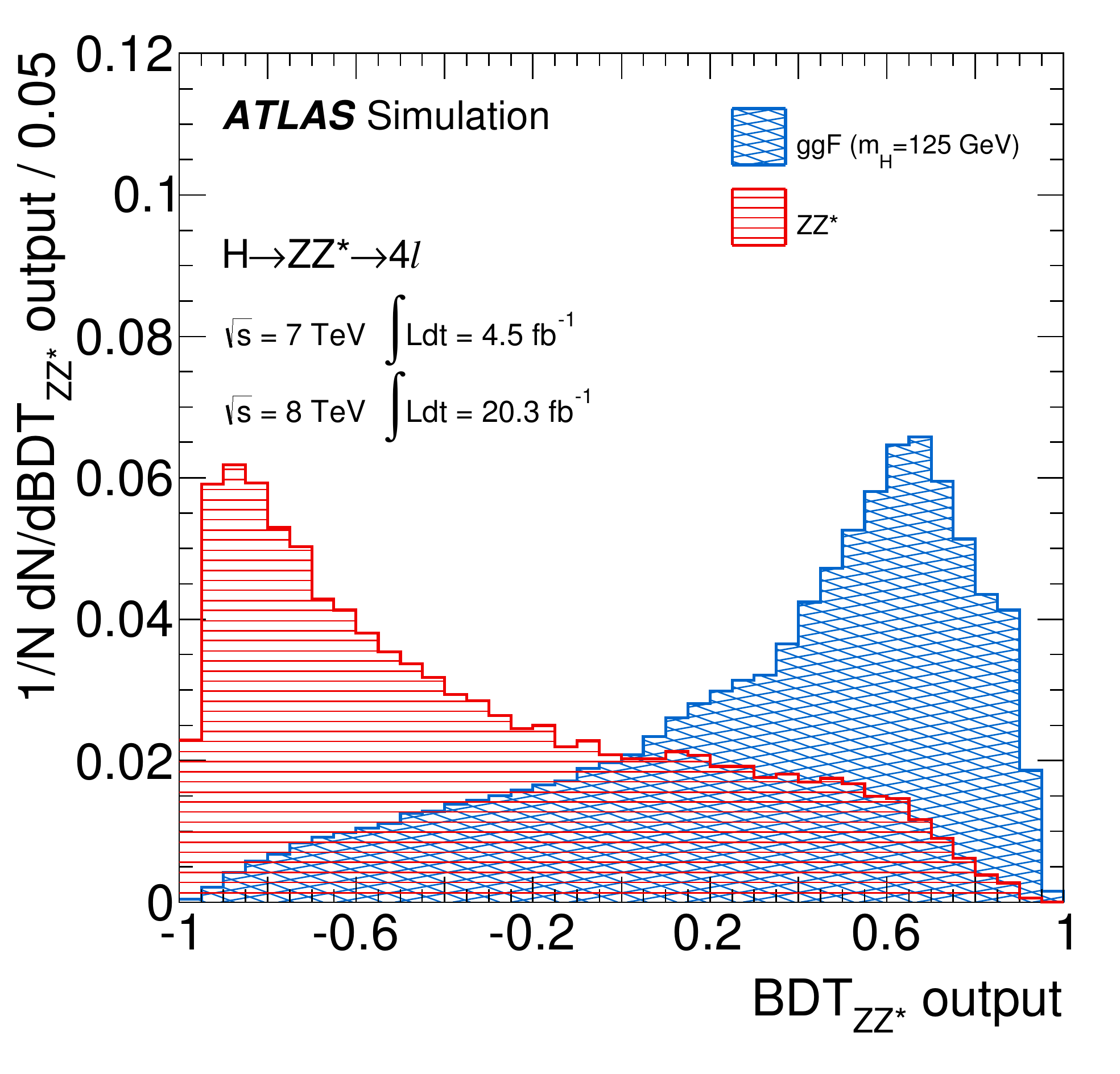}}
  \caption{Distributions for signal (blue) and \zzstar\ background (red) events, showing
    \subref{fig:variables_KD} $D_{ZZ^{*}}$ output, \subref{fig:variables_pt}
    $p^{4\ell}_{\mathrm{T}}$ and \subref{fig:variables_eta} $\eta^{4\ell}$ after the inclusive
    analysis selection in the mass range $115<m_{4\ell}<130$ \gev\ used for the training of the
    \bdtzz\ classifier.  \subref{fig:variables_BDT} \bdtzz\ output distribution for the signal
    (blue) and \zzstar\ background (red) in the mass range $115<m_{4\ell}<130$ \gev.  All histograms
    are normalized to the same area. \label{fig:variables}}
\end{figure*}

The differences in the kinematics of the \htollllbrief\ decay and the \zzstar\ background are
incorporated into a BDT discriminant (\bdtzz).  The training is done using fully simulated
\htollllbrief\ signal events, generated with $m_{H}=$125 \gev\ for ggF production, and
$qq\rightarrow$\zzstar\ background events.  Only events satisfying the inclusive event selection
requirements and with 115 $<m_{4\ell}<$ 130 \gev\ are considered.  This range contains 95\% of the
signal and is asymmetric around 125 \gev\ to include the residual effects of FSR and bremsstrahlung.
The discriminating variables used in the training are: the transverse momentum of the four-lepton
system ($p^{4\ell}_{\mathrm{T}}$), the pseudorapidity of the four-lepton system ($\eta^{4\ell}$),
correlated to the $p^{4\ell}_{\mathrm{T}}$, and a matrix-element-based kinematic discriminant
($D_{ZZ^{*}}$). The discriminant $D_{ZZ^{*}}$ is defined as:

\begin{linenomath}
  \begin{equation} 
    D_{ZZ^{*}}=\ln\left(\frac{\left|{\cal{M}}_{\rm sig}\right|^{2}}{\left|{\cal{M}}_{ZZ}\right|^{2}}\right) ,
  \end{equation}
\end{linenomath}
where ${\cal{M}}_{\rm sig}$ corresponds to the matrix element for the signal process, while
${{\cal{M}}_{ZZ}}$ is the matrix element for the \zzstar\ background process.  The matrix elements
for both signal and background are computed at leading order using MadGraph5~\cite{Alwall:2011uj}.
The matrix element for the signal is evaluated according to the SM hypothesis of a scalar boson with
spin-parity $J^{P}=0^{+}$~\cite{Aad:2013xqa} and under the assumption that $m_{H}=m_{4\ell}$.
Figures \ref{fig:variables}\subref{fig:variables_KD}-\subref{fig:variables_eta} show the
distributions of the variables used to train the \bdtzz\ classifier for the signal and the
\zzstar\ background. The separation between a SM Higgs signal and the \zzstar\ background can be
seen in \figref{fig:variables_BDT}.

As discussed in \secref{sec:SignalModel}, the \bdtzz\ output is exploited in the two-dimensional
model built to measure the Higgs boson mass, the inclusive signal strength and the signal strength
in the \ggfcat\ category.

\subsection{BDT for categorization}
\label{sec:BDTCat}
\begin{figure*}
  \centering
  \subfigure[\label{fig:dijet_mass}]{\includegraphics[width=\smallDoublePlotSize]{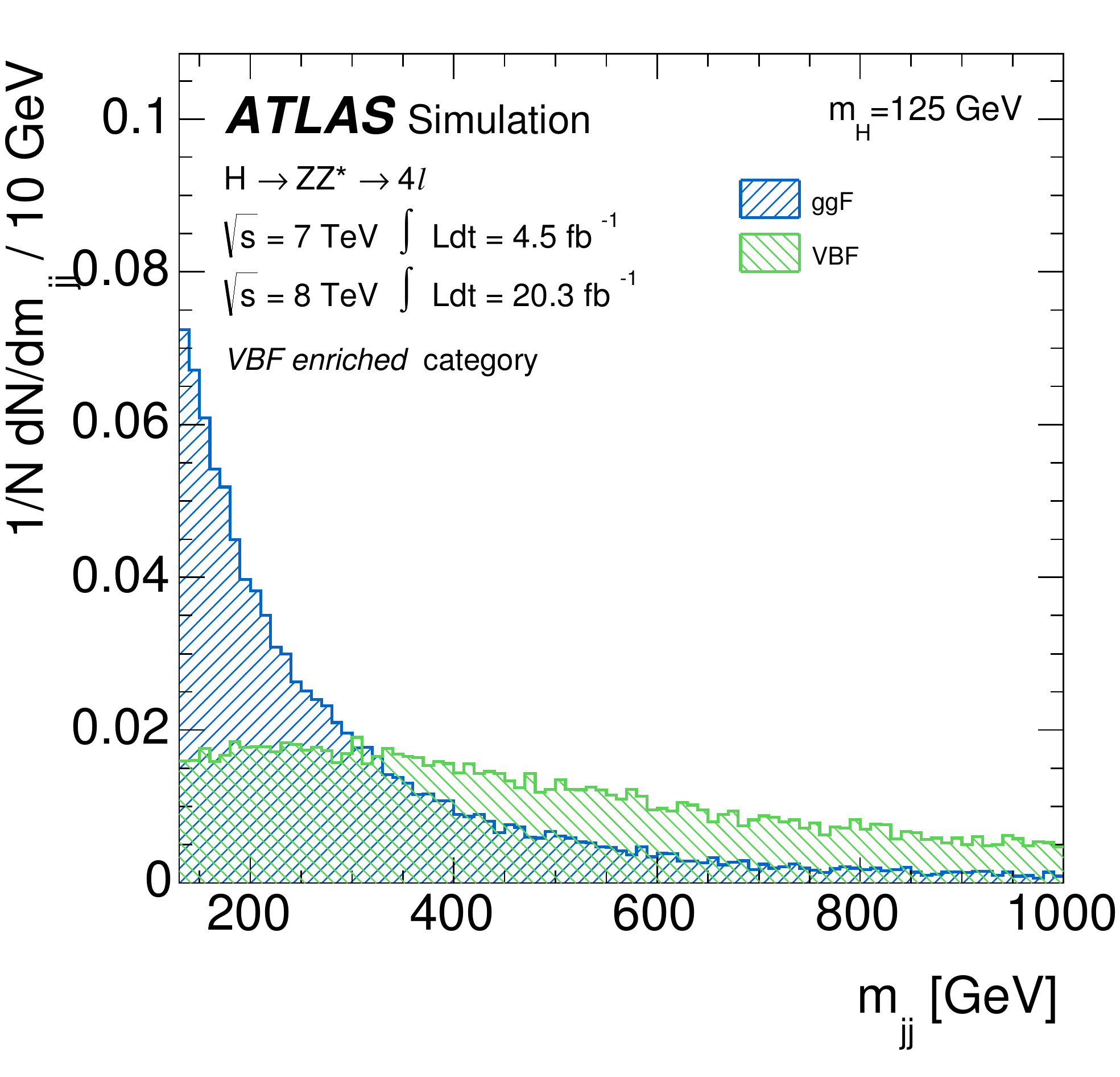}}
  \subfigure[\label{fig:dijet_eta}]{\includegraphics[width=\smallDoublePlotSize]{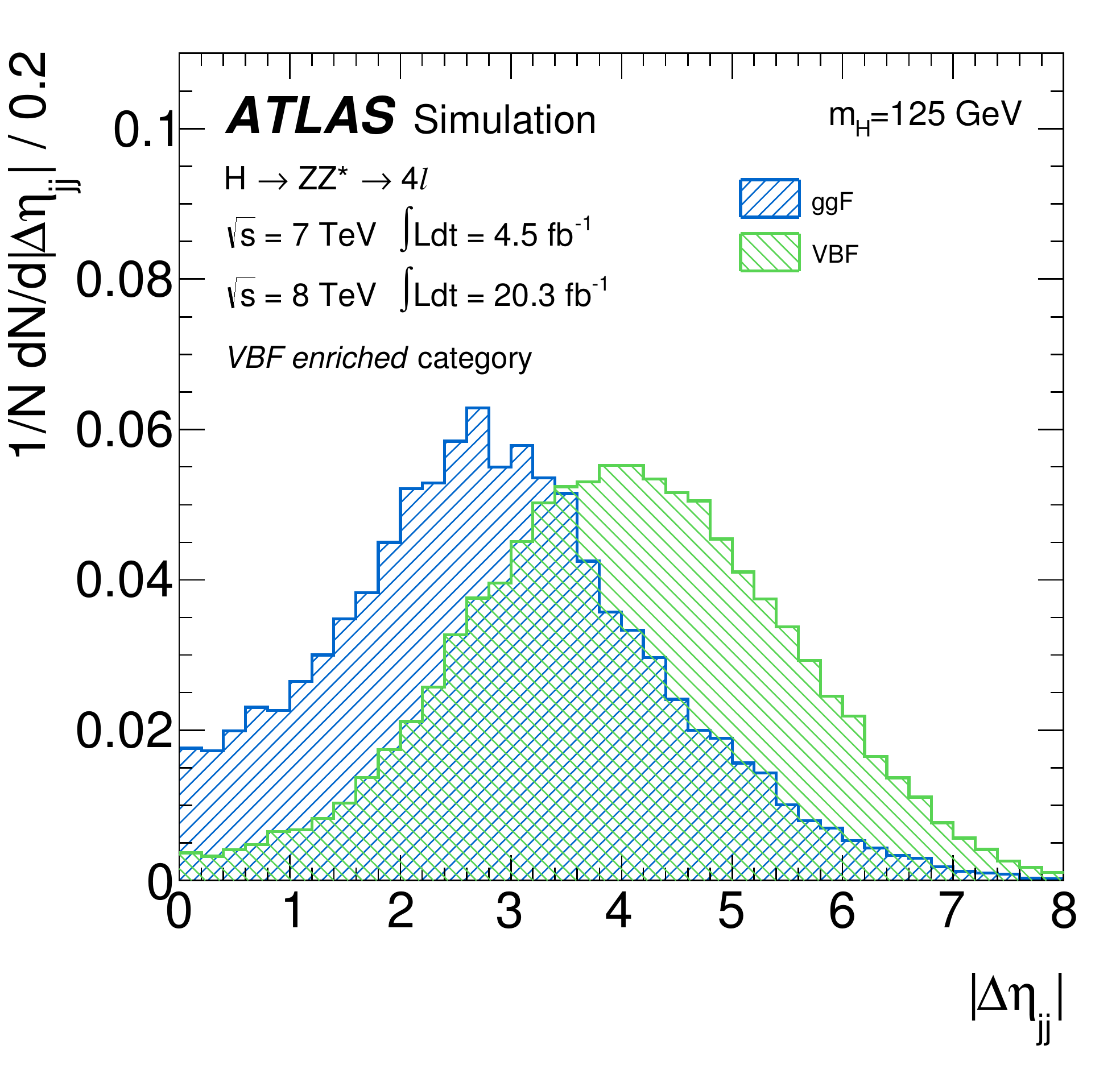}}
  \subfigure[\label{fig:leadjet_pt}]{\includegraphics[width=\smallDoublePlotSize]{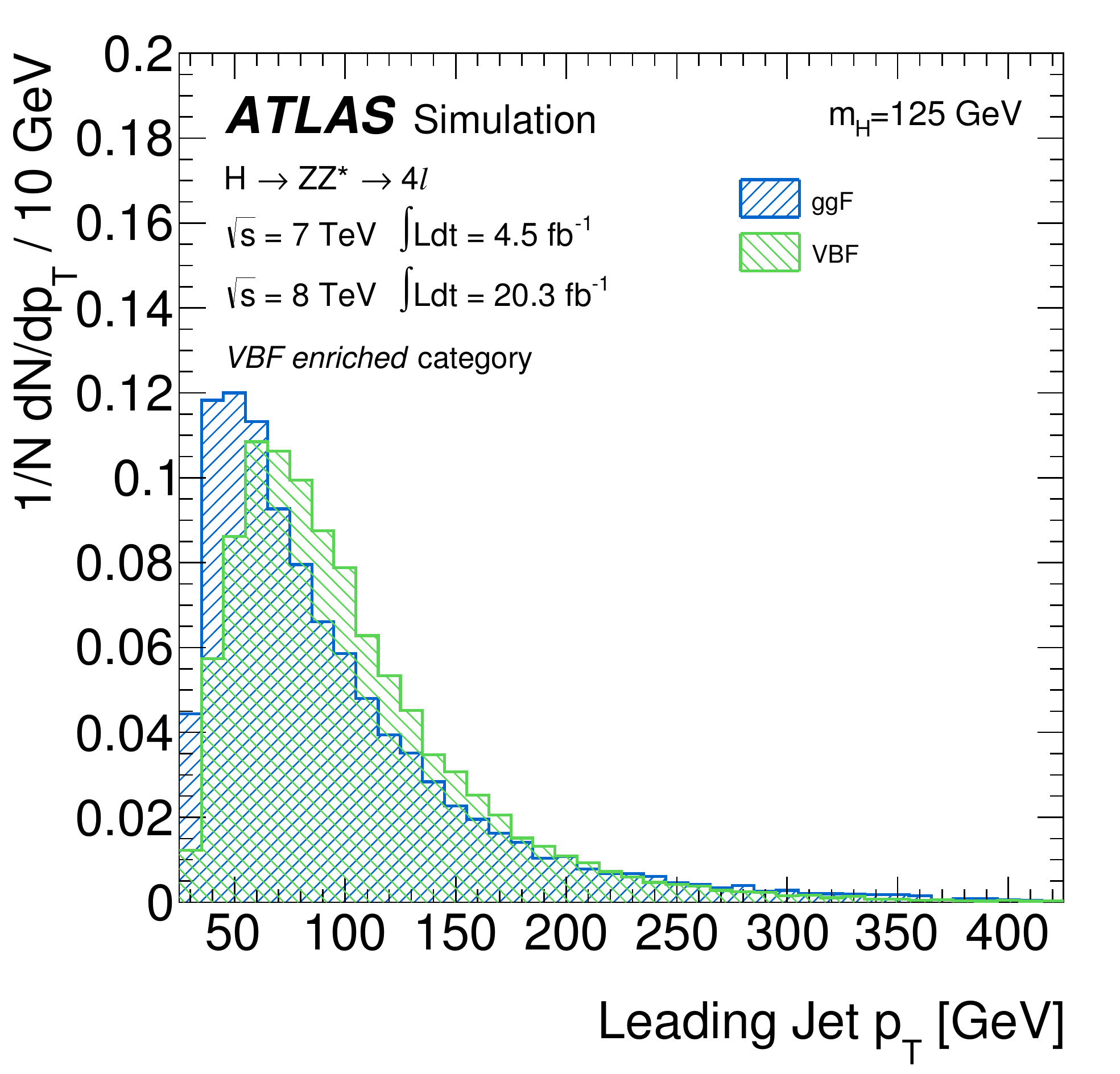}}
  \subfigure[\label{fig:subleadjet_pt}]{\includegraphics[width=\smallDoublePlotSize]{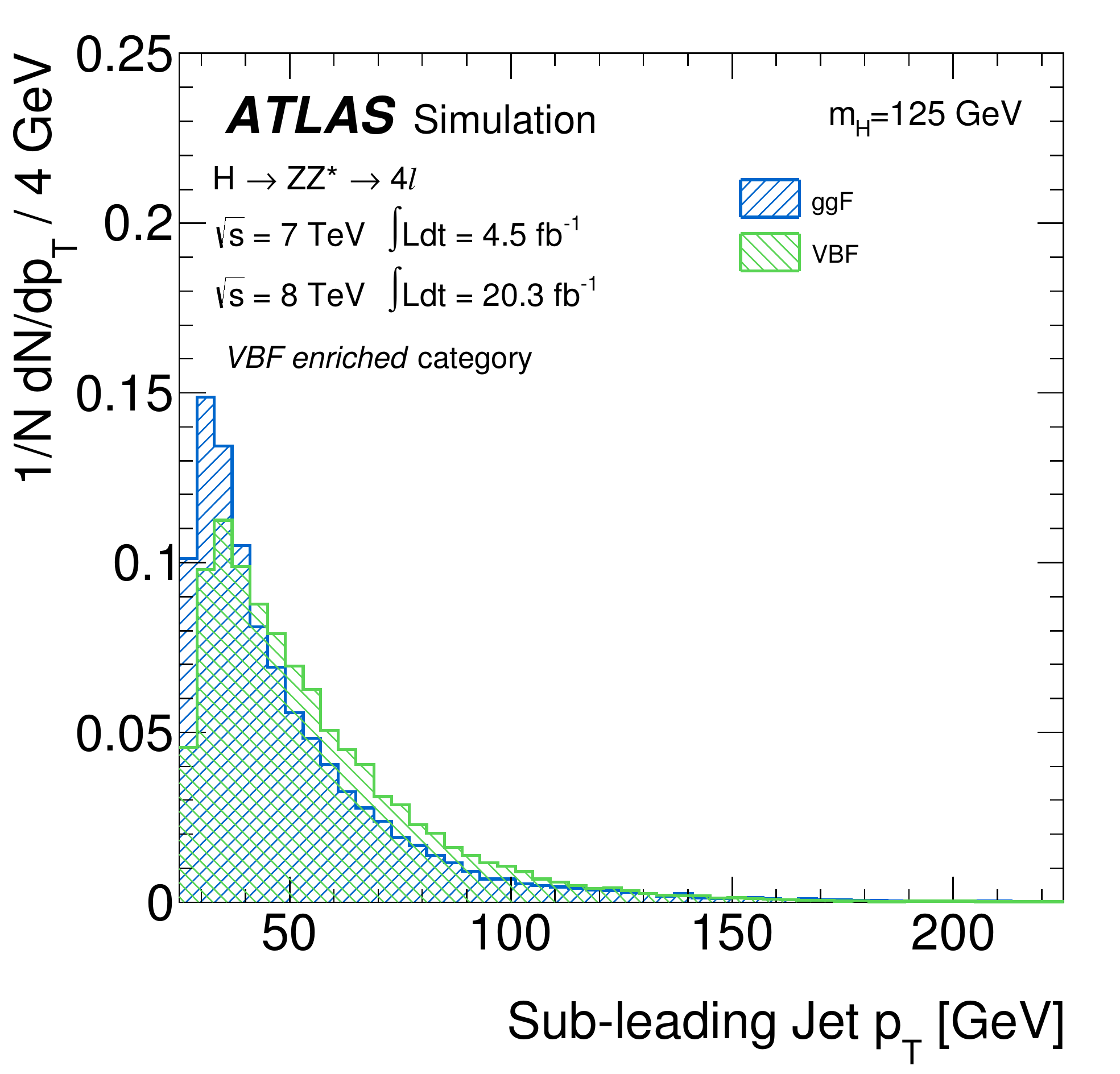}}
  \subfigure[\label{fig:subleadjet_eta}]{\includegraphics[width=\smallDoublePlotSize]{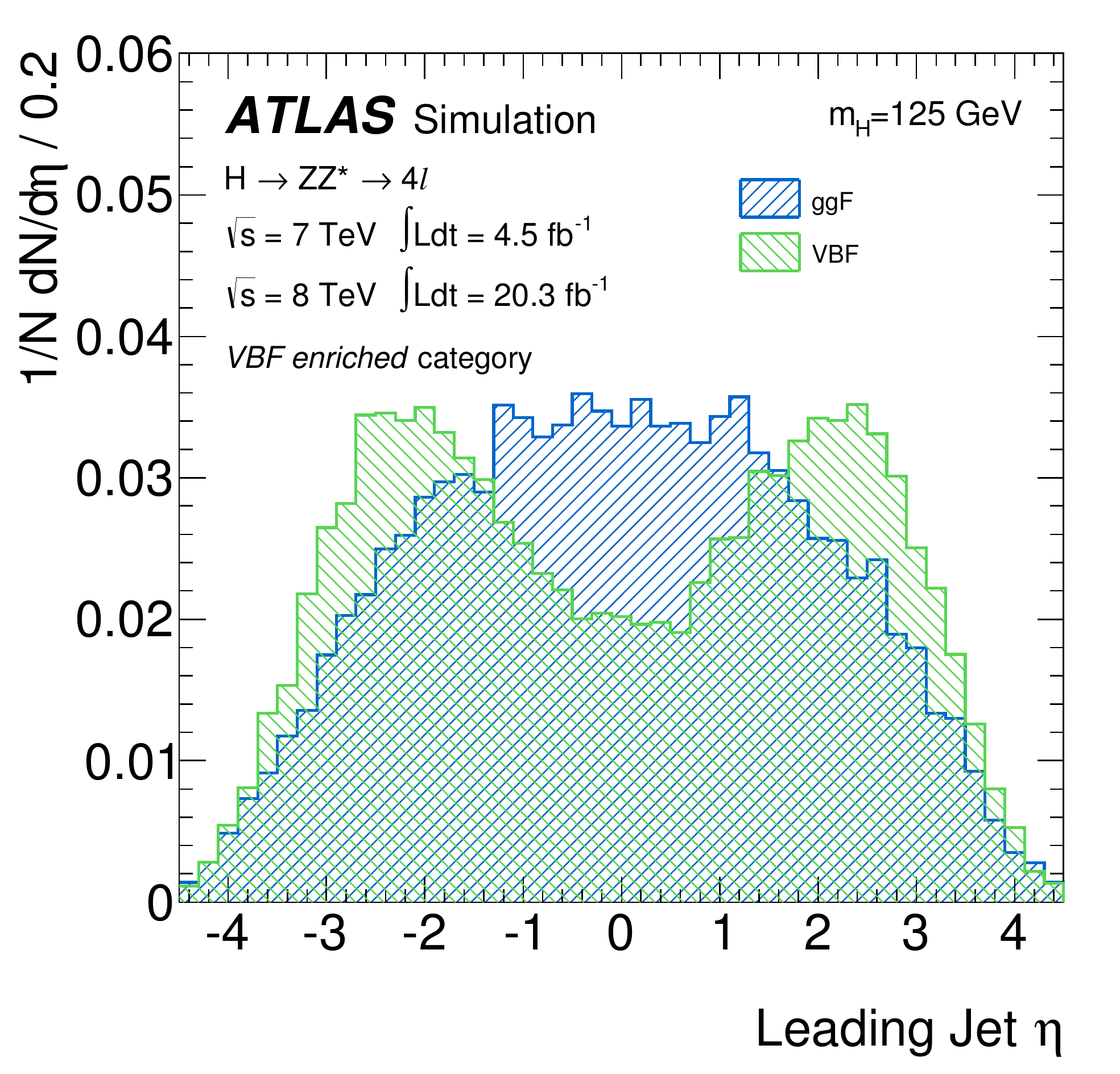}}
  \subfigure[\label{fig:bdt_output_two_cases}]{\includegraphics[width=\smallDoublePlotSize]{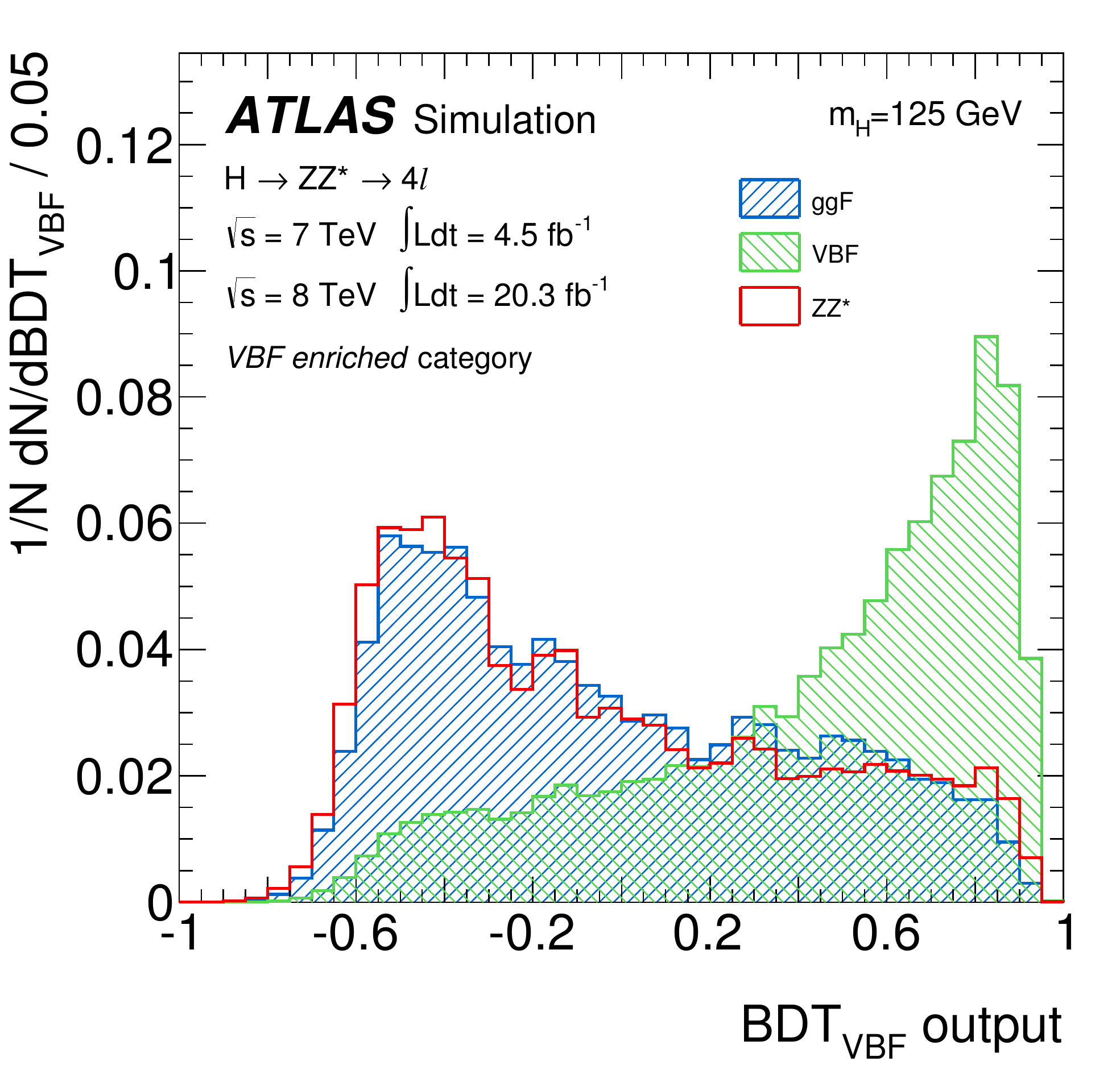}}
  \caption{Distribution of kinematic variables for signal (VBF events, green) and background (ggF
    events, blue) events used in the training of the VBF boosted decision tree:
    \subref{fig:dijet_mass} dijet invariant mass, \subref{fig:dijet_eta} dijet $\eta$ separation,
    \subref{fig:leadjet_pt} leading jet \pt, \subref{fig:subleadjet_pt} subleading jet \pt\ and
    \subref{fig:subleadjet_eta} leading jet $\eta$.  \subref{fig:bdt_output_two_cases} Output
    distributions of \bdtvbf\ for VBF and ggF events as well as for the \zzstar\ background
    (red). All histograms are normalized to the same area. \label{fig:bdt_input_vbf} }
\end{figure*}

\begin{figure}[!h]
  \centering
  \subfigure[]{\includegraphics[width=\singlePlotSize]{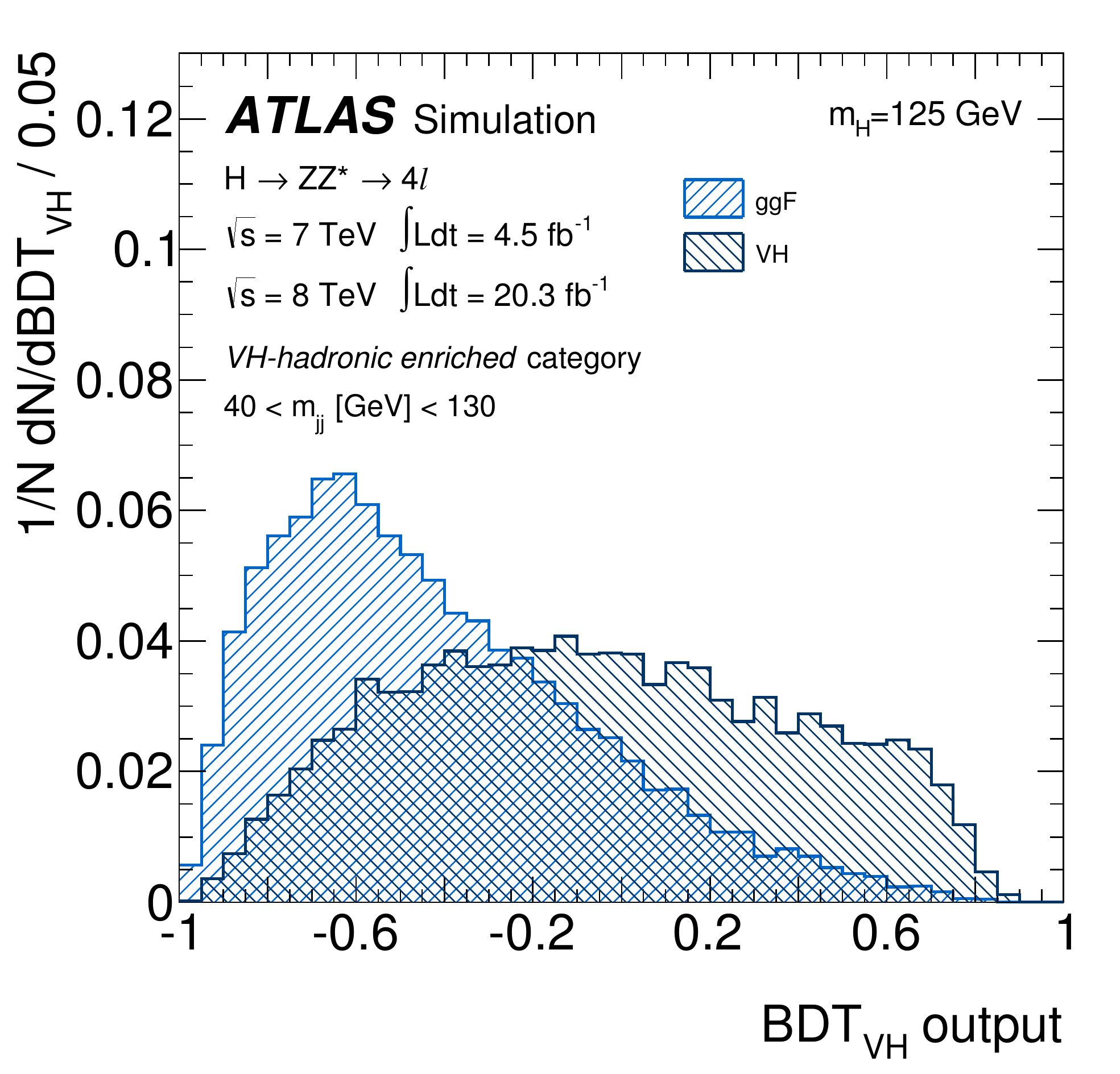}}
  \caption{Final \bdtvh\ discriminant output for the \vhhadcat\ category for signal (VH events, dark
    blue) and background (ggF events, blue) events.
  \label{fig:hadvh_bdt_discriminant} }
\end{figure}

For event categorization, two separate BDT classifiers were developed to discriminate against ggF
production: one for VBF production (\bdtvbf) and another for the vector boson hadronic decays of VH
production (\bdtvh). In the first case the BDT output is used as an observable together with
$m_{4\ell}$ in a maximum likelihood fit for the VBF category, while in the latter case the BDT
output value is used as a selection requirement for the event to be classified in the
\vhhadcat\ category, as discussed in \secref{sec:eventCategorisation}.  In both cases the same five
discriminating variables are used.  In order of decreasing separation power between the two
production modes, the variables are:
\begin{inparaenum}[\itshape a\upshape)] 
\item invariant mass of the dijet system, 
\item pseudorapidity separation between the two jets ($|\Delta\eta_{jj}|$),
\item transverse momentum of each jet, and
\item pseudorapidity of the leading jet.  
\end{inparaenum}
For the training of the BDT discriminant, fully simulated four-lepton Higgs boson signal events
produced through ggF and VBF production and hadronically decaying vector boson events for VH
production are used.  The distributions of these variables for \bdtvbf\ are presented in
Figs.\ \ref{fig:bdt_input_vbf}\subref{fig:dijet_mass}-\subref{fig:subleadjet_eta}, where all the
expected features of the VBF production of a Higgs boson can be seen: the dijet system has a high
invariant mass and the two jets are emitted in opposite high-$|\eta|$ regions with a considerable
$\Delta\eta$ separation between them. The jets of ggF events, on the other hand, are more centrally
produced and have a smaller invariant mass and $\Delta\eta$ separation.  The separation between VBF
and ggF can be seen in the output of \bdtvbf\ in \figref{fig:bdt_output_two_cases}, where the
separation between VBF and \zzstar\ is found to be similar.  The output of \bdtvbf\ is unchanged for
various mass points around the main training mass of $m_{H}=125$ \gev.  For variables entering the
\bdtvh\ discriminant, the invariant mass of the dijet system, which peaks at the $Z$ mass, exhibits
the most important difference between ggF and VH production modes.  The other variables have less
separation power.  The corresponding separation for \bdtvh\ is shown in
\figref{fig:hadvh_bdt_discriminant}.  As described in \secref{sec:eventCategorisation}, the
\vhhadcat\ category applies a selection on the \bdtvh\ discriminant ($< -0.4$) which optimizes the
signal significance.

\section{Signal and Background Modeling} 
\label{sec:SignalModel}


\subsection{Signal and background modeling for the inclusive analysis}
\label{sec:SignalModelInclusive}
\begin{figure*}
  \centering 
  \subfigure[\label{fig:bdtsig}]{\includegraphics[width=\doublePlusPlotSize]{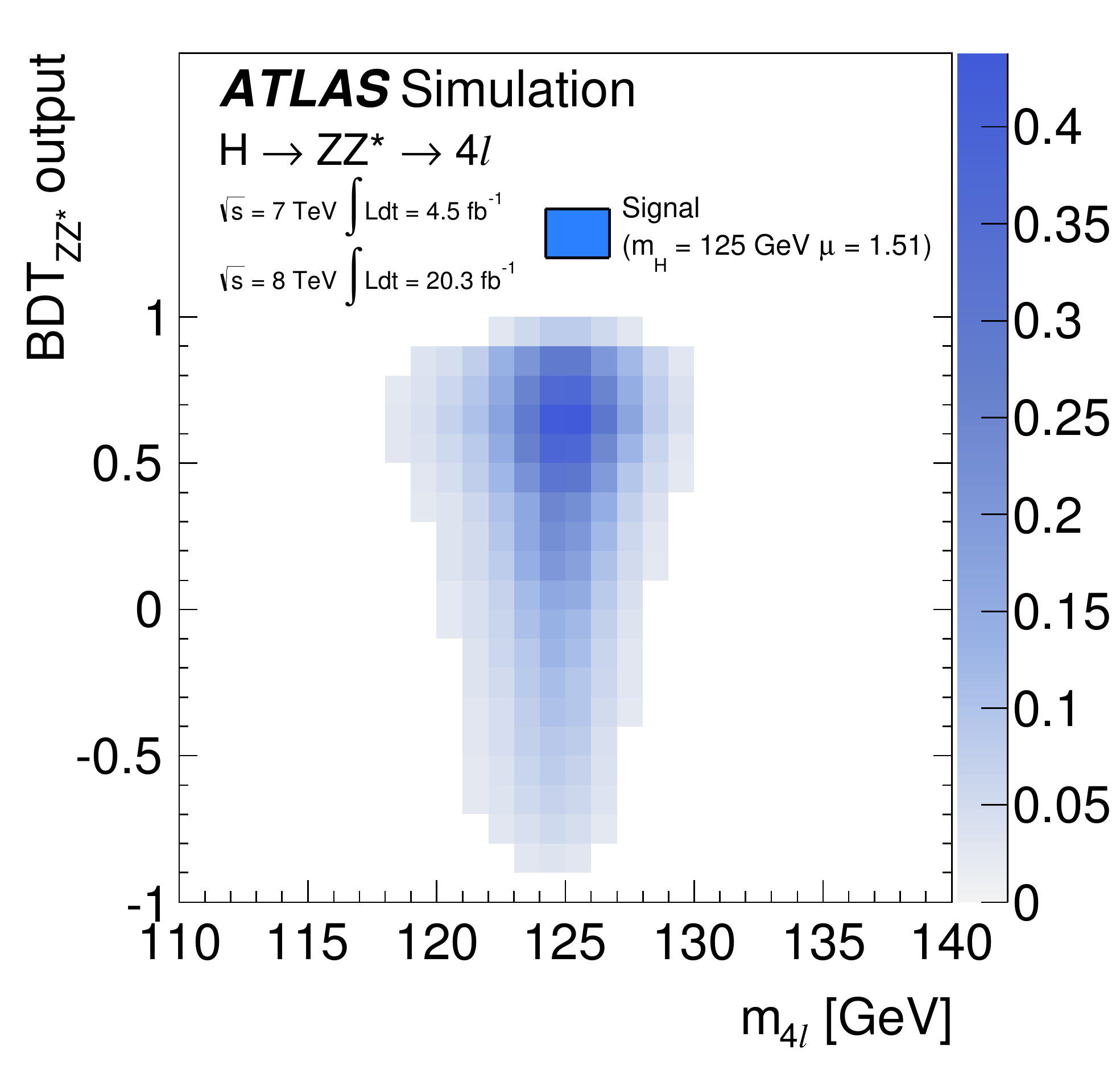}}
  \subfigure[\label{fig:bdtzz}]{\includegraphics[width=\doublePlotSize]{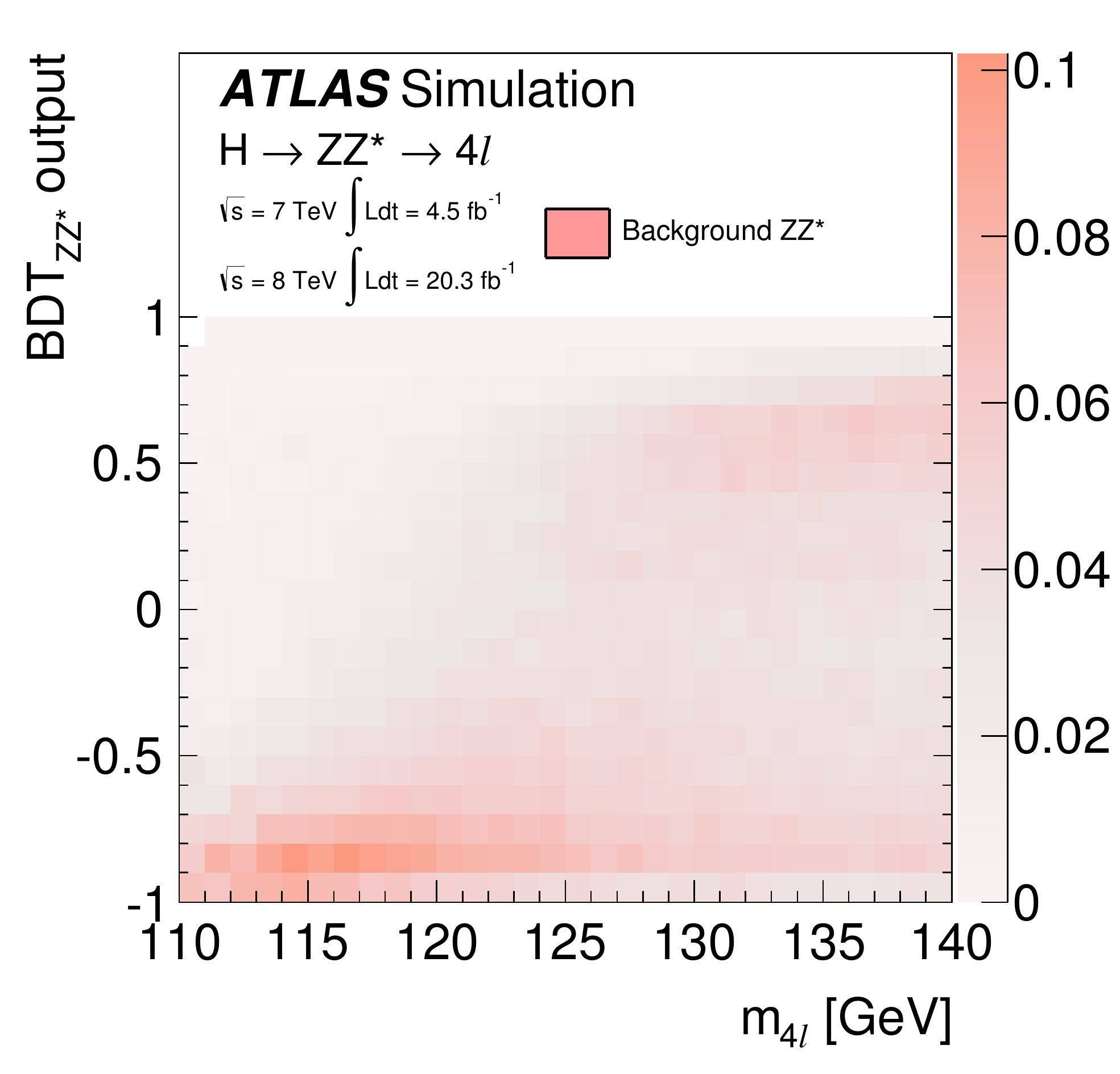}}
  \subfigure[\label{fig:bdtred}]{\includegraphics[width=\doublePlotSize]{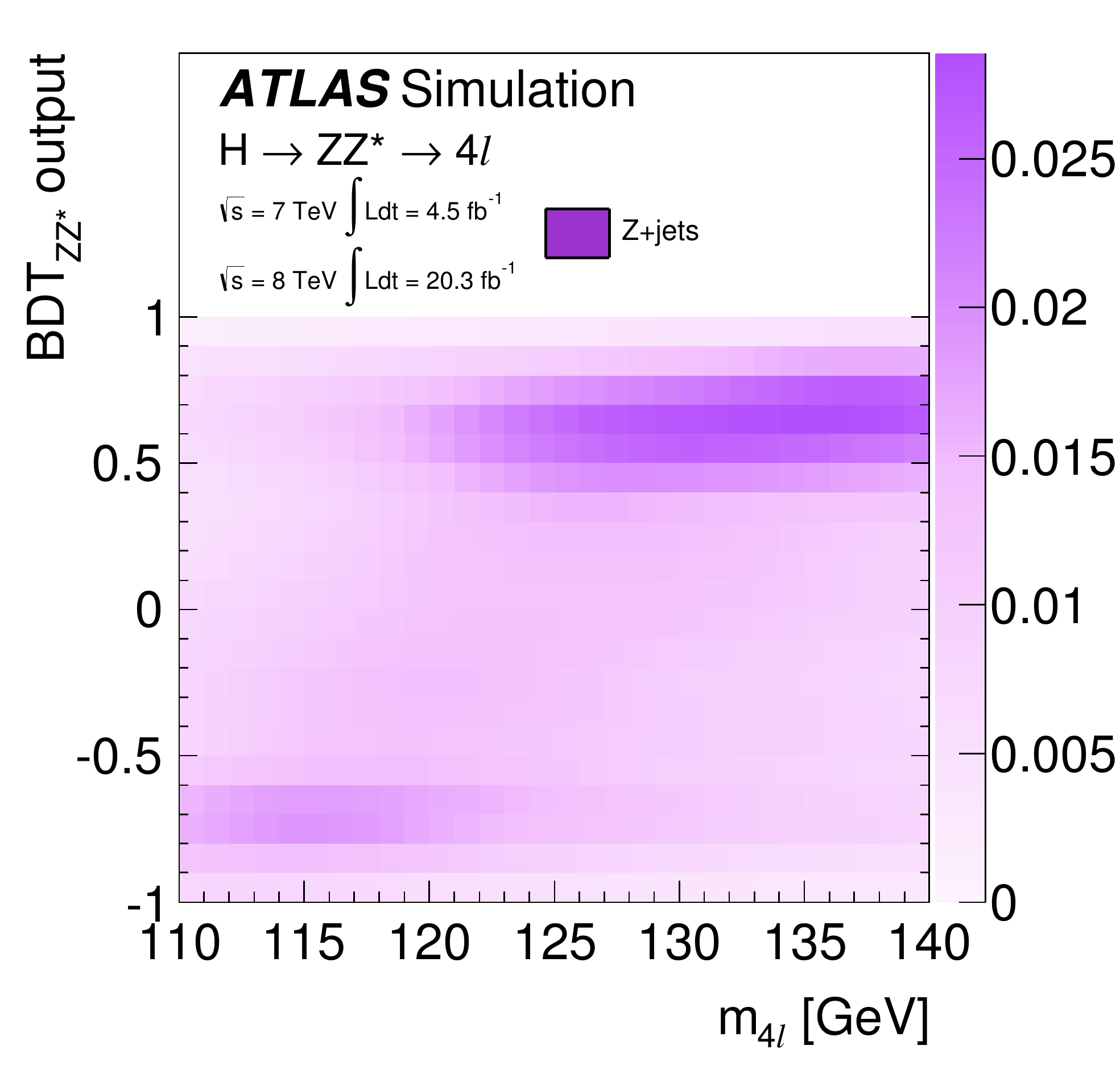}}
  \caption{Probability density for the signal and the different backgrounds normalized to the
    expected number of events for the 2011 and 2012 data sets, summing over all the final states:
    \subref{fig:bdtsig} $\mathcal{P}(m_{4\ell},\bdtzz~|~m_{H})$ for the signal assuming $m_{H}=125$
    \gev, \subref{fig:bdtzz} probability density $\mathcal{P}(m_{4\ell},\bdtzz)$ for the
    \zzstar\ background and \subref{fig:bdtred} $\mathcal{P}(m_{4\ell},\bdtzz)$ for the reducible
    background.
    \label{fig:bdts}}
\end{figure*}

\begin{figure*}
\centering 
\subfigure[]{\includegraphics[width=\doublePlotSize]{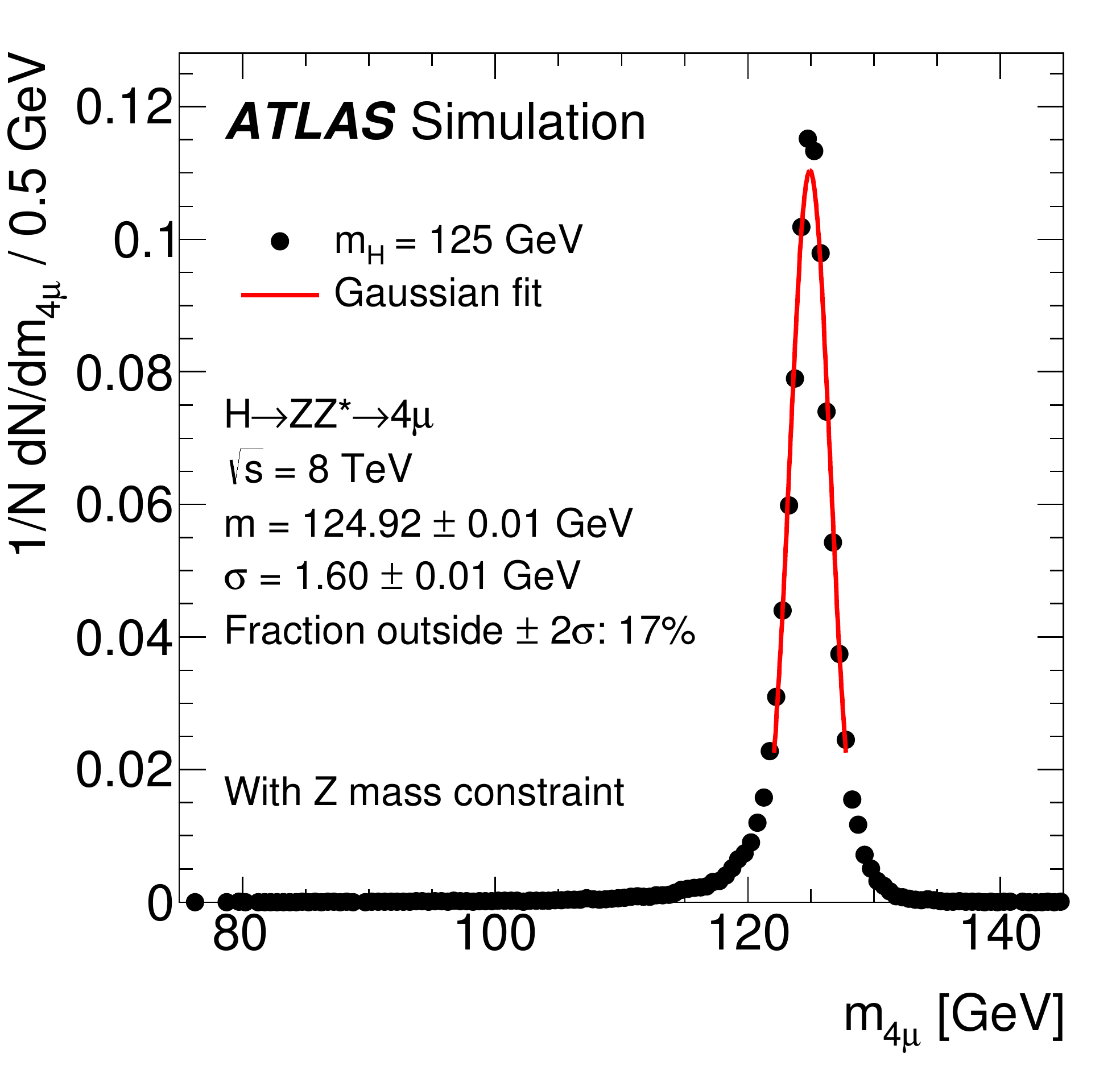}} 
\subfigure[]{\includegraphics[width=\doublePlotSize]{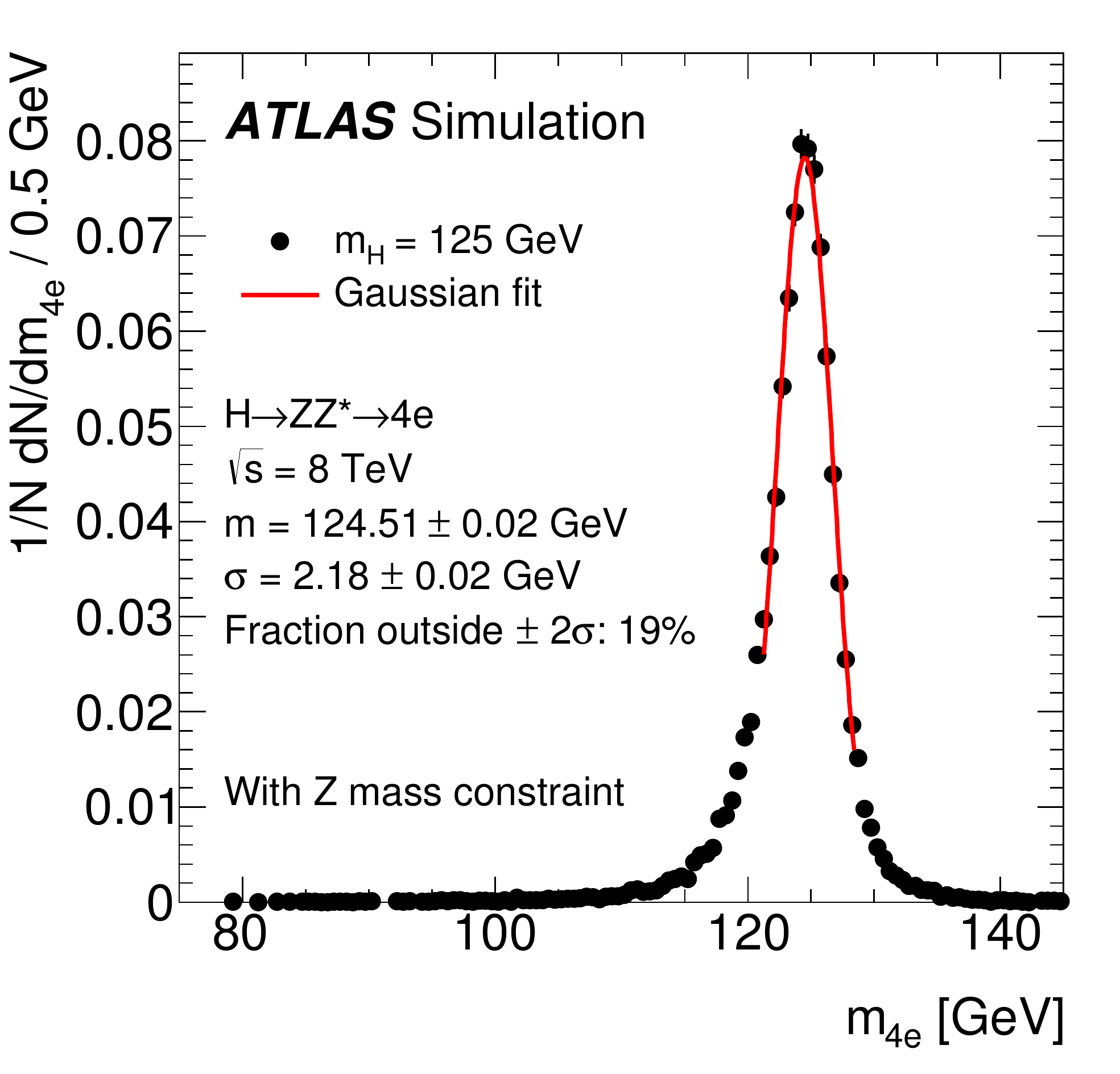}} 
\subfigure[]{\includegraphics[width=\doublePlotSize]{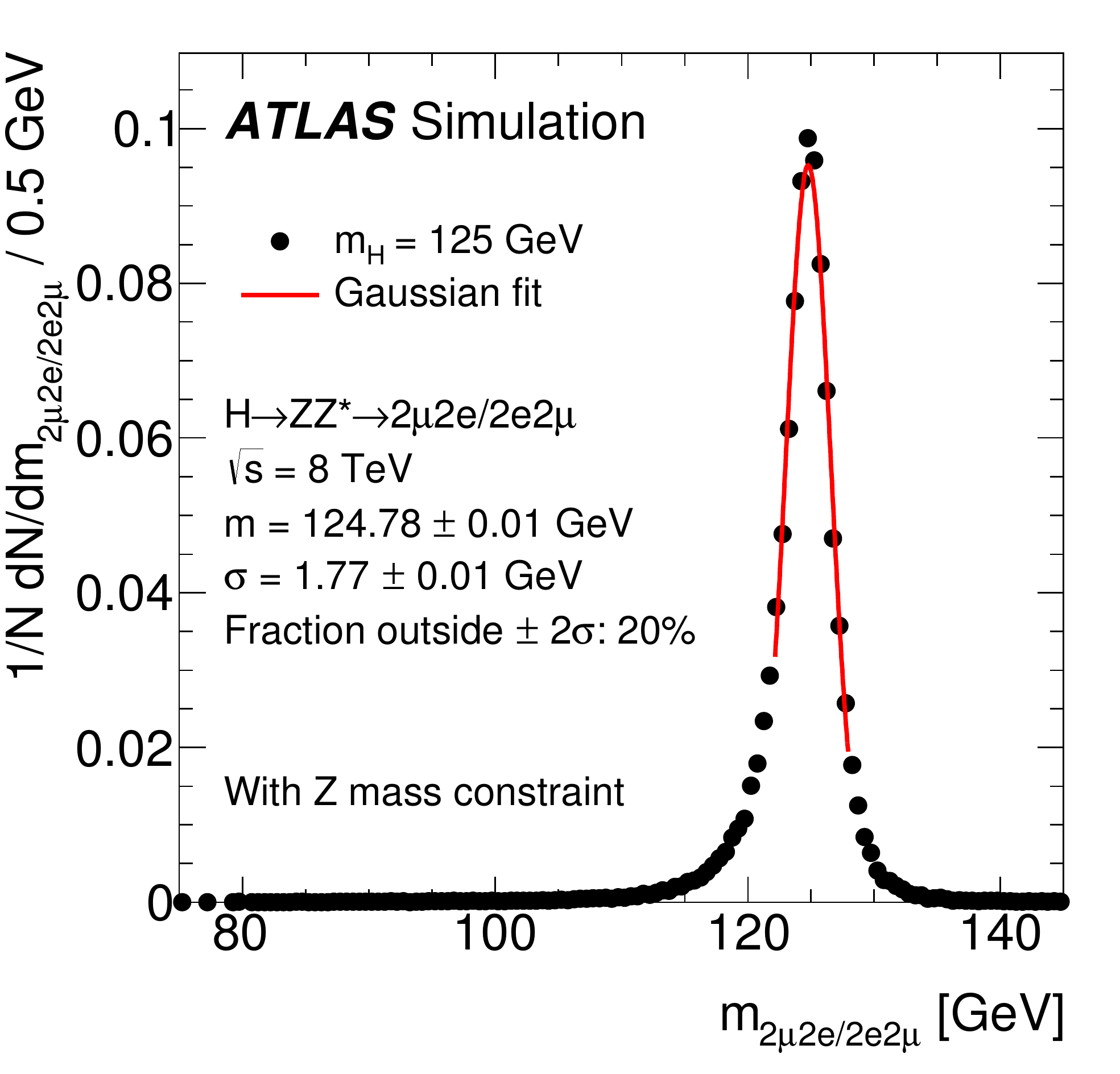}} 
\caption{Invariant mass distribution for a simulated signal sample with $m_{H}=125$ \gev;
  superimposed is the Gaussian fit to the $m_{4\ell}$ peak after the correction for final-state
  radiation and the $Z$-mass constraint. \label{fig:m4lresoall}}
\end{figure*}

For the measurements of the Higgs boson mass, of its natural width and of the inclusive production
rate relative to the SM expectation (the signal strength denoted as $\mu$) in the
\htollllbrief\ channel, three different parametrizations of the signal and background were
developed as described in Ref.~\cite{combmasspaper}, where the Higgs boson mass measurement is
reported.  The baseline method is a two-dimensional (2D) fit to $m_{4\ell}$ and the \bdtzz\ output
($O_{{\rm BDT}_{ZZ^{*}}}$).  This method provides the smallest expected uncertainties for both the
mass and inclusive signal strength measurements.  The one-dimensional (1D) fit to the $m_{4\ell}$
distribution that was used in the previous measurements~\cite{ATLAS:2012af,Aad:2013wqa} is used as a
cross-check.  A third method, using per-event resolution, is discussed after a description of the 1D
and 2D models.  The $m_{4\ell}$ range used in the fit for all of the methods is 110--140 \gev.  A
kernel density estimation method~\cite{Cranmer:2000du} uses fully simulated events to obtain smooth
distributions for both the 1D and 2D signal models.  These templates are produced using samples
generated at 15 different $m_{H}$ values in the range 115--130 \gev\ and parametrized as functions
of $m_H$ using B-spline interpolation~\cite{NURBS:1997}.  These simulation samples at different
masses are normalized to the expected SM $\sigma \times B$
~\cite{LHCHiggsCrossSectionWorkingGroup:2011ti} to derive the expected signal yields after
acceptance and selection.  The probability density function for the signal in the 2D fit is:
\iftoggle{isPRD} {
  \begin{widetext}
    \begin{linenomath}
      \begin{equation}
        \begin{aligned}
          \mathcal{P}(m_{4\ell},O_{{\rm BDT}_{ZZ^{*}}}~|~m_{H})&=\mathcal{P}(m_{4\ell}~|~O_{{\rm BDT}_{ZZ^{*}}},~m_{H})~ \mathcal{P}(O_{{\rm BDT}_{ZZ^{*}}}~|~m_{H})\\
          &\simeq  \left( \sum_{n=1}^{4} \mathcal{P}_n(m_{4\ell}~|~m_{H}) \theta_n (O_{{\rm BDT}_{ZZ^{*}}}) \right) \mathcal{P}(O_{{\rm BDT}_{ZZ^{*}}}~|~m_{H}) \\
        \end{aligned} \label{eq:sigpdf}
      \end{equation}
    \end{linenomath}
  \end{widetext}
}{ 
  \begin{linenomath}
    \begin{equation}
      \begin{aligned}
        \mathcal{P}(m_{4\ell},O_{{\rm BDT}_{ZZ^{*}}}~|~m_{H})&=\mathcal{P}(m_{4\ell}~|~O_{{\rm BDT}_{ZZ^{*}}},~m_{H})~ \mathcal{P}(O_{{\rm BDT}_{ZZ^{*}}}~|~m_{H})\\
        &\simeq  \left( \sum_{n=1}^{4} \mathcal{P}_n(m_{4\ell}~|~m_{H}) \theta_n (O_{{\rm BDT}_{ZZ^{*}}}) \right) \mathcal{P}(O_{{\rm BDT}_{ZZ^{*}}}~|~m_{H}) \\
      \end{aligned} \label{eq:sigpdf}
    \end{equation}
  \end{linenomath}
} where $\theta_{n}$ defines four equal-sized bins for the value of the \bdtzz\ output, and
$\mathcal{P}_n$ represents the 1D probability density function of the signal in the corresponding
\bdtzz\ bin.  The variation of the $m_{4\ell}$ shape is negligible within a single \bdtzz\ bin, so
no bias is introduced in the mass measurement.  The background model, $\mathcal{P}_{\rm
  bkg}(m_{4\ell},O_{{\rm BDT}_{ZZ^{*}}})$, is described using a two-dimensional probability
density. For the \zzstar\ and reducible $\ell\ell+\mu\mu$ backgrounds, the two-dimensional
probability density distributions are derived from simulation, where the $\ell\ell+\mu\mu$
simulation was shown to agree well with data in the control region.  For the $\ell\ell+ee$
background model, the two-dimensional probability density can only be obtained from data, which is
done using the $3\ell + X$ data control region weighted with the transfer factor to match the
kinematics of the signal region.  Figure~\ref{fig:bdts} shows the probability density in the \bdtzz
-- $m_{4\ell}$ plane, for the signal with $m_{H}$ = 125 \gev, the \zzstar\ background from
simulation and the reducible background from the data control region. The visible separation between
the signal and the background using the \bdtzz\ discriminant is exploited in the fit.
With respect to the 1D approach, there is an expected reduction of the statistical uncertainty for
the mass and inclusive signal strength measurements, which is estimated from simulation to be
approximately 8\% for both measurements.  Both the 1D and the 2D models are built using $m_{4\ell}$
after applying a $Z$-mass constraint to $m_{12}$ during the fit, as described in
\secref{sec:inclusiveAnalysis}.  \Figref{fig:m4lresoall} shows the $m_{4\ell}$ distribution for a
simulated signal sample with $m_{H}=125$ \gev, after applying the correction for final-state
radiation and the $Z$-mass constraint for the $4\mu$, $4e$ and $2e2\mu/2\mu2e$ final states.  The
width of the reconstructed Higgs boson mass for $m_{H}=125$ \gev\ ranges between 1.6 \gev\ ($4\mu$
final state) and 2.2 \gev\ ($4e$ final state) and is expected to be dominated by the experimental
resolution since, for $m_{H}$ of about 125 \gev, the natural width in the Standard Model is
approximately 4 MeV.

In addition to the 1D and 2D fit methods described above, the signal probability density for
$m_{4\ell}$ is also modeled on a per-event basis using both the \bdtzz\ information and the energy
resolution of the individual leptons. This method is referred to as the per-event-resolution model
and is used both as a cross-check for the mass measurement and as the baseline method to set an
upper limit on the Higgs boson total width $\Gamma_{H}$, which is discussed
elsewhere~\cite{combmasspaper}.  The detector-level $m_{4\ell}$ distribution for the signal is
obtained for each event through the convolution of an analytic description of the single-lepton
detector response with a Breit-Wigner function that describes the Higgs boson mass line shape.  The
$Z$-mass constraint is not applied in this fit because this introduces a correlation between the two
leptons of the leading $Z$ which must be included in their detector response functions.  The
parametrization of the muon and electron response function is performed in bins of $\eta$ and
\pt\ of the leptons and consists of the sum of two or three normal distributions.  This
parametrization takes into account the tails of the single-lepton responses.  A broad range of
cross-checks were performed to validate all the models described above~\cite{combmasspaper}.

A likelihood function ${\cal L}$ that depends on $m_{H}$ and $\mu$ is constructed using the signal
and background models defined above and is defined as
\iftoggle{isPRD} {
  \begin{widetext}
    \begin{linenomath}
      \begin{equation}
        {\cal L}(m_{H},\mu,\vecth) = \prod_{i}^{\mathrm{year}}\prod_{j}^{\mathrm{^{final}_{state}}} \mathrm{Poisson}(N_{ij}|\mu\cdot S_{ij} (m_{H},\vecth)+B_{ij}(\vecth) ) \cdot\prod_{k=1}^{N_{ij}}{\cal F}_{ij}((m_{4\ell},O_{{\rm BDT}_{ZZ^{*}}})_{k} ,m_{H},\mu,\vecth).
      \end{equation}
    \end{linenomath}
  \end{widetext}
}{ 
  \begin{linenomath}
    \begin{equation}
      {\cal L}(m_{H},\mu,\vecth) = \prod_{i}^{\mathrm{year}}\prod_{j}^{\mathrm{^{final}_{state}}} \mathrm{Poisson}(N_{ij}|\mu\cdot S_{ij} (m_{H},\vecth)+B_{ij}(\vecth) ) \cdot\prod_{k=1}^{N_{ij}}{\cal F}_{ij}((m_{4\ell},O_{{\rm BDT}_{ZZ^{*}}})_{k} ,m_{H},\mu,\vecth).
    \end{equation}
  \end{linenomath}
}
This likelihood function corresponds to the product of the Poisson probability of observing $N_{ij}$
events in the 2011 and 2012 data sets and each of the four final states, given the expectation for
the signal $S_{ij}$ and background $B_{ij}$, and is multiplied with the product of the values of the
probability density ${\cal F}_{ij}$, for $(m_{4\ell}, O_{{\rm BDT}_{ZZ^{*}}})_{k}$ of all
events. ${\cal F}_{ij}$ is constructed by using both the signal and background probability density
described above.  The symbol $\vecth$ represents the set of nuisance parameters used to model the
effect of the systematic uncertainties described in \secref{sec:systematics}.

The statistical procedure used to interpret the data is described in
Refs.~\cite{paper2012prd,Cowan:2010js} . The confidence intervals are based on the profile
likelihood ratios $\Lambda(\vecalpha)$ that depend on one or more parameters of interest
$\vecalpha$ (i.e. the Higgs boson mass or the signal strength) and on the nuisance parameters
\vecth\ :
\begin{linenomath}
  \begin{equation}
    \Lambda(\vecalpha) = { {\cal L}(\vecalpha, \hat{\hat{\vecth}}(\vecalpha))\over 
      {\cal L}(\hat{\vecalpha},\hat{\vecth})}.
  \end{equation}
\end{linenomath}
The likelihood fit to the data is then performed for the parameters of interest; 
$\hat{\hat{\vecth}}$ corresponds to the value of $\vecth$ which maximizes $\cal L$ for the specified
$\vecalpha$, and $\hat{\vecth}$ denotes the unconditional maximum likelihood estimate of the nuisance
parameters, i.e. where the likelihood is maximized for both $\vecth$  and $\vecalpha$.
In particular, the profile likelihood ratios $\Lambda(m_{H})$ and $\Lambda(\mu)$, used
for the Higgs boson mass and the inclusive signal strength measurements, respectively, are:
\iftoggle{isPRD} {
  \begin{linenomath}
    \begin{equation}
      \begin{aligned}
        \Lambda(m_{H}) &= {\frac{ {\cal L}(m_{H}, \hat{\hat{\mu}}(m_{H}),
            \hat{\hat{\vecth}}(m_{H}))}{{\cal L}(\hat{m}_{H},\hat{\mu},\hat{\vecth})}} 
        \hspace{0.5cm} {\rm~and} \hspace{0.5cm} \\
        \Lambda(\mu)  &= {\frac{ {\cal L}(\mu  , \hat{\hat{\vecth}}(\mu))}{ {\cal L}(\hat{\mu},\hat{\vecth})}},
      \end{aligned} 
    \end{equation}  
  \end{linenomath}
}{ 
  \begin{linenomath}
    \begin{equation}
        \Lambda(m_{H}) = {\frac{ {\cal L}(m_{H}, \hat{\hat{\mu}}(m_{H}),
            \hat{\hat{\vecth}}(m_{H}))}{{\cal L}(\hat{m}_{H},\hat{\mu},\hat{\vecth})}} 
        \hspace{0.5cm} {\rm~and}  \hspace{0.5cm} 
        \Lambda(\mu)  = {\frac{ {\cal L}(\mu  , \hat{\hat{\vecth}}(\mu))}{ {\cal L}(\hat{\mu},\hat{\vecth})}},
    \end{equation}  
  \end{linenomath}
} 
where the profile likelihood ratio for $m_H$ has the signal strength treated as a parameter of
interest in the fit, while that for $\mu$ is evaluated for a fixed value of $m_H$.

\subsection{Signal and background modeling for the categorized analysis}
\label{sec:SignalModelCat}

The model developed for the categorized analysis allows the measurement of the signal strength for
the different production modes.  Since no direct $t\bar{t}H$ and $b\bar{b}H$ production is observed,
a common signal strength $\mu_{{\rm ggF}+t\bar{t}H+b\bar{b}H}$ is assigned to gluon fusion,
$t\bar{t}H$ and $b\bar{b}H$ production.  This simplification is also justified by the fact that in
the SM the two production modes scale with the $q\bar{q}H$ ($q=b,t$) coupling.  Similarly, a common
signal strength $\mu_{\rm VBF+VH}$ is assigned to the VBF and VH production modes since in the SM
they scale with the $WH/ZH$ gauge couplings.

For the categorized analysis, all of the candidates are grouped into four separate categories to
have better sensitivity to the different production mechanisms, as described in
\secref{sec:eventSelection}.  In the \vbfcat\ category, where the \bdtvbf\ discriminant is
introduced to separate the ggF-like events from VBF-like events, the two-dimensional probability
density $\mathcal{P}(m_{4\ell},\bdtvbf)$ is constructed by factorizing the \bdtvbf\ and
$m_{4\ell}$ distributions.  This factorization is justified by the negligible dependence of the
\bdtvbf\ on $m_{4\ell}$ for both signal and background. The \bdtvbf\ dependence on the Higgs boson
mass is negligible and is neglected in the probability density.  Adding the \bdtvbf\ in the
\vbfcat\ category reduces the expected uncertainty on the signal strength of the VBF and VH
production mechanisms $\mu_{\rm VBF+VH}$ by about 25\%.  The improvement in the expected uncertainty
on $\mu_{\rm VBF+VH}$ reaches approximately 35\% after adding the leptonic and hadronic VH
categories to the model.  In these two VH categories, a simple one-dimensional fit to the
$m_{4\ell}$ observable is performed, since for the \vhhadcat\ category, a selection on the
\bdtvh\ output is included in the event selection, while for the \vhlepcat\ category, no BDT is
used. Finally, in the \ggfcat\ category, the 2D model defined in Eq.~(\ref{eq:sigpdf}), including
the \bdtzz\ trained as specified in~\secref{sec:zz_bdt}, is used. These procedures allow a further
reduction of the expected uncertainty on $\mu_{\rm VBF+VH}$ ($\mu_{{\rm ggF}+t\bar{t}H+b\bar{b}H}$)
by 6\% (8\%).

\section{Systematic Uncertainties} 
\label{sec:systematics}



The uncertainties on the lepton reconstruction and identification efficiency, and on the lepton
energy or momentum resolution and scale, are determined using samples of $W$, \Zboson~and $J/\psi$
decays.  The description of these systematic uncertainties, as well as of the uncertainties
associated with the event categorizations, is separated into three parts.  A brief overview of the
systematic uncertainties that affect the mass measurement is given in \secref{sec:systematicsMass}.
The description of the systematic uncertainties related to the measurement of the signal rate and
event categorizations is provided in Secs.\ \ref{sec:systematicsEff} and
\ref{sec:systematicsCategories}, respectively.

\subsection{Systematic uncertainties in the mass measurement}
\label{sec:systematicsMass}
For the \htollllbrief\ decay modes involving electrons, the electron energy scale uncertainty,
determined from \Zee\ and $J/\psi\rightarrow ee$ decays, is propagated as a function of the
pseudorapidity and the transverse energy of the electrons. The precision of the energy scale is
better than 0.1\% for $|\eta|<1.2$ and $1.8<|\eta|<2.47$, and a few per mille for
$1.2<|\eta|<1.8$~\cite{ref:run1-egamma-calib}.  The uncertainties on the measured Higgs boson mass
due to the electron energy scale uncertainties are $\pm 0.04\%$, $\pm 0.025\%$ and $\pm 0.04\%$ for
the $4e$, $2e2\mu$ and $2\mu2e$ final states, respectively.

Similarly, for the \htollllbrief\ decay modes involving muons, the various components of the
systematic uncertainty on the muon momentum scale are determined using large samples of $J/\psi
\rightarrow \mu\mu$ and $Z \rightarrow \mu\mu$ decays and validated using $\Upsilon \rightarrow
\mu\mu$, $J/\psi \rightarrow \mu\mu$ and $Z \rightarrow \mu\mu$ decays.  In the muon transverse
momentum range of 6--100 \gev, the systematic uncertainties on the scales are about $\pm 0.04\%$ in
the barrel region and reach $\pm 0.2\%$ in the region $|\eta|>2$~\cite{MCPpaper2014}. The
uncertainties on the measured Higgs boson mass due to the muon energy scale uncertainties are
estimated to be $\pm 0.04\%$, $\pm 0.015\%$ and $\pm 0.02\%$ for the $4\mu$, $2e2\mu$ and $2\mu2e$
final states, respectively.

Uncertainties on the measured Higgs boson mass related to the background contamination and
final-state QED~radiation modeling are negligible compared to the other sources described
above. 

The weighted contributions to the uncertainty in the mass measurement, when all the final states are
combined, are $\pm 0.01\%$ for the electron energy scale uncertainty and $\pm 0.03\%$ for the muon
momentum scale uncertainty.  The larger impact of the muon momentum scale uncertainty is due to the
fact that the muon final states have a greater weight in the combined mass fit.

\subsection{Systematic uncertainties in the inclusive signal strength measurement}
\label{sec:systematicsEff}
The efficiencies to trigger, reconstruct and identify electrons and muons are studied using $Z
\rightarrow \ell\ell$ and $J/\psi \rightarrow\ell\ell$
decays~\cite{ElectronEff2012,ElectronEff2011,ATLAS-CONF-2013-088,MCPpaper2014}.  The expected impact
from simulation of the associated systematic uncertainties on the signal yield is presented in
Table~\ref{tab:efficiency_syst}.  The impact is presented for the individual final states and for
all channels combined.

The level of agreement between data and simulation for the efficiency of the isolation and impact
parameter requirements of the analysis is studied using a tag-and-probe method. As a result, a small
additional uncertainty on the isolation and impact parameter selection efficiency is applied for
electrons with \et\ below 15 \gev.  The effect of the isolation and impact parameter uncertainties
on the signal strength is given in Table~\ref{tab:efficiency_syst}.  The corresponding uncertainty
for muons is found to be negligible.
 
The uncertainties on the data-driven estimates of the background yields are are discussed in
\secref{sec:Background} and are summarized in Tables~\ref{tab:fitSR} and \ref{tab:bkg_overviewllee},
and their impact on the signal strength is given in Table~\ref{tab:efficiency_syst}.

The overall uncertainty on the integrated luminosity for the complete 2011 data set is
$\pm$1.8\%~\cite{ref:AtlasLuminosity-2011-final}.  The uncertainty on the integrated luminosity for
the 2012 data set is $\pm$2.8\%; this uncertainty is derived following the methodology used for the
2011 data set, from a preliminary calibration of the luminosity scale with beam-separation scans
performed in November 2012.

The theory-related systematic uncertainty for both the signal and the \zzstar\ background is
discussed in \secref{sec:SandBSim}.  The three most important theoretical uncertainties, which
dominate the signal strength uncertainty, are given in Table~\ref{tab:efficiency_syst}.
Uncertainties on the predicted Higgs boson \pt\ spectrum due to those on the PDFs and higher-order
corrections are estimated to affect the signal strength by less than $\pm$1\%.  The systematic
uncertainty of the \zzstar\ background rate is around $\pm$4\% for $m_{4\ell}$ = 125 \gev\ and
increases for higher mass, averaging to around $\pm$6\% for the \zzstar\ production above 110 \gev.

\begin{table*}[h]
  \caption{The expected impact of the systematic uncertainties on the signal yield, derived from
    simulation, for $m_H=125$ \gev, are summarized for each of the four final states for the
    combined 4.5 \ifb\ at $\sqrt{s}=7$ \tev\ and 20.3 \ifb\ at $\sqrt{s}=8$ \tev. The symbol ``--''
    signifies that the systematic uncertainty does not contribute to a particular final state. The
    last three systematic uncertainties apply equally to all final states. All uncertainties have
    been symmetrized.
    \label{tab:efficiency_syst}}
  \vspace{0.2cm}
  \begin{center}
    \scalebox{1.00}{
      \begin{tabular}{lccccc}
	\hline \hline
        \noalign{\vspace{0.05cm}}
        Source of uncertainty          & $4\mu$   & $2e2\mu$ & $2\mu2e$ & $4e$   & combined \\

        \noalign{\vspace{0.05cm}}
        \hline
        \noalign{\vspace{0.05cm}}
        Electron reconstruction and identification efficiencies 
                                       & --       & \ 1.7\%  & 3.3\%    & 4.4\%  &  1.6\%   \\
        Electron isolation and impact parameter selection                                
                                       & --       & 0.07\%   & 1.1\%    & 1.2\%  &  0.5\%   \\
        Electron trigger efficiency    & --       & 0.21\%   & 0.05\%   & 0.21\% &$<$0.2\%  \\
        $\ell\ell + ee$ backgrounds    & --       & --       & 3.4\%    & 3.4\%  &  1.3\%   \\
        \noalign{\vspace{0.05cm}}
        \hline                                                                           
        \noalign{\vspace{0.05cm}}
        Muon reconstruction and identification efficiencies                              
                                       & 1.9\%    & 1.1\%\   & 0.8\%    & --     &  1.5\%   \\
        Muon trigger efficiency        & 0.6\%    & 0.03\%   & 0.6\%    & --     &  0.2\%   \\
        $\ell\ell + \mu\mu$ backgrounds& 1.6\%    & 1.6\%\   & --       & --     &  1.2\%   \\
        \noalign{\vspace{0.05cm}}
        \hline                                                                           
        \noalign{\vspace{0.05cm}}
        QCD scale uncertainty            &        &          &          &        &  6.5\%   \\
        PDF, $\alpha_s$ uncertainty      &        &          &          &        &  6.0\%   \\
        $H\rightarrow ZZ^{*}$ branching ratio uncertainty &        &          &          &        &  4.0\%   \\
        \noalign{\vspace{0.05cm}}
        \hline \hline 
    \end{tabular}}
  \end{center}
\end{table*}

\subsection{Systematic uncertainties in the event categorization }
\label{sec:systematicsCategories}

The systematic uncertainties on the expected yields (as in Table~\ref{tab:categoryexpected}) from
different processes contributing to the \vbfcat, \vhhadcat, \vhlepcat\ and \ggfcat\ categories are
reported in Table~\ref{tab:categoryesyst2012}, expressed as the fractional uncertainties on the
yields. The uncertainties on the theoretical predictions for the cross sections for the different
processes arise mainly from the requirement on the jet multiplicity used in the event
categorization~\cite{Stewart:2011cf,Heinemeyer:2013tqa}.  Because of event migrations, this also
affects the \vhlepcat\ and \ggfcat\ categories, where no explicit requirement on jets is applied.
The uncertainty accounting for a potential mismodeling of the underlying event is conservatively
estimated with $Z \to \mu\mu$ simulated events by applying the selection for the \vbfcat\ (or
\vhhadcat) category and taking the difference of the efficiencies with and without multiparton
interactions.

The main experimental uncertainty is related to the jet energy scale determination, including the
uncertainties associated with the modeling of the absolute and relative \textit{in situ} jet
calibrations, as well as the flavor composition of the jet sample. The impact on the yields of the
various categories is anticorrelated because a variation of the jet energy scale results primarily
in the migration of events among the categories. The impact of the jet energy scale uncertainty
results in an uncertainty of about $\pm$10\% for the \vbfcat\ category, $\pm$8\% for the
\vhhadcat\ category, $\pm$1.5\% for the \vhlepcat\ category and $\pm$1.5\% for the
\ggfcat\ category.

The uncertainty on the jet energy resolution is also taken into account, even though its impact is
small compared to that of the jet energy scale uncertainty, as reported in
Table~\ref{tab:categoryesyst2012}. Finally, the uncertainties associated with the additional leptons
in the \vhlepcat\ category are the same as already described in Sec.~\ref{sec:systematicsEff} for
the four leptons of the Higgs boson decay.

\begin{table*}[h]
  \centering
  \caption{ Systematic uncertainties on the yields expected from various processes contributing to
    the \vbfcat, \vhlepcat, \vhhadcat\ and \ggfcat\ categories expressed as percentages of the
    yield.  The various uncertainties are added in quadrature.  Uncertainties that are negligible
    are denoted by a ``$-$''. All uncertainties have been symmetrized.
    \vspace{0.2cm}
    \label{tab:categoryesyst2012}}
  \begin{tabular}{lcccc}
    \hline
    \hline
    \noalign{\vspace{0.05cm}}
    Process                & $gg \to H,q\bar{q}/gg\to b\bar{b}H/t\bar{t}H$  & $qq'\to Hqq'$    &   $q\bar{q}\to W/ZH$ & \zzstar \\
    \noalign{\vspace{0.05cm}}
    \hline
    \noalign{\vskip 0.2mm}
    \multicolumn{5}{c}{\vbfcat\  category} \\
    \noalign{\vskip 0.2mm}
    \hline
    \noalign{\vspace{0.05cm}}
    Theoretical cross section & 20.4\%                               &   4\%            &   4\%                & 8\%  \\
    Underlying event          & 6.6\%                                &  1.4\%           &   --                 & --     \\
    Jet energy scale          & 9.6\%                                &  4.8\%           &   7.8\%              & 9.6\%  \\
    Jet energy resolution     & 0.9\%                                &  0.2\%           &   1.0\%              & 1.4\% \\
    Total                     & 23.5\%                               &  6.4\%           &   8.8\%              & 12.6\%  \\
    \noalign{\vspace{0.05cm}}
    \hline
    \noalign{\vskip 0.2mm}
    \multicolumn{5}{c}{\vhhadcat\  category} \\
    \noalign{\vskip 0.2mm}
    \hline
    \noalign{\vspace{0.05cm}}
    
    Theoretical cross section & 20.4\%                               &   4\%            &   4\%                & 2\%  \\
    Underlying event          & 7.5\%                                &  3.1\%           &   --                  & --     \\
    Jet energy scale          & 9.4\%                                &  9.3\%           &   3.7\%              & 12.6\%  \\
    Jet energy resolution     & 1.0\%                                &  1.7\%           &   0.6\%              & 1.8\% \\
    Total                     & 23.7\%                               & 10.7\%           &   5.5\%              & 12.9\% \\
    \noalign{\vspace{0.05cm}}
    \hline
    \noalign{\vskip 0.2mm}
    \multicolumn{5}{c}{\vhlepcat\  category} \\
    \noalign{\vskip 0.2mm}
    \hline
    \noalign{\vspace{0.05cm}}
    
    Theoretical cross section & 12\%                             &   4\%            &   4\%                & 5\% \\ 
    Leptonic VH-specific cuts & 1\%                              &   1\%            &   5\%                & --   \\
    Jet energy scale          & 8.8\%                            & 9.9\%            &   1.7\%              & 3.2\% \\ 
    Total                     & 14.9\%                           & 10.7\%           &   6.6\%              & 5.9\% \\
    \noalign{\vspace{0.05cm}}
    \hline
    \noalign{\vskip 0.2mm}
    \multicolumn{5}{c}{\ggfcat\  category} \\
    \noalign{\vskip 0.2mm}
    \hline
    \noalign{\vspace{0.05cm}}
    
    Theoretical cross section & 12\%                                  &  4\%             &   4\%                & 4\%  \\
    Jet energy scale          & 2.2\%                                 &  6.6\%           &   4.0\%              & 1.0\%  \\
    Total                     & 12.2\%                                &  7.7\%           &   5.7\%              & 4.1\% \\
    \noalign{\vspace{0.05cm}}
    \hline
    \hline
  \end{tabular}
\end{table*}


\clearpage
\section{Results}
\label{sec:results}

\subsection{Results of the inclusive analysis}
\label{sec:inclres}

As described in \secref{sec:inclusiveAnalysis}, the inclusive selection is used to measure the Higgs
boson mass.  In addition, the inclusive signal strength measurement, described below, allows a
direct comparison with the predicted total production cross section times branching ratio of the
Standard Model Higgs boson at the measured mass.  This inclusive analysis is the same as that used
for the combined mass measurement~\cite{combmasspaper}; in the following more details and new
comparisons of the data and expectations are provided in view of the inclusive mass and signal
strength measurements.

\subsubsection{Signal and background yields}
The number of observed candidate events for each of the four decay channels in a mass window of
120--130 \gev\ and the signal and background expectations are presented in Table~\ref{tab:yields}.
The signal and \zzstar\ background expectations are normalized to the SM expectation while the
reducible background is normalized to the data-driven estimate described in
\secref{sec:Background}. Three events in the mass range $120<m_{4\ell}<130$ \gev\ are corrected for
FSR: one 4$\mu$ event and one $2\mu2e$ are corrected for noncollinear FSR, and one $2\mu2e$ event
is corrected for collinear FSR. In the full mass spectrum, there are 8 (2) events corrected for
collinear (noncollinear) FSR, in good agreement with the expected number of 11 events.

\begin{figure*}[!htbp]
  \centering 
  \subfigure[\label{fig:finalMassesSignala}]{\includegraphics[width=\doublePlotSize]{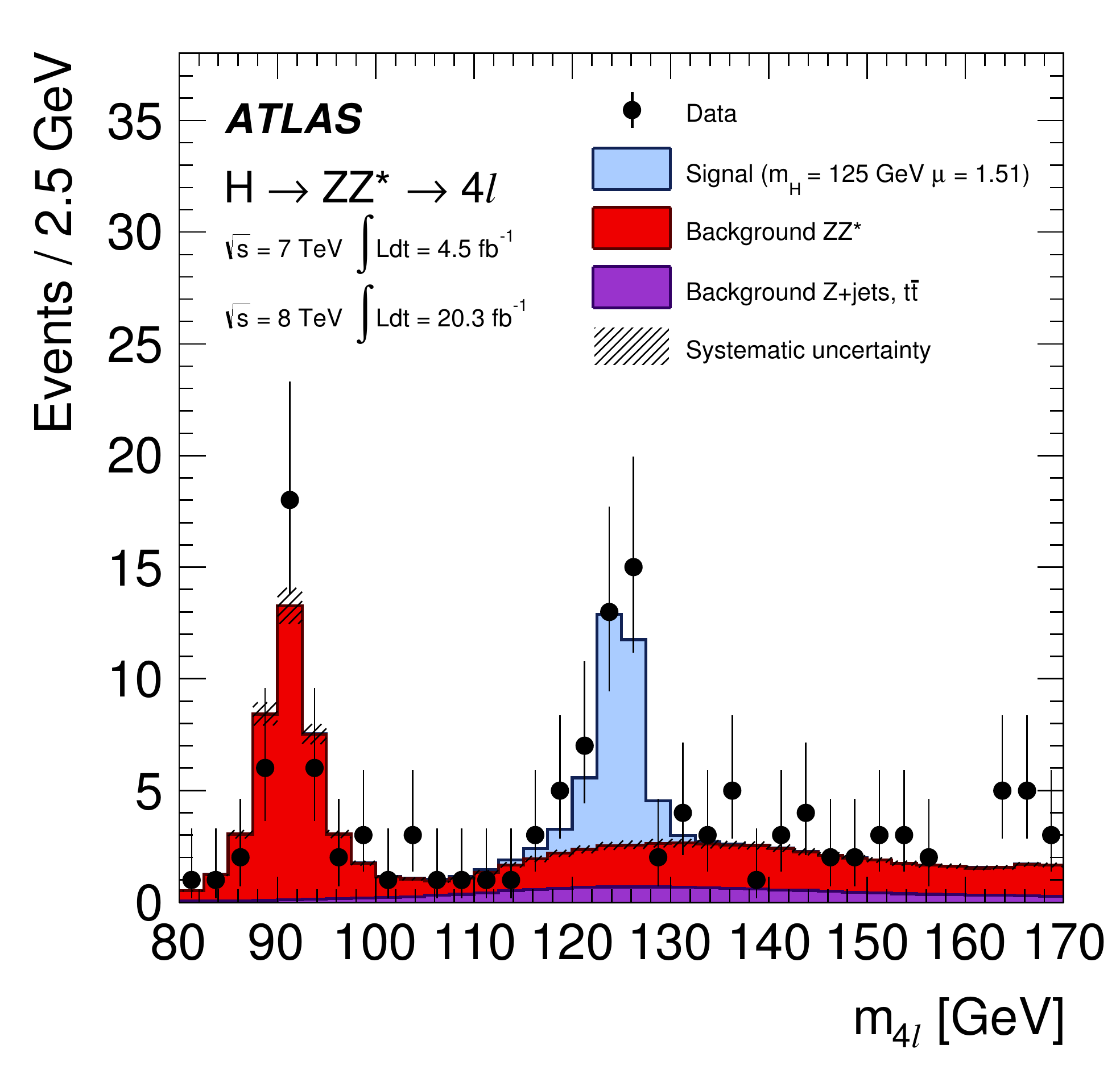}} 
  \subfigure[\label{fig:finalMassesSignalb}]{\includegraphics[width=\doublePlotSize]{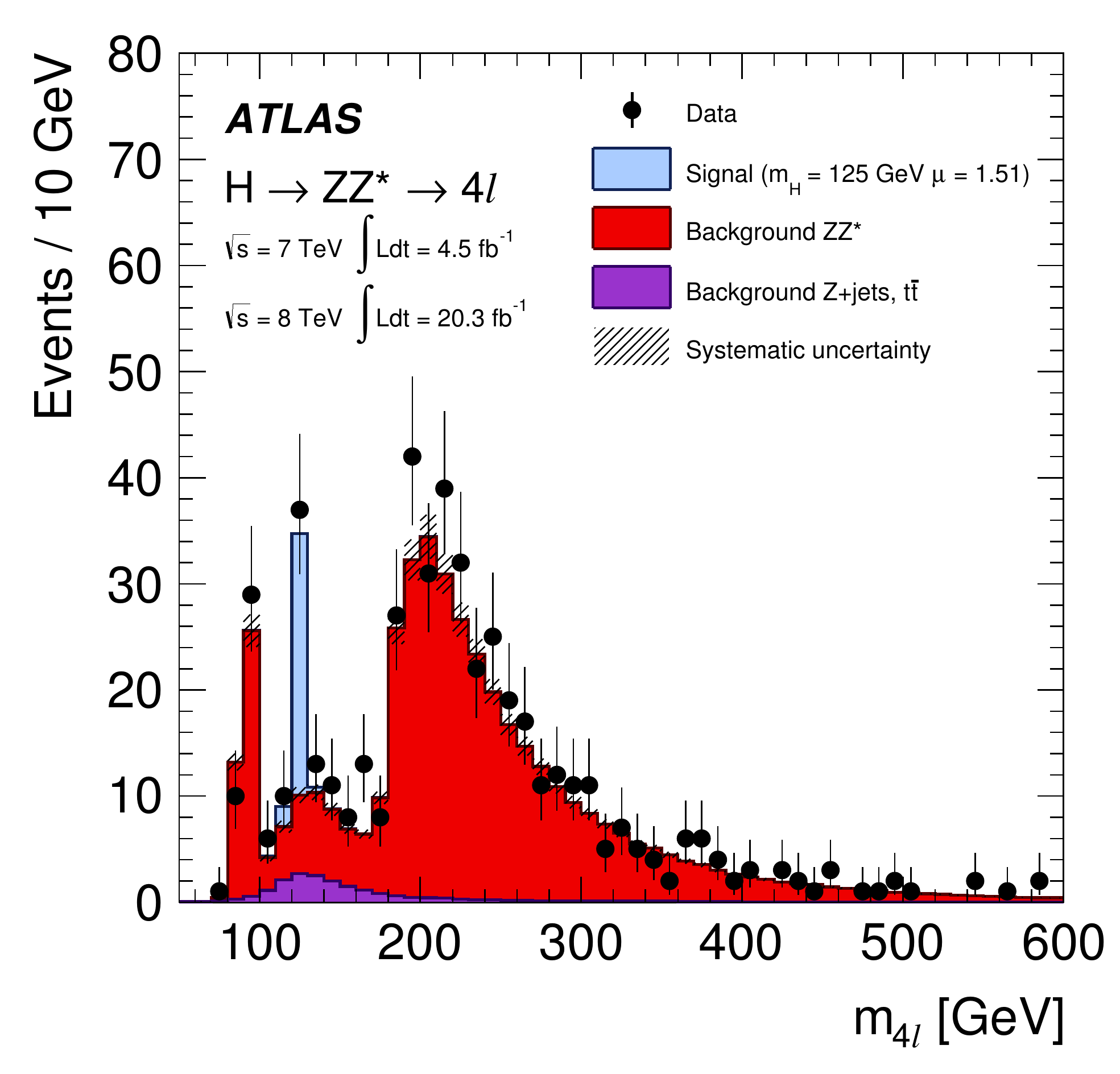}} 
  \caption{The distribution of the four-lepton invariant mass, $m_{4\ell}$, for the selected
    candidates (filled circles) compared to the expected signal and background contributions (filled
    histograms) for the combined $\sqrt{s}=7$ \tev\ and $\sqrt{s}=8$ \tev\ data for the mass ranges:
    \subref{fig:finalMassesSignala} 80--170 \gev, and \subref{fig:finalMassesSignalb} 80--600 \gev.
    The signal expectation shown is for a mass hypothesis of $m_{H}=125$ \gev\ and normalized to
    $\mu=1.51$ (see text).  The expected backgrounds are shown separately for the \zzstar\ (red
    histogram), and the reducible $Z+\rm jets$ and $t\bar{t}$ backgrounds (violet histogram); the
    systematic uncertainty associated to the total background contribution is represented by the
    hatched areas.  \label{fig:finalMasses}}
\end{figure*}

\begin{figure*}
  \centering 
  \subfigure[\label{fig:M34M12}]        {\includegraphics[width=\doublePlusPlotSize]{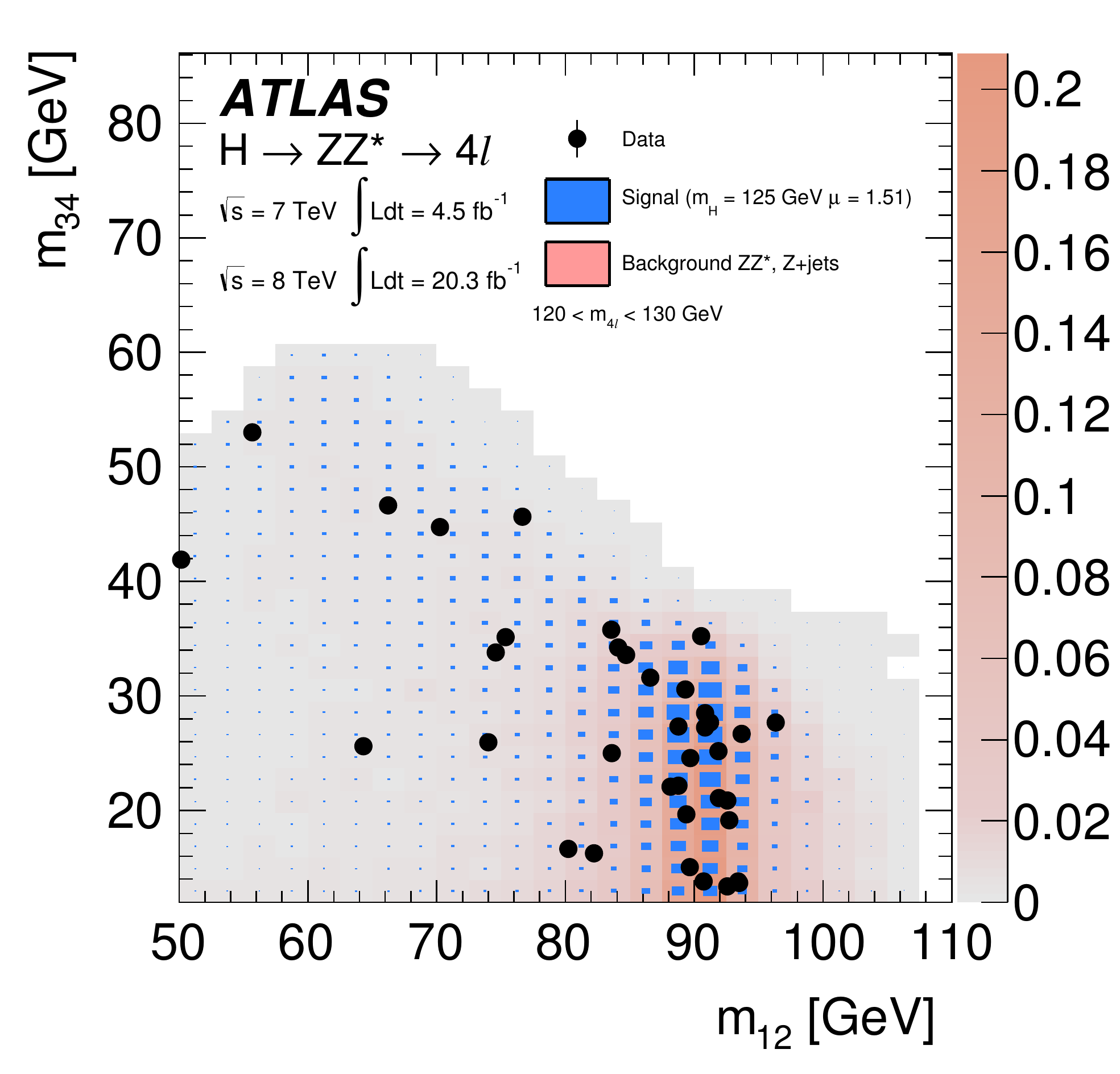}}
  \subfigure[\label{fig:finalMassesm12}]{\includegraphics[width=\doublePlotSize]{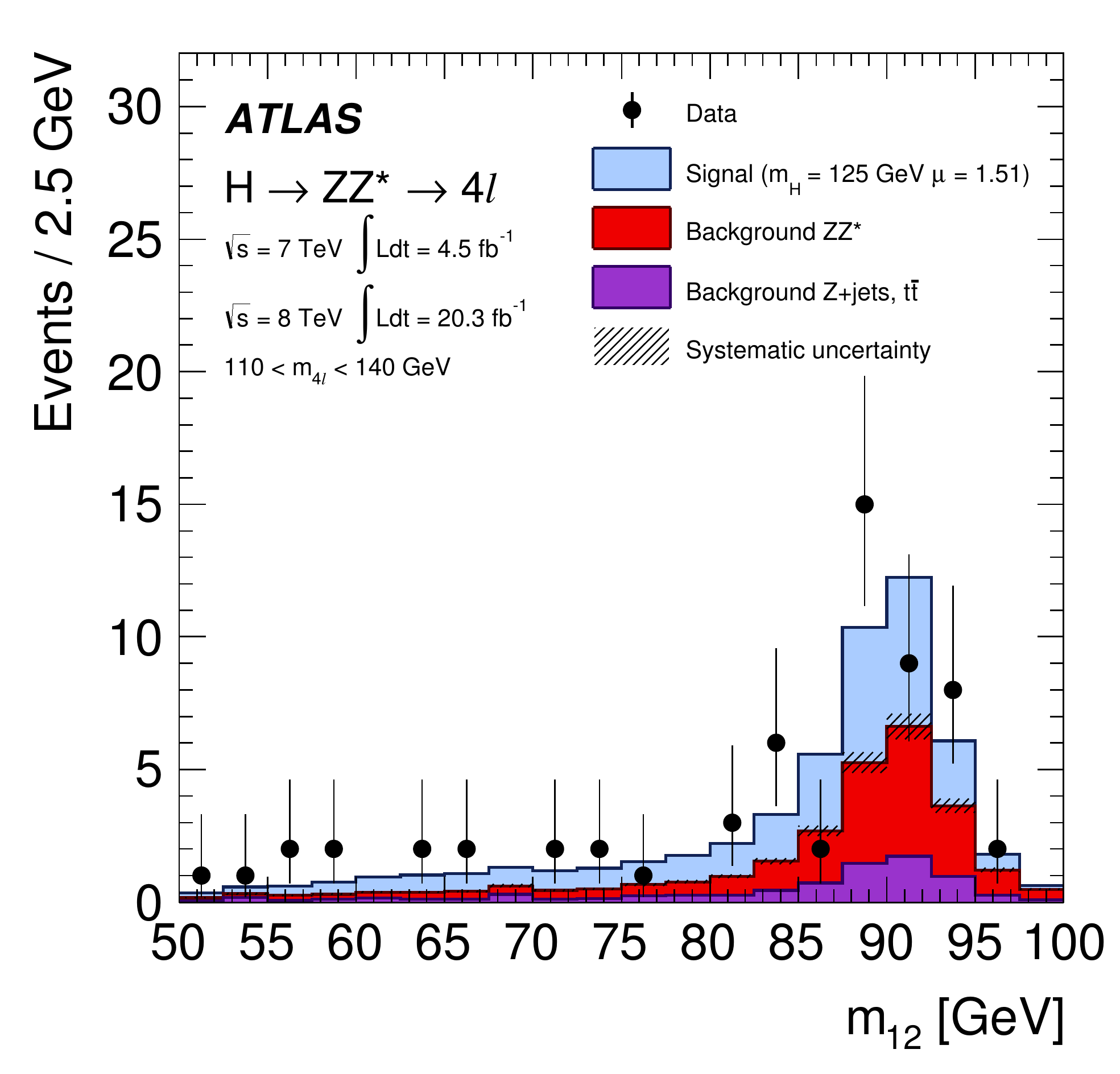}}
  \subfigure[\label{fig:finalMassesm34}]{\includegraphics[width=\doublePlotSize]{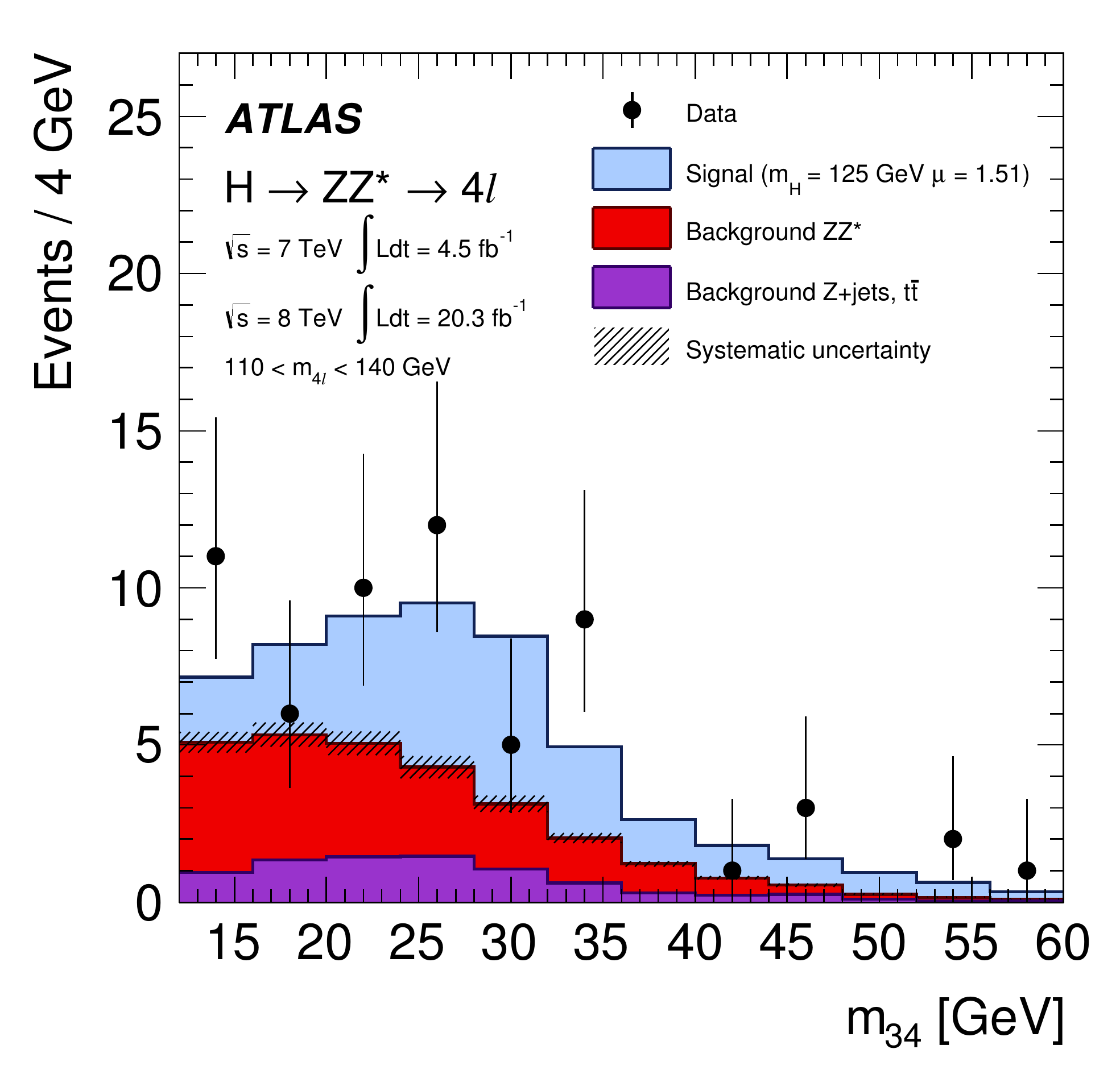}}
  \caption{Distributions of data (filled circles) and the expected signal and backgrounds events in
    \subref{fig:M34M12} the $m_{34}$ -- $m_{12}$ plane with the requirement of $m_{4\ell}$ in
    120--130 \gev.  The projected distributions for \subref{fig:finalMassesm12} $m_{12}$ and
    \subref{fig:finalMassesm34} $m_{34}$ are shown for $m_{4\ell}$ in 110--140 \gev, the fit range.
    The signal contribution is shown for $\mH = 125$ \gev\ and normalized to $\mu = 1.51$ (see text)
    as blue histograms in \subref{fig:finalMassesm12} and \subref{fig:finalMassesm34}.  The expected
    background contributions, \zzstar\ (red histogram) and $Z+$ jets plus $t\bar{t}$ (violet
    histogram), are shown in \subref{fig:finalMassesm12} and \subref{fig:finalMassesm34}; the
    systematic uncertainty associated to the total background contribution is represented by the
    hatched areas. The expected distributions of the Higgs signal (blue) and total background (red)
    are superimposed in \subref{fig:M34M12}, where the box size (signal) and color shading
    (background) represent the relative density. In every case, the combination of the 7 \tev\ and 8
    \tev\ results is shown.  \label{fig:finalMassesSignal1234}}

\end{figure*}

The expected $m_{4\ell}$ distribution for the backgrounds and the signal hypothesis are compared
with the combined $\sqrt{s}=7$ \tev\ and $\sqrt{s}=8$ \tev\ data in \figref{fig:finalMassesSignala}
for the $m_{4\ell}$ range 80--170 \gev, and in \figref{fig:finalMassesSignalb} for the invariant
mass range 80--600 \gev.  In \figref{fig:finalMasses} one observes the single $Z\rightarrow 4\ell$
resonance~\cite{Aad:2014wra,CMS:2012bw}, the threshold of the $ZZ$ production above 180 \gev\ and a
narrow peak around 125 \gev.  \Figref{fig:finalMassesSignal1234} shows the distribution of the
$m_{12}$ versus $m_{34}$ invariant masses, as well as their projections, for the candidates with
$m_{4\ell}$ within 120--130 \gev. The $Z$-mass constrained kinematic fit is not applied for these
distributions.  The Higgs signal is shown for $m_H = 125$ \gev\ with a value of $\mu = 1.51$,
corresponding to the combined $\mu$ measurement for the \htollllbrief\ final state, discussed below
in \secref{sec:couplingres}, scaled to this mass by the expected variation in the SM Higgs boson
cross section times branching ratio.

\begin{figure*}
  \centering 
  \subfigure[\label{fig:BDTvsM4l}]   {\includegraphics[width=\doublePlusPlotSize]{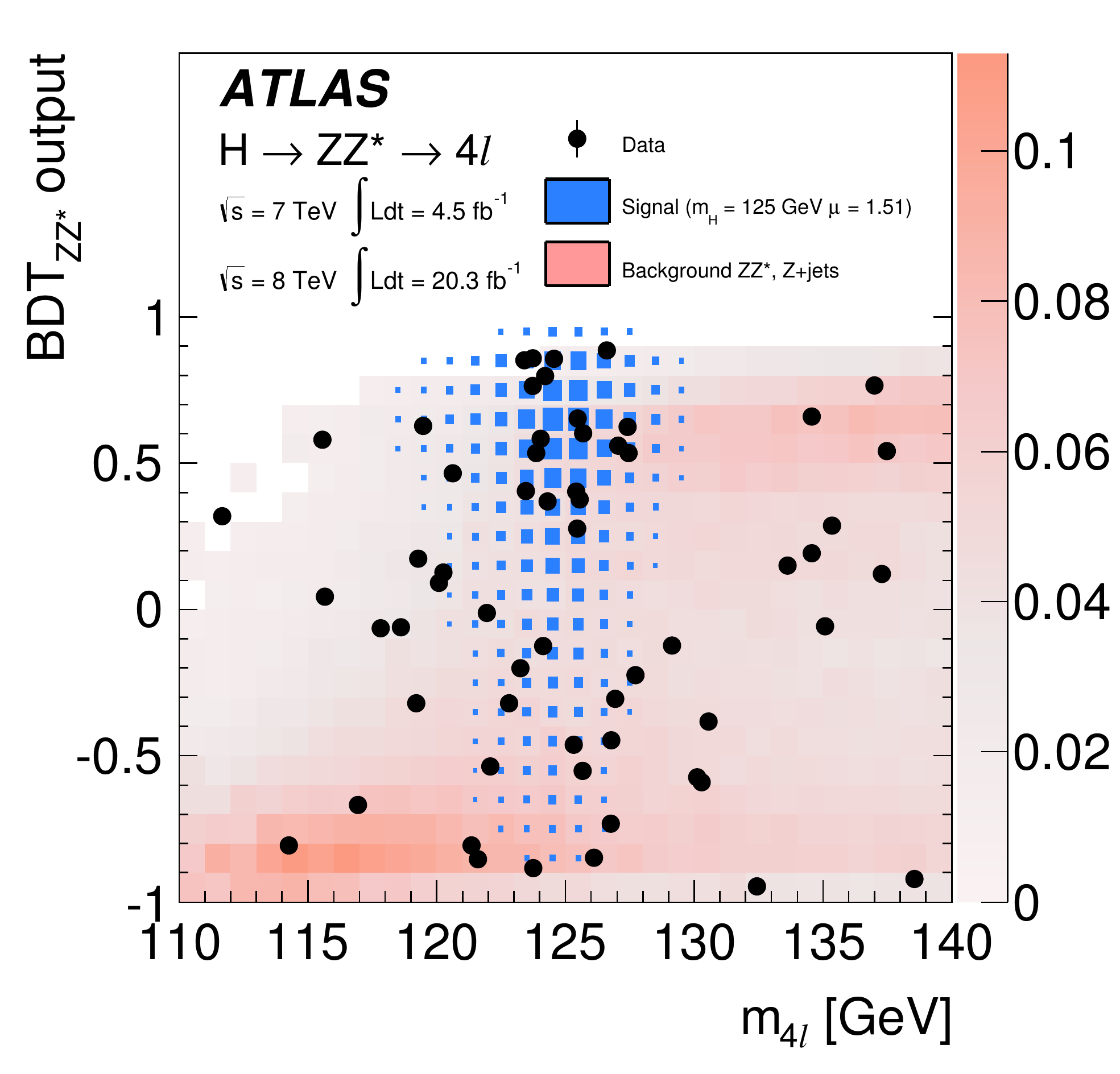}}
  \subfigure[\label{fig:BDTr}]       {\includegraphics[width=\doublePlotSize]{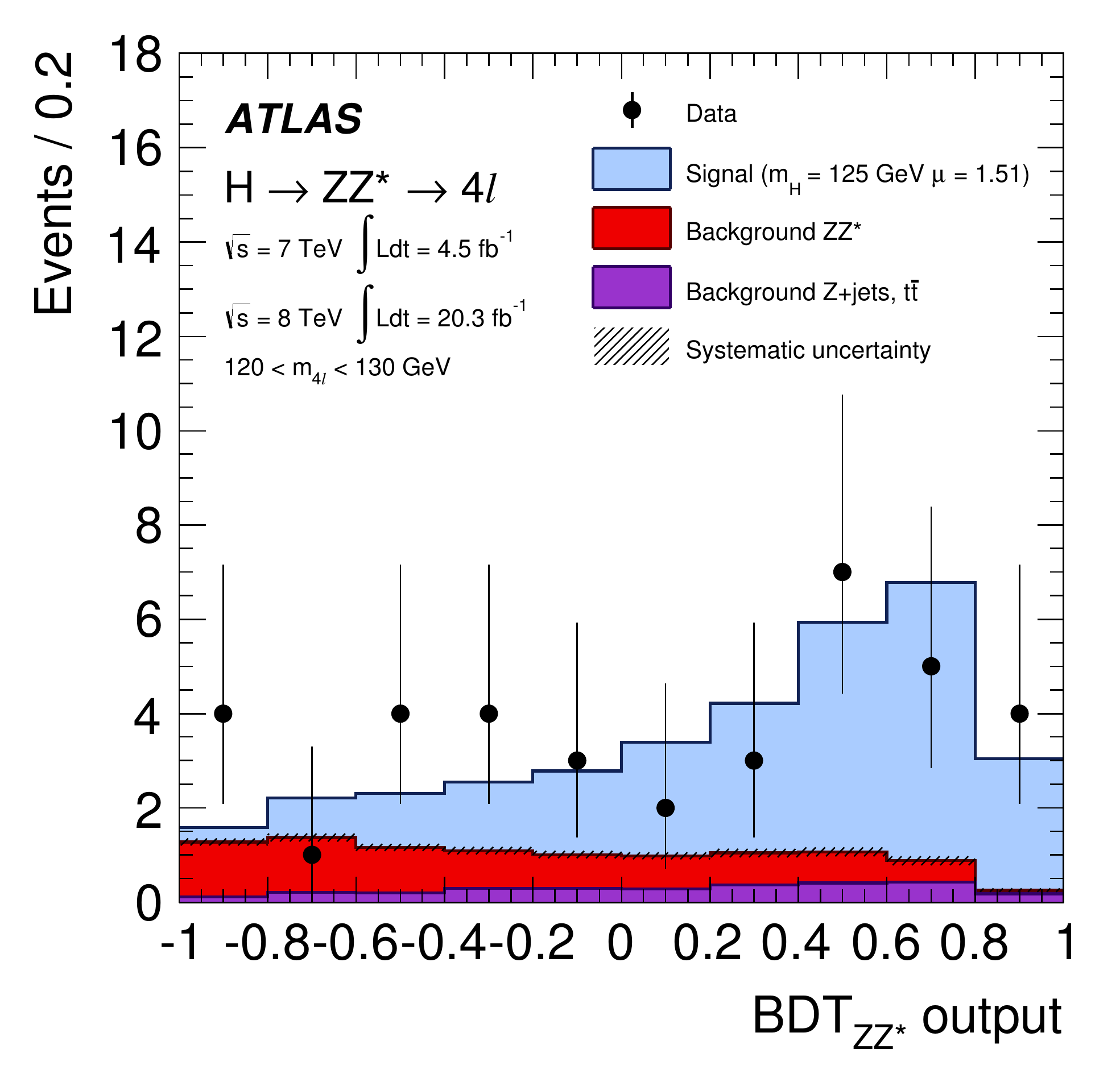}}
  \subfigure[\label{fig:MassBDTgood}]{\includegraphics[width=\doublePlotSize]{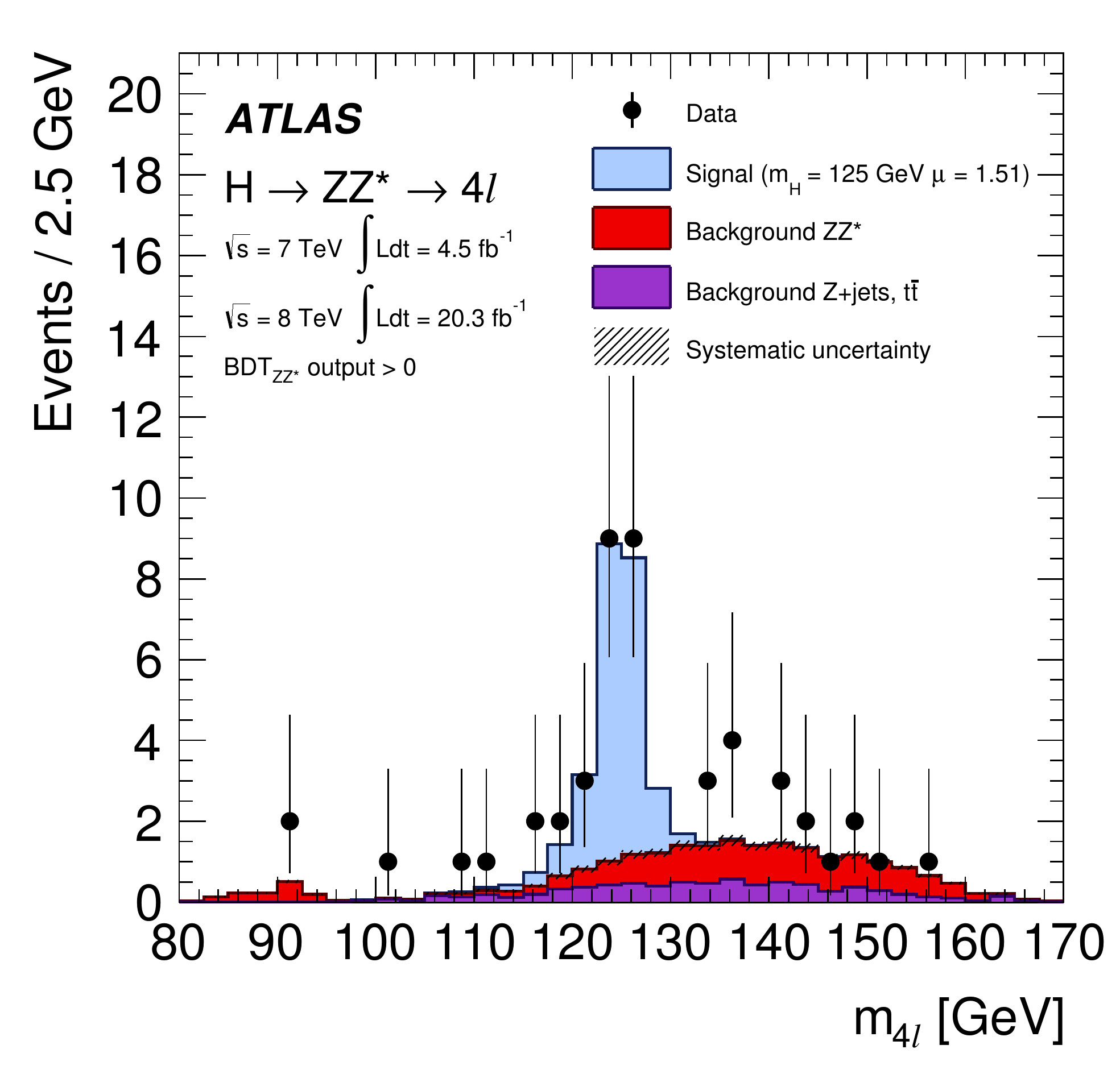}}

  \caption{Distributions of data (filled circles) and the expected signal and background events in
    \subref{fig:BDTvsM4l} the \bdtzz\ -- $m_{4\ell}$ plane, \subref{fig:BDTr} \bdtzz\ with the
    restriction $120 < m_{4\ell} < 130$ \gev, and \subref{fig:MassBDTgood} $m_{4\ell}$ with the
    additional requirement that the \bdtzz\ be positive.  The expected Higgs signal contribution is
    shown for $m_H=125$ GeV and normalized to $\mu=1.51$ (see text) as blue histograms in
    \subref{fig:BDTr} and \subref{fig:MassBDTgood}. The expected background contributions,
    \zzstar\ (red histogram) and $Z+ \rm jets$ plus $t\bar{t}$ (violet histogram), are shown in
    \subref{fig:BDTr} and \subref{fig:MassBDTgood}; the systematic uncertainty associated to the
    total background contribution is represented by the hatched areas. The expected distributions of
    the Higgs signal (blue) and total background (red) are superimposed in \subref{fig:BDTvsM4l},
    where the box size (signal) and color shading (background) represent the relative density. In
    every case, the combination of the 7 TeV and 8 TeV results is shown. \label{fig:2D} }

\end{figure*}

The distribution of the \bdtzz\ output versus $m_{4\ell}$ is shown in \figref{fig:BDTvsM4l} for the
reconstructed candidates with $m_{4\ell}$ within the fitted mass range 110--140 \gev.  An excess of
events with high-\bdtzz\ output is present for values of $m_{4\ell}$ close to 125 \gev, compatible
with the Higgs signal hypothesis at that mass.  The compatibility of the data with the expectations
shown in \figref{fig:BDTvsM4l} is checked using pseudoexperiments generated according to the
expected two-dimensional distribution and good agreement is found.
\Figref{fig:BDTr} shows the distribution of the \bdtzz\ output for the candidates in the $m_{4\ell}$
range 120--130 \gev\ compared with signal and background expectations.  In \figref{fig:MassBDTgood}
the distribution of the invariant mass of the four leptons is presented for candidates satisfying
the requirement that the value of the \bdtzz\ output be greater than zero, which maximizes the
expected significance for a SM Higgs boson with a mass of about 125 \gev.

\begin{table*}[htbb!]
  \centering 

  \caption{The number of events expected and observed for a $m_{H}$=$125$ \gev\ hypothesis for the
    four-lepton final states in a window of $120 < m_{4\ell} < 130$ \gev.  The second column shows
    the number of expected signal events for the full mass range, without a selection on
    $m_{4\ell}$.  The other columns show for the 120--130 \gev\ mass range the number of expected
    signal events, the number of expected \ZZbkg\ and reducible background events, and the
    signal-to-background ratio ({\it S/B}), together with the number of observed events, for 4.5
    \ifb\ at $\sqrt{s}=7$ \tev\ and 20.3 \ifb\ at $\sqrt{s}=8$ \tev\ as well as for the combined
    sample.  \label{tab:yields}}

  \vspace{0.1cm}
  \begin{tabular}{*{8}{c}}
      \hline\hline
    \noalign{\vspace{0.05cm}}
    Final state & Signal & Signal & \ZZbkg & $Z+\rm jets$,~$t\bar{t}$ & {\it S/B } & Expected & Observed \\
    & full mass range     \\

    \hline                                                     
    \noalign{\vspace{0.05cm}}
    \multicolumn{8}{c}{$\sqrt{s}=7$ \tev}\\
    \noalign{\vspace{0.05cm}}
    \hline
    $4\mu$    & 1.00 $\pm$  0.10  & 0.91 $\pm$  0.09 &  0.46 $\pm$  0.02 & 0.10 $\pm$  0.04 &  1.7 & 1.47 $\pm$  0.10 & 2 \\
    $2e2\mu$  & 0.66 $\pm$  0.06  & 0.58 $\pm$  0.06 &  0.32 $\pm$  0.02 & 0.09 $\pm$  0.03 &  1.5 & 0.99 $\pm$  0.07 & 2 \\
    $2\mu2e$  & 0.50 $\pm$  0.05  & 0.44 $\pm$  0.04 &  0.21 $\pm$  0.01 & 0.36 $\pm$  0.08 &  0.8 & 1.01 $\pm$  0.09 & 1 \\
    $4e$      & 0.46 $\pm$  0.05  & 0.39 $\pm$  0.04 &  0.19 $\pm$  0.01 & 0.40 $\pm$  0.09 &  0.7 & 0.98 $\pm$  0.10 & 1 \\
    \noalign{\vspace{0.05cm}}
    Total     & 2.62 $\pm$  0.26  & 2.32 $\pm$  0.23 &  1.17 $\pm$  0.06 & 0.96 $\pm$  0.18 &  1.1 & 4.45 $\pm$  0.30 & 6 \\

    \hline    					    
    \noalign{\vspace{0.05cm}}
    \multicolumn{8}{c}{$\sqrt{s}=8$ \tev}\\
    \noalign{\vspace{0.05cm}}
    \hline
    $4\mu$    & 5.80 $\pm$  0.57  &  5.28 $\pm$  0.52  & 2.36 $\pm$  0.12   & 0.69 $\pm$  0.13 &  1.7 &  8.33 $\pm$  0.6 & 12 \\
    $2e2\mu$  & 3.92 $\pm$  0.39  &  3.45 $\pm$  0.34  & 1.67 $\pm$  0.08   & 0.60 $\pm$  0.10 &  1.5 &  5.72 $\pm$  0.37 &  7 \\
    $2\mu2e$  & 3.06 $\pm$  0.31  &  2.71 $\pm$  0.28  & 1.17 $\pm$  0.07   & 0.36 $\pm$  0.08 &  1.8 &  4.23 $\pm$  0.30 &  5 \\
    $4e$      & 2.79 $\pm$  0.29  &  2.38 $\pm$  0.25  & 1.03 $\pm$  0.07   & 0.35 $\pm$  0.07 &  1.7 &  3.77 $\pm$  0.27 &  7 \\
    \noalign{\vspace{0.05cm}}
    Total     & 15.6 $\pm$  1.6    & 13.8  $\pm$  1.4   & 6.24 $\pm$  0.34   & 2.00 $\pm$  0.28 &  1.7 & 22.1 $\pm$  1.5 & 31 \\

    \hline
    \noalign{\vspace{0.05cm}}
    \multicolumn{8}{c}{$\sqrt{s}=7$ \tev\ and $\sqrt{s}=8$ \tev}\\
    \noalign{\vspace{0.05cm}}
    \hline
    $4\mu$    & 6.80 $\pm$  0.67  &  6.20 $\pm$  0.61  & 2.82 $\pm$  0.14  & 0.79 $\pm$  0.13 &  1.7 &  9.81 $\pm$  0.64 & 14 \\
    $2e2\mu$  & 4.58 $\pm$  0.45  &  4.04 $\pm$  0.40  & 1.99 $\pm$  0.10  & 0.69 $\pm$  0.11 &  1.5 &  6.72 $\pm$  0.42 &  9 \\
    $2\mu2e$  & 3.56 $\pm$  0.36  &  3.15 $\pm$  0.32  & 1.38 $\pm$  0.08  & 0.72 $\pm$  0.12 &  1.5 &  5.24 $\pm$  0.35 &  6 \\
    $4e$      & 3.25 $\pm$  0.34  &  2.77 $\pm$  0.29  & 1.22 $\pm$  0.08  & 0.76 $\pm$  0.11 &  1.4 &  4.75 $\pm$  0.32 &  8 \\
    \noalign{\vspace{0.05cm}}
    Total     &18.2  $\pm$  1.8   & 16.2  $\pm$  1.6   & 7.41 $\pm$  0.40  & 2.95 $\pm$  0.33 &  1.6 & 26.5  $\pm$  1.7 & 37 \\
    \hline\hline 
  \end{tabular}

\end{table*}

The local $p_0$-value of the observed signal, representing the significance of the excess relative
to the background-only hypothesis, is obtained with the asymptotic approximation~\cite{Cowan:2010js}
using the 2D fit without any selection on \bdtzz\ and is shown as a function of $m_{H}$ in
\figref{fig:Combp0}.  The local $p_{0}$-value at the measured mass for this channel,
\hllllmassvalue\ \gev\ (see below), is 8.2 standard deviations.  At the value of the Higgs boson
mass, \atlasmassveryshort, obtained from the combination of the \htollllbrief\ and \htogg\ mass
measurements~\cite{combmasspaper}, the local $p_{0}$-value decreases to 8.1 standard deviations.
The expected significance at these two masses is 5.8 and 6.2 standard deviations, respectively.

\begin{figure}[!htb]
  \centering
  \includegraphics[width=\singlePlotSize]{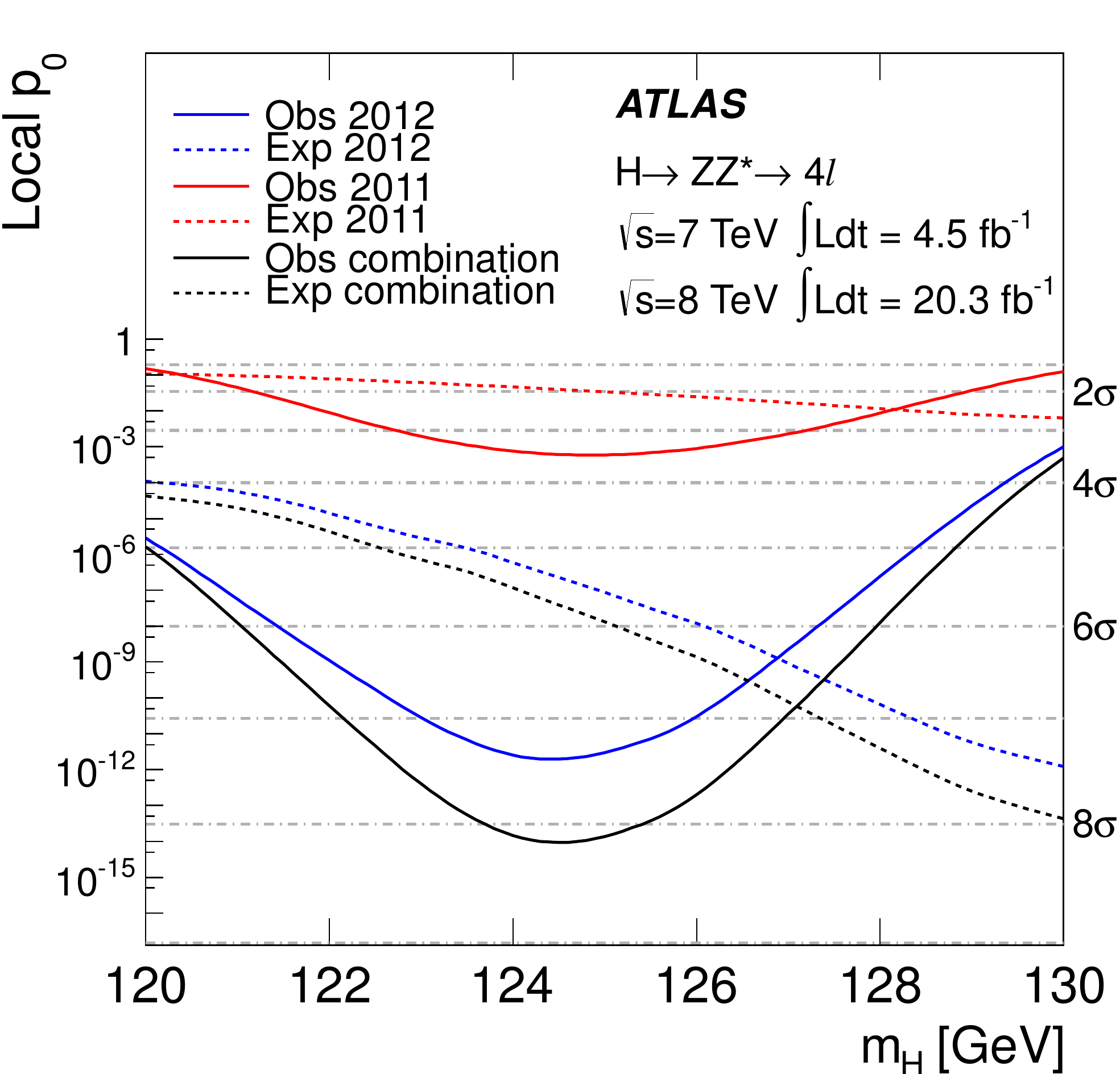}
  \caption{The observed local \pval-value for the combination of the 2011 and 2012 data sets (solid
    black line) as a function of $m_H$; the individual results for $\sqrt{s}=7$\,\tev\ and
    8\,\tev\ are shown separately as red and blue solid lines, respectively.  The dashed curves show
    the expected median of the local \pval-value for the signal hypothesis with signal strength $\mu = 1$,
    when evaluated at the corresponding \mH. The horizontal dot-dashed lines indicate the
    \pval-values corresponding to local significances of 1--8$\sigma$. \label{fig:Combp0} }
\end{figure}

\subsubsection{Mass and inclusive signal strength}
\begin{figure*}
  \centering
  \subfigure[\label{fig:muvsmh}]{\includegraphics[width=\doublePlotSize]{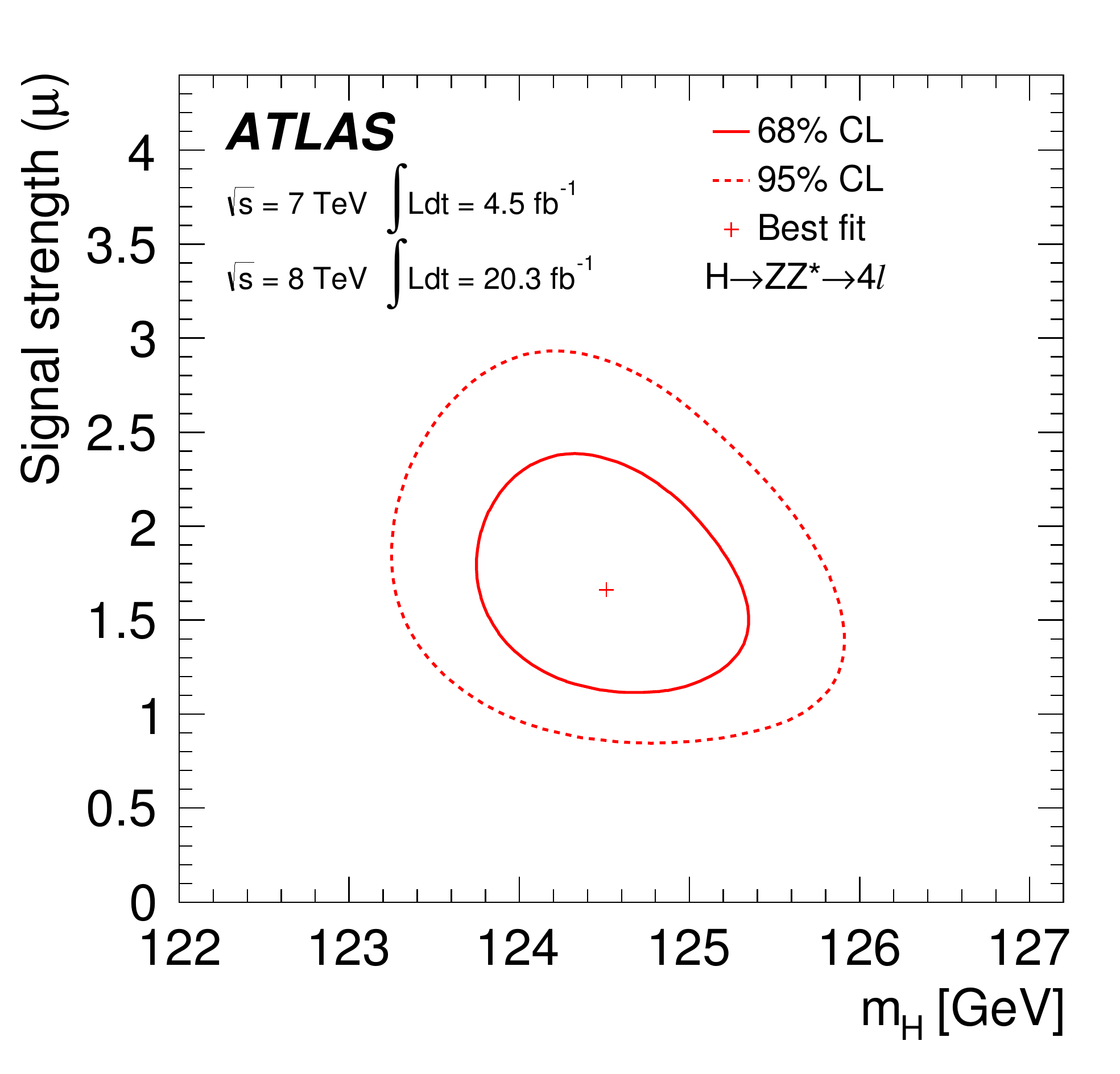}}
  \subfigure[\label{fig:muScan}]{\includegraphics[width=\doublePlotSize]{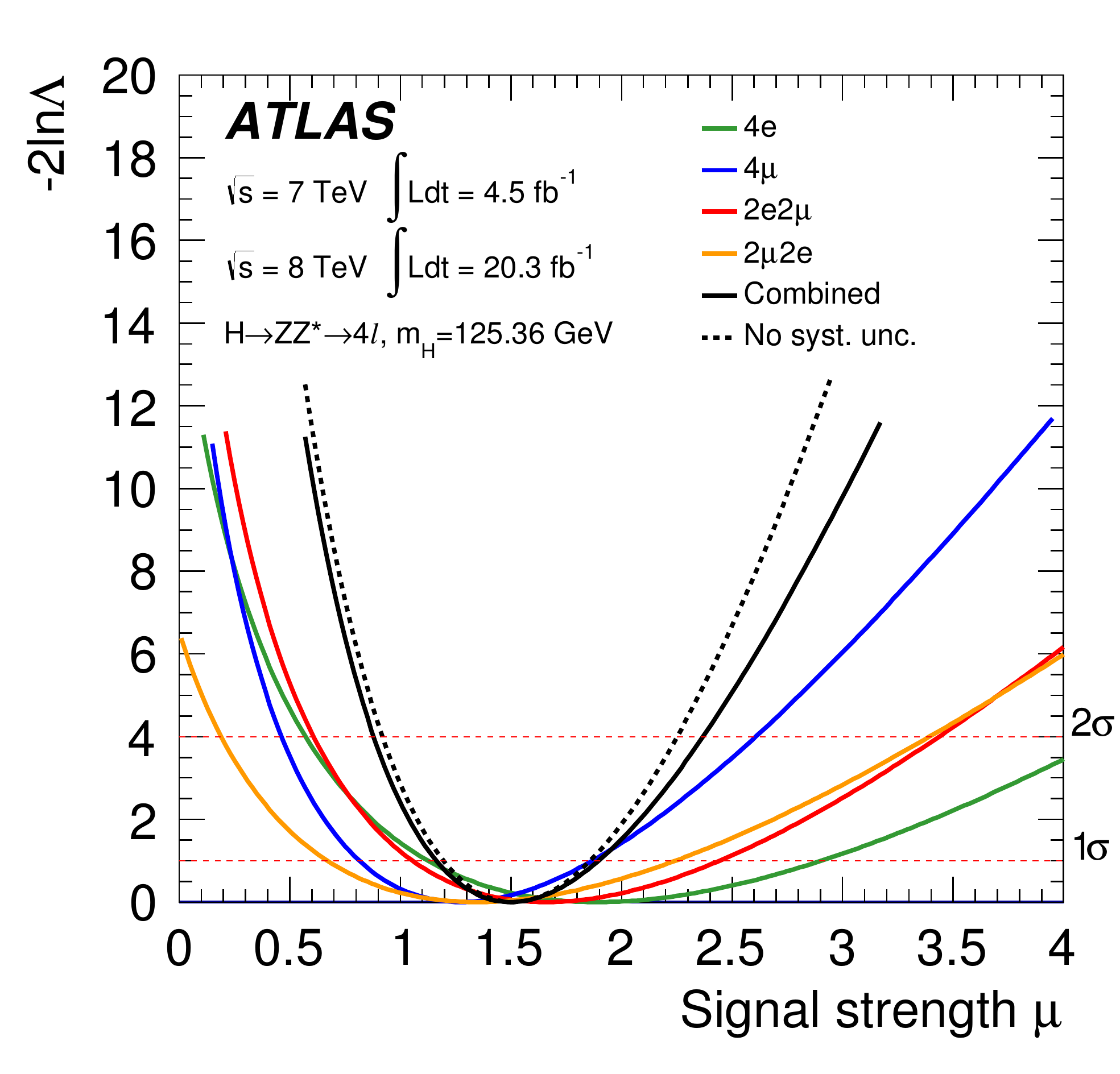}}
  \caption{~\subref{fig:muvsmh} The 68\% and 95\% confidence level (CL) contours in the $\mu$ --
    $m_H$ plane for the inclusive analysis. ~\subref{fig:muScan} The profile likelihood as a
    function of the inclusive signal strength $\mu$ for the individual channels ($4e$, green line;
    $4\mu$, blue line; $2e2\mu$, red line; $2\mu2e$, yellow line) as well as for their combination
    (black lines); the scan for the combination of all channels is shown both with (solid line) and
    without (dashed line) systematic uncertainties. The value of $m_{H}$ is fixed to
    \atlasmassvalue\ \gev\ while all the other nuisance parameters are profiled in the fit. In every
    case, the combination of the 7 \tev\ and 8 \tev\ results is shown.}
\end{figure*}

The models described in \secref{sec:SignalModelInclusive} are used to perform the inclusive mass and
signal strength measurements.  The measured Higgs boson mass obtained with the baseline 2D method is
\hllllmassshort.  The signal strength at this value for $m_{H}$ is \hllllmu.  The other methods of
\secref{sec:SignalModelInclusive}, 1D and per-event resolution, yield similar results for the Higgs
boson mass~\cite{combmasspaper}.  \Figref{fig:muvsmh} shows the best fit values of $\mu$ and $m_H$
as well as the profile likelihood ratio contours in the ($m_H$,$\mu$) plane corresponding to the
68\% and 95\% confidence level intervals. Finally, the best fit value for $m_{H}$ obtained using the
model developed for the categorized analysis, described in \secref{sec:SignalModelCat}, is within 90
MeV of the value found with the inclusive 2D method.

At the combined ATLAS measured value of the Higgs boson mass, \atlasmassveryshort, the signal
strength is found to be \hllllmuCombmass.  The scan of the profile likelihood,
$-2\ln{\Lambda(\mu)}$, as a function of the inclusive signal strength $\mu$ for each one of the four
channels separately, as well as for their combination, is shown in \figref{fig:muScan}.

  \clearpage 
\subsection{Coupling studies}
\label{sec:couplingres}

\begin{figure}[!htbp]
  \centering
  \includegraphics[width=\singlePlotSize]{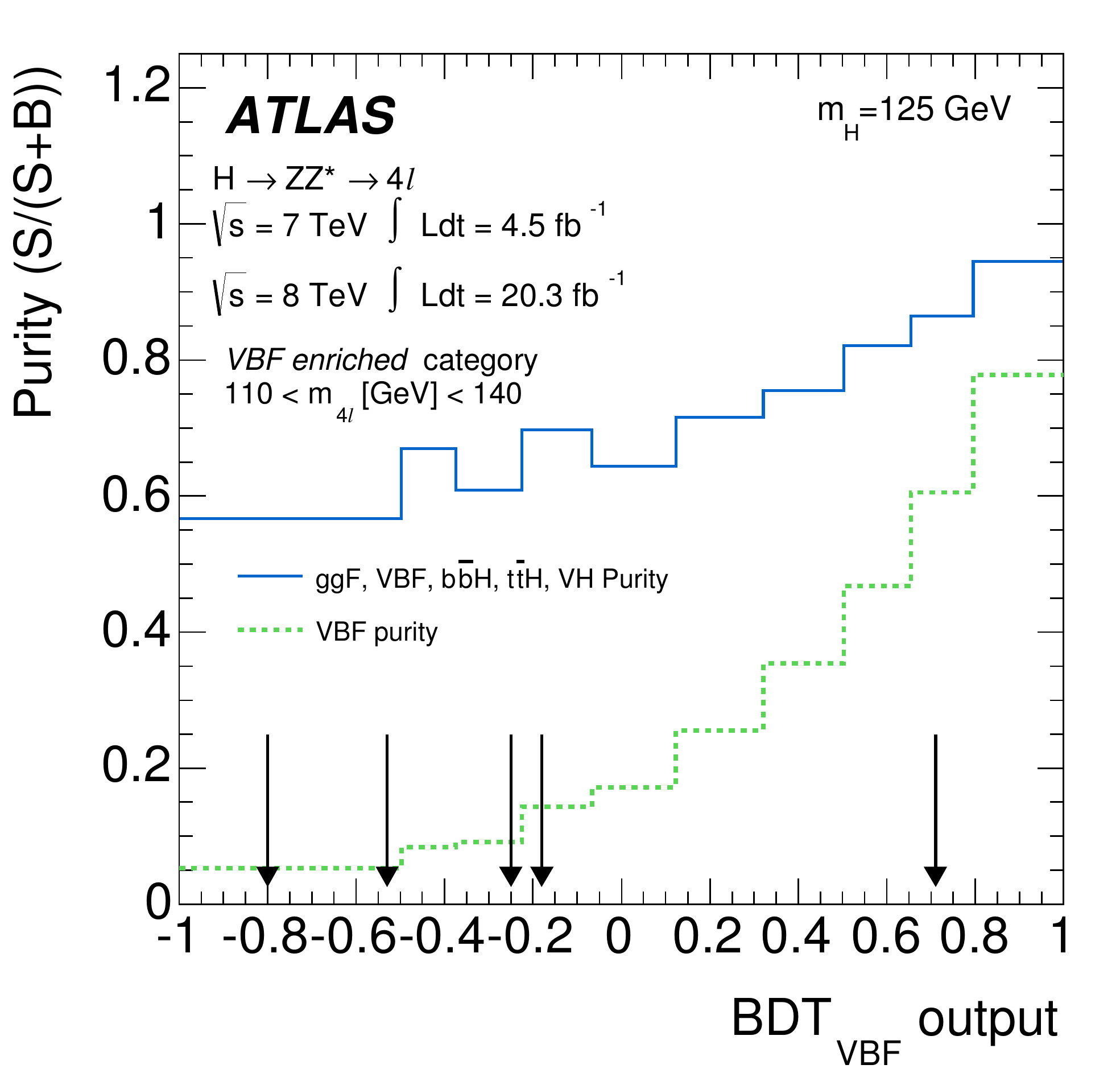}
  \caption{Signal purity, defined as $S/(S+B)$, as a function of the \bdtvbf\ output. The solid blue
    line shows the purity for all Higgs signal production mechanisms relative to the \zzstar\ and
    reducible backgrounds.  The dashed green line shows the purity for VBF events relative to the
    other Higgs boson production mechanisms, for the fit region $110 < m_{4\ell} < 140$ \gev.  The
    binning is chosen so that each bin contains 10\% of the total expected signal events.
    The five VBF candidates observed in data in the signal region
    are indicated with the black arrows.  \label{fig:vbf_bdt_purity}}
\end{figure}

The numbers of expected and observed events in each of the categories described in
\secref{sec:eventCategorisation} are summarized in \tabref{tab:couplingResultsYields}.  The expected
yield in each enriched category is given for each of the production modes, where the ggF,
$b\bar{b}H$ and $t\bar{t}H$ yields are combined.  The expected and observed numbers of events are
given for two $m_{4\ell}$ mass ranges: 120--130 \gev\ and above 110 \gev.  Three of the VBF
candidates are found in the mass region 120--130 \gev\ with invariant masses of 123.2 \gev, 123.4
\gev\ and 125.7 \gev.  Only one VBF candidate has a \bdtvbf\ output above zero: $m_{4\ell}=$123.4
\gev\ and a \bdtvbf\ output value of 0.7.  In this mass window, the expected number of VBF
candidates with \bdtvbf\ output above zero is 1.26 $\pm$ 0.15, where half of this is expected to be
from a true VBF signal, about 35\% from ggF production and the rest is mostly from \zzstar\ and
reducible backgrounds.  The distributions of $m_{4\ell}$ and the \bdtvbf\ output for the
\vbfcat\ category in the full mass range and in the fit range of 110--140 \gev\ are shown in
\figref{fig:vbf_yields}.  The signal purity, defined as $S/(S+B)$, as a function of the
\bdtvbf\ output is shown in \figref{fig:vbf_bdt_purity} for Higgs events relative to the backgrounds
and for VBF events relative to the other Higgs boson production mechanisms for $110 < m_{4\ell} <
140$ \gev.  There is no VH candidate in the 120--130 \gev\ mass range for either the hadronic or
leptonic categories. For the full mass range above 110 \gev\ all categories are dominated by
\zzstar\ background, and the observed number of events agrees well with the expectation as can be
seen in \tabref{tab:couplingResultsYields}.

\begin{table*}[h]
  \centering
  \footnotesize 
  \caption{Expected and observed yields in the \vbfcat, \vhhadcat, \vhlepcat\ and
    \ggfcat\ categories. The yields are given for the different production modes and the \zzstar\ and
    reducible background for 4.6 $\rm fb^{-1}$ at $\sqrt{s}=7$ \tev\ and 20.3 $\rm fb^{-1}$ at
    $\sqrt{s}=8$ \tev. The estimates are given for both the $m_{4\ell}$ mass range 120--130 \gev\ and
    the mass range above 110 \gev.  \label{tab:couplingResultsYields}}
  \vspace{0.1cm}
  \begin{tabular}{* {9} {@{\hspace{1.5pt}}c@{\hspace{1.5pt}}}}
    \hline
    \noalign{\vspace{0.05cm}}
    Enriched                       & \multicolumn{4}{c}{Signal }                                                       & \multicolumn{2}{c}{Background}      & Total           & Observed \\
    category                       &  $ggF + b\bar{b}H  + t\bar{t}H$ & VBF      & VH-hadronic          & VH-leptonic       & \zzstar          &$Z+\rm jets$,~$t\bar{t}$& expected  &  \\
    \noalign{\vspace{0.05cm}}
    \hline
    \noalign{\vspace{0.05cm}}
    \multicolumn{9}{c}{${\bf 120 < m_{4\ell} < 130}$ {\bf GeV} } \\
    \noalign{\vspace{0.05cm}}
    \hline
    \noalign{\vspace{0.05cm}}
    {\it VBF }                     & 1.18 $\pm$ 0.37   & 0.75 $\pm$ 0.04     & 0.083 $\pm$ 0.006   & 0.013 $\pm$ 0.001 & 0.17 $\pm$ 0.03  & 0.25 $\pm$ 0.14  & 2.4 $\pm$ 0.4   & 3  \\
    {\scriptsize(\bdtvbf\ $>$ 0)}  & 0.48 $\pm$ 0.15   & 0.62 $\pm$ 0.04     & 0.023 $\pm$ 0.002   & 0.004 $\pm$ 0.001 & 0.06 $\pm$ 0.01  & 0.10 $\pm$ 0.05  & 1.26 $\pm$ 0.15 & 1  \\
    \noalign{\vspace{0.05cm}}
    \hline
    \noalign{\vspace{0.05cm}}
    {\it VH-hadronic }             & 0.40 $\pm$ 0.12   & 0.034 $\pm$ 0.004   & 0.20 $\pm$ 0.01     & 0.009 $\pm$ 0.001 & 0.09 $\pm$ 0.01  & 0.09 $\pm$ 0.04  & 0.80 $\pm$ 0.12 & 0  \\
    {\it VH-leptonic }             & 0.013 $\pm$ 0.002 & $< 0.001$           &      $< 0.001$      & 0.069 $\pm$ 0.004 & 0.015 $\pm$ 0.002& 0.016 $\pm$ 0.019&0.11 $\pm$ 0.02  & 0  \\
    {\it ggF }                     & 12.8 $\pm$ 1.3    & 0.57 $\pm$ 0.02     & 0.24 $\pm$ 0.01     & 0.11  $\pm$ 0.01  & 7.1 $\pm$ 0.2    & 2.7 $\pm$ 0.4    & 23.5 $\pm$ 1.4  & 34 \\
    \noalign{\vspace{0.05cm}}
    \hline
    \noalign{\vspace{0.05cm}}
    \multicolumn{9}{c}{${\bf m_{4\ell}} > {\bf 110}$ {\bf GeV} } \\
    \noalign{\vspace{0.05cm}}
    \hline
    \noalign{\vspace{0.05cm}}
    {\it VBF }                     & 1.4 $\pm$ 0.4     & 0.82 $\pm$ 0.05     & 0.092 $\pm$ 0.007   & 0.022 $\pm$ 0.002 & 20 $\pm$ 4     & 1.6 $\pm$ 0.9     &  24. $\pm$ 4.   & 32 \\
    {\scriptsize(\bdtvbf\ $>$ 0)}  & 0.54 $\pm$ 0.17   & 0.68 $\pm$ 0.04     & 0.025 $\pm$ 0.002   & 0.007 $\pm$ 0.001 & 8.2 $\pm$ 1.6    & 0.6 $\pm$ 0.3    & 10.0 $\pm$ 1.6  & 12 \\
    \noalign{\vspace{0.05cm}}
    \hline
    \noalign{\vspace{0.05cm}}
    {\it VH-hadronic }             & 0.46 $\pm$ 0.14   & 0.038 $\pm$ 0.004   & 0.23 $\pm$ 0.01     & 0.015 $\pm$ 0.001 & 9.0 $\pm$ 1.2    & 0.6 $\pm$ 0.2    & 10.3 $\pm$ 1.2  & 13 \\
    {\it VH-leptonic }             & 0.026 $\pm$ 0.004 &     $< 0.002$       &       $< 0.002$     & 0.15 $\pm$ 0.01   &0.63 $\pm$ 0.04   &0.11 $\pm$ 0.14   & 0.92 $\pm$ 0.16 & 1  \\
    {\it ggF }                     & 14.1 $\pm$ 1.5    & 0.63 $\pm$ 0.02     & 0.27  $\pm$ 0.01    & 0.17  $\pm$ 0.01  & 351. $\pm$ 20   & 16.6 $\pm$ 2.2   & 383. $\pm$ 20  & 420 \\
    \noalign{\vspace{0.05cm}}
    \hline
    \end{tabular} 
\end{table*}

\begin{figure*}[h]
  \centering
  \subfigure[\label{fig:vbf_m4l}]{\includegraphics[width=\doublePlotSize]{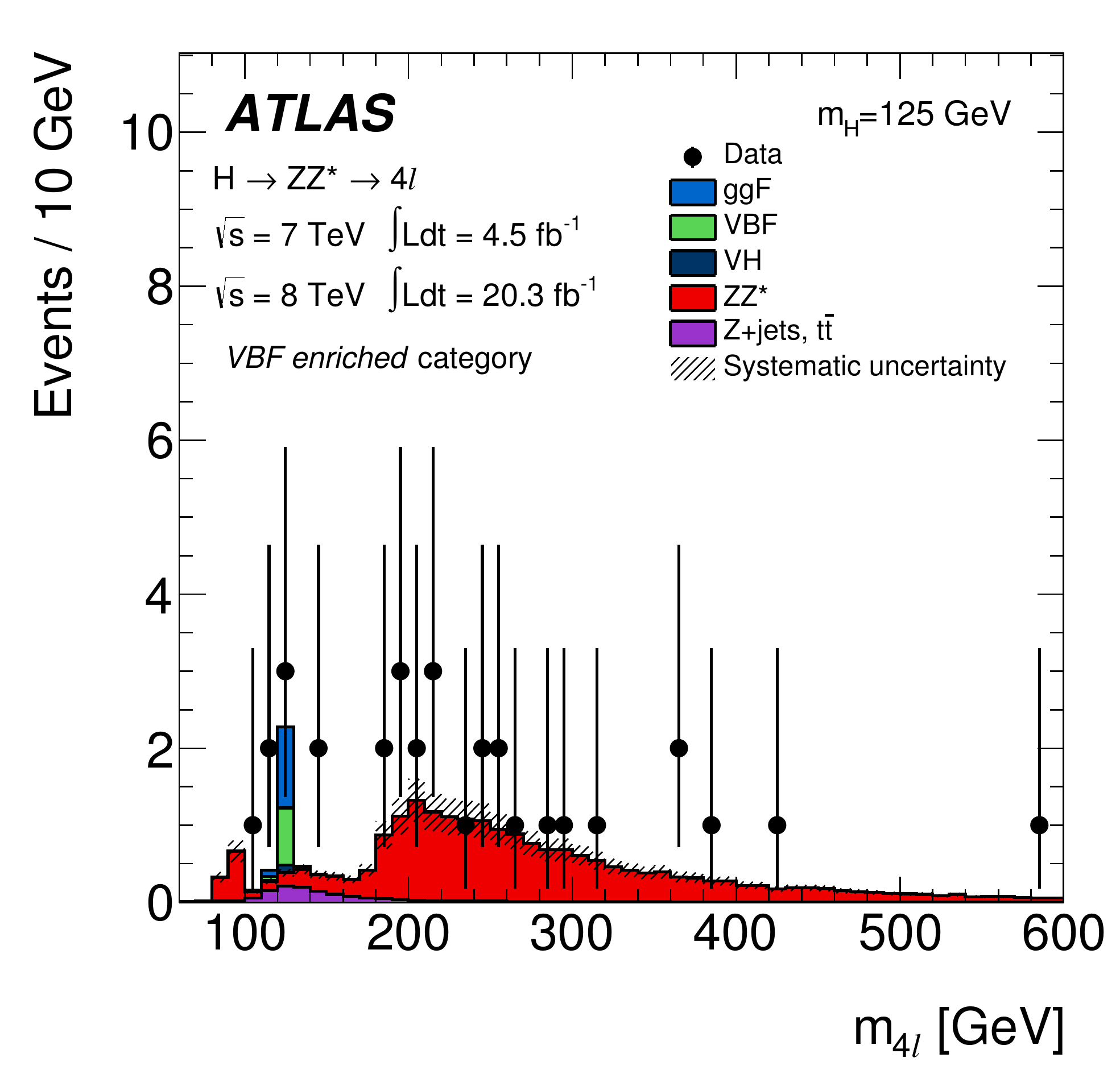}}
  \subfigure[\label{fig:vbf_bdt}]{\includegraphics[width=\doublePlotSize]{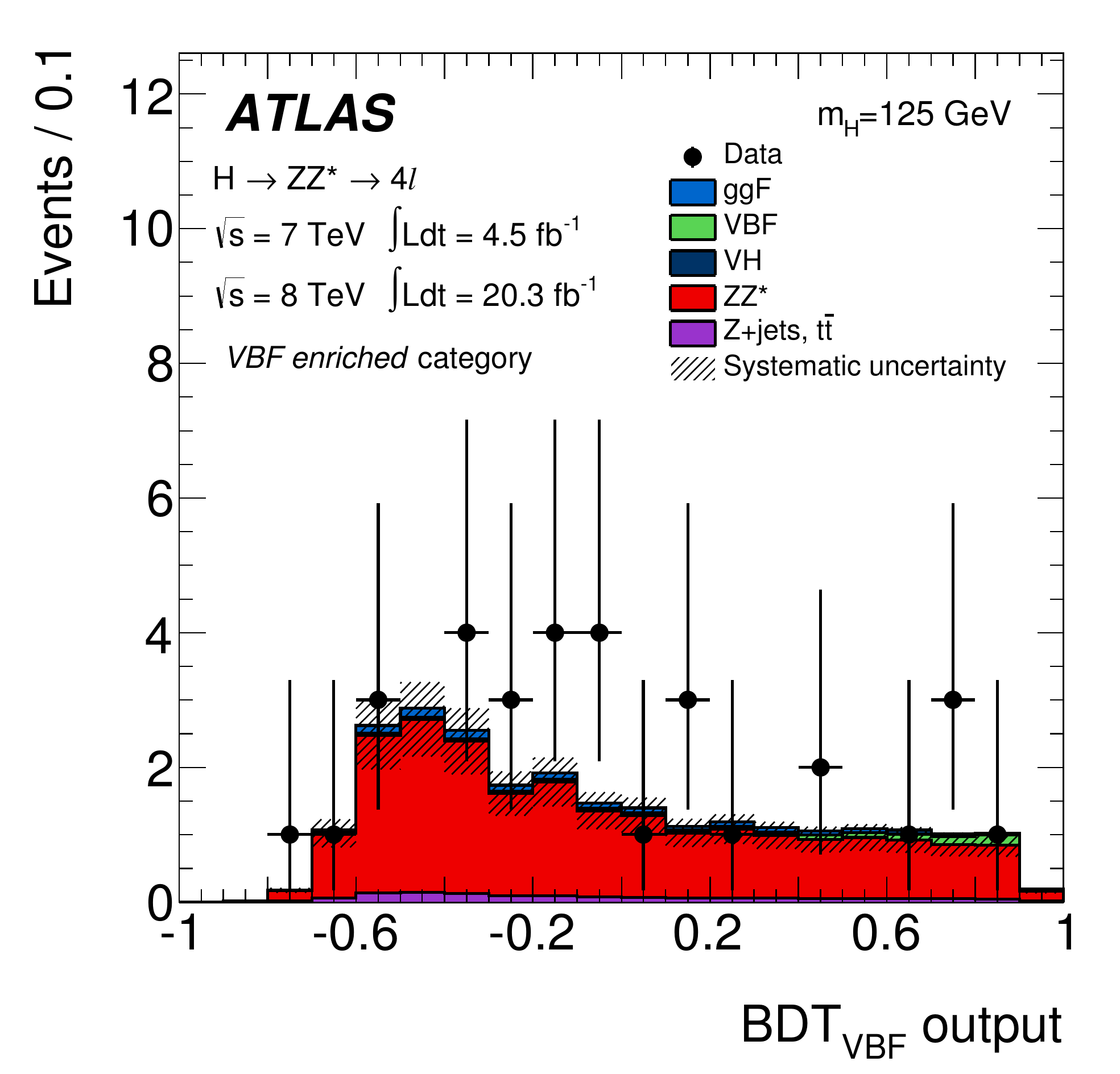}}
  \subfigure[\label{fig:m4l_VBF}]{\includegraphics[width=\doublePlotSize]{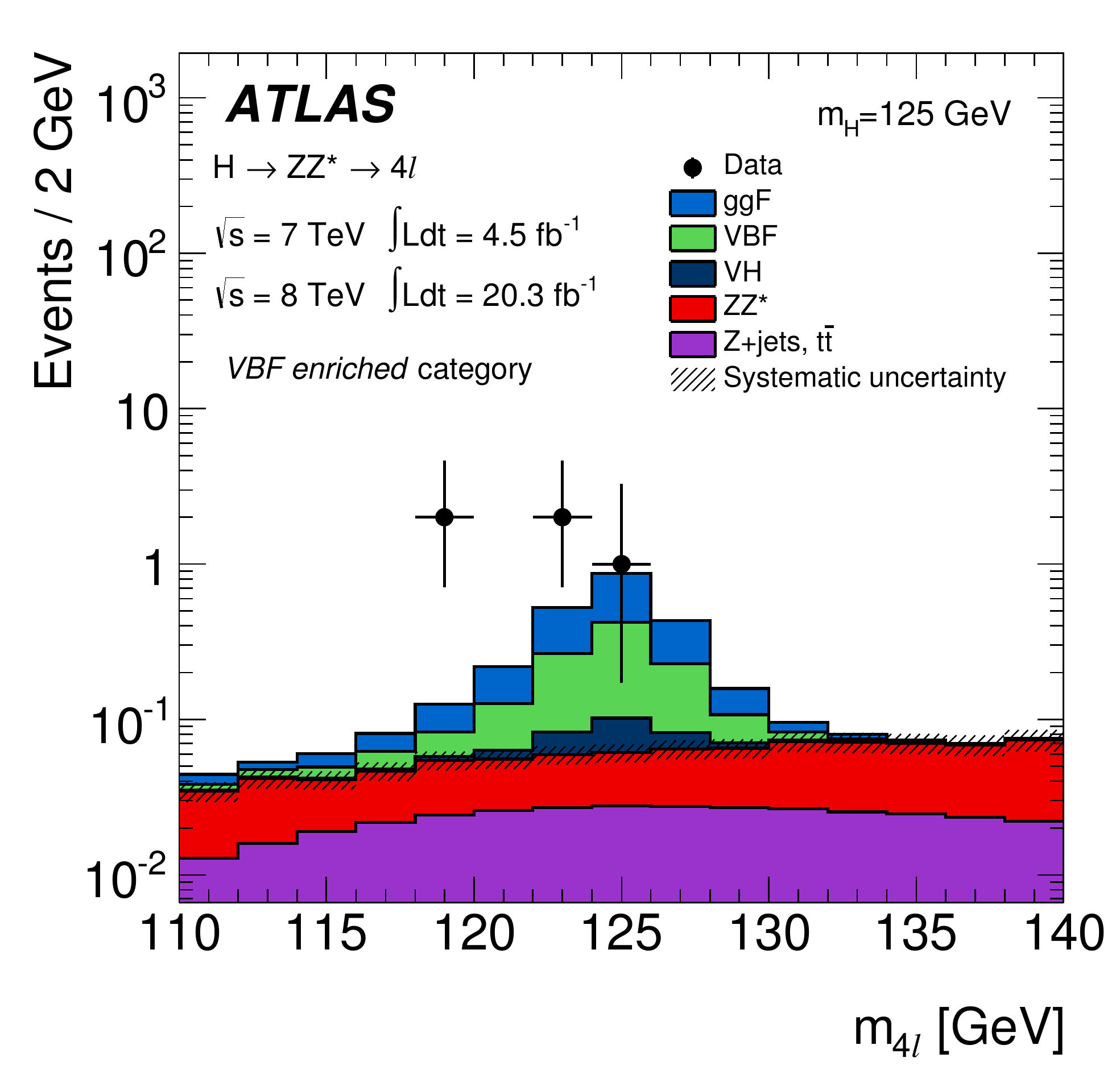}}
  \subfigure[\label{fig:BDT_VBF}]{\includegraphics[width=\doublePlotSize]{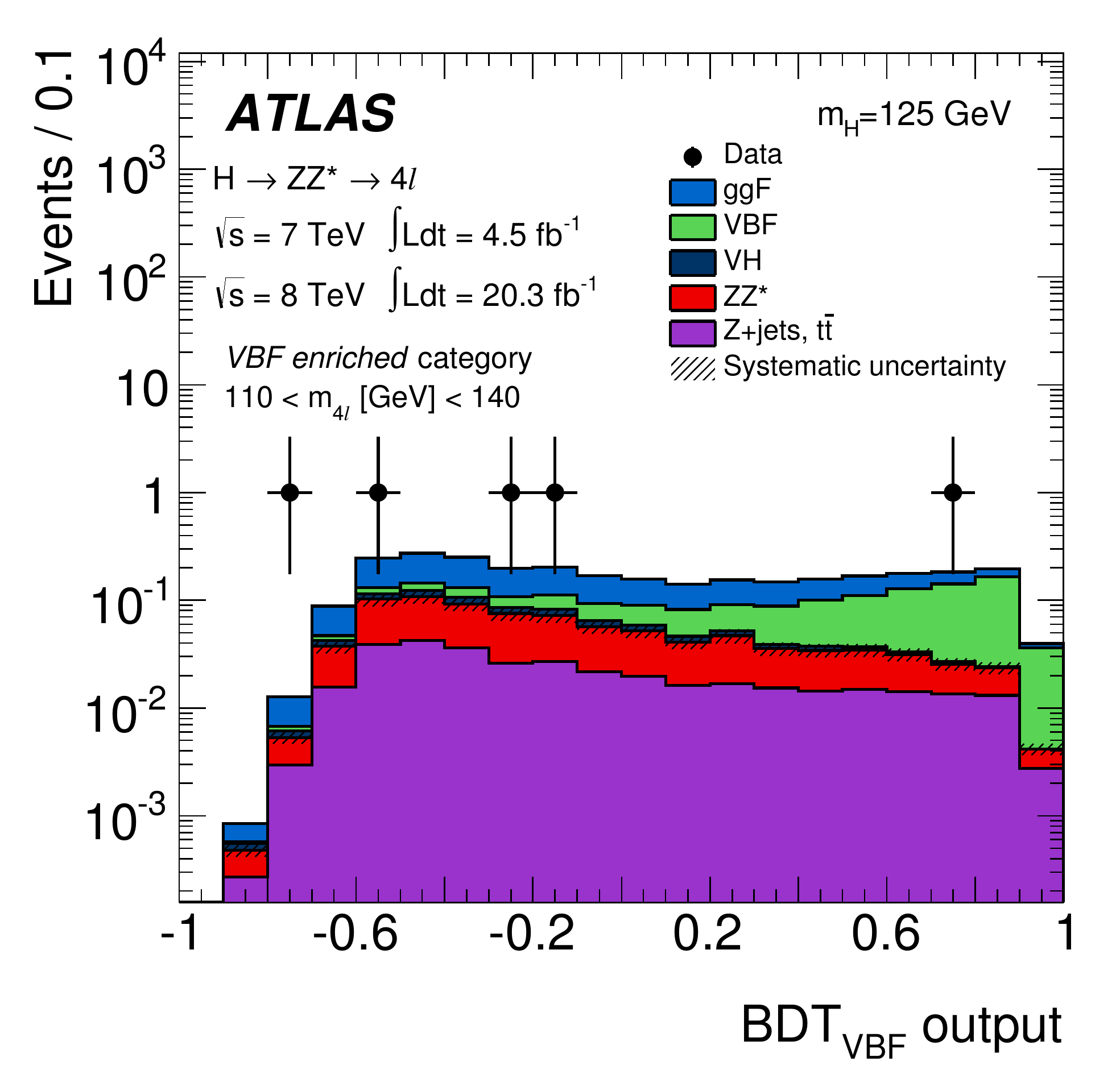}}  
  \caption{Distributions of the selected events and expected signal and background yields for the
    \vbfcat\ category for \subref{fig:vbf_m4l} $m_{4\ell}$ and \subref{fig:vbf_bdt} the
    \bdtvbf\ output in the full mass range\label{fig:vbf_yields}, and for \subref{fig:m4l_VBF}
    $m_{4\ell}$ and \subref{fig:BDT_VBF} the \bdtvbf\ output in the fit mass range
    $110<m_{4\ell}<140$ \gev.  The expected Higgs signal contributions, assuming $m_H=125$ \gev,
    from the ggF (blue histogram), VBF (green histogram) and VH (dark-blue histogram) production
    modes are included.  The expected background contributions, \zzstar\ (red histogram) and $Z+ \rm
    jets$ plus $t\bar{t}$ (violet histogram), are also shown; the systematic uncertainty associated
    to the total background contribution is represented by the hatched areas. In every case, the
    combination of the 7 \tev\ and 8 \tev\ results is shown. \label{fig:vbf_yields}}
\end{figure*}

In the following, measurements of the production strengths and couplings are discussed.  They are
all evaluated assuming the ATLAS combined mass \atlasmassveryshort.  The measurement of a global
signal strength factor, discussed in \secref{sec:inclres}, can be extended to a  measurement of  the signal
strength factors for specific production modes.

\begin{figure*}
  \centering
  \subfigure[\label{fig:mu_vbfvsggf}]{\includegraphics[width=\doublePlotSize]{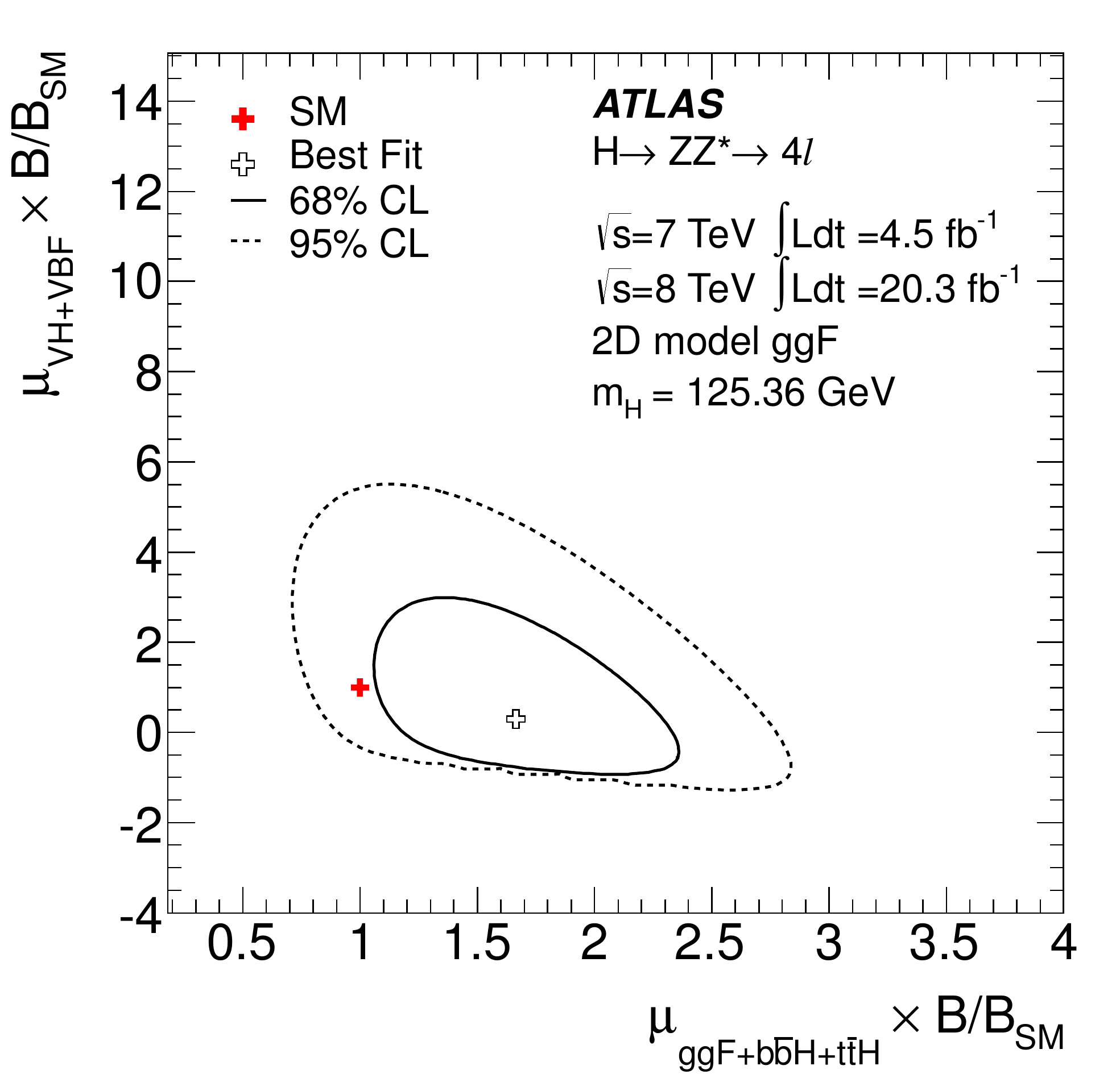}}
  \subfigure[\label{fig:mu_vbfoverggf}]{\includegraphics[width=\doublePlotSize]{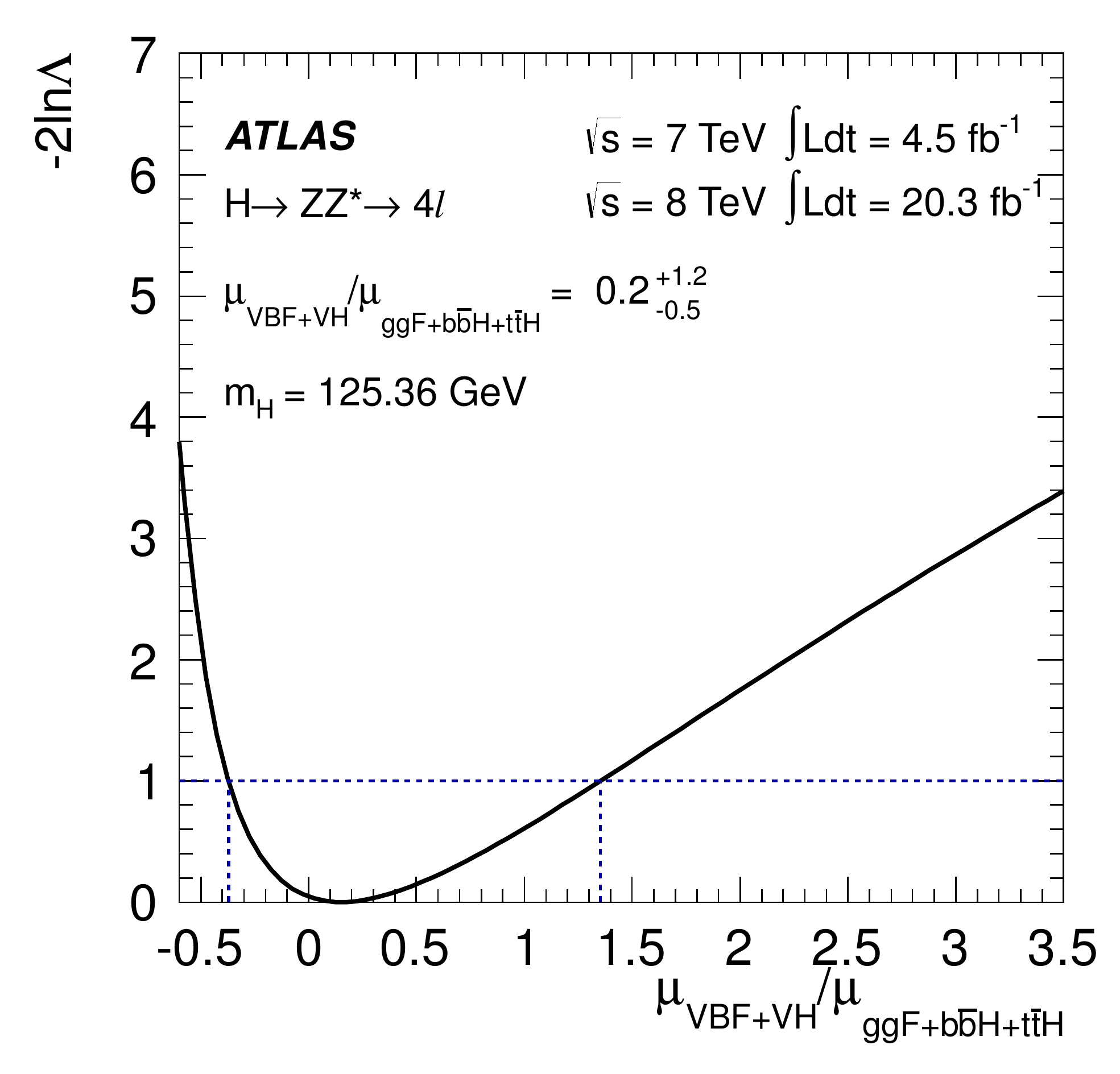}} 
  \caption{\subref{fig:mu_vbfvsggf} Likelihood contours in the
    $(\mu_{\mathrm{ggF}+b\bar{b}H+t\bar{t}H},\mu_{\mathrm{VBF}+\mathrm{VH}})$ plane including the
    branching ratio factor $B/B_{\mathrm{SM}}$.  Only the part of the plane where the expected
    number of signal events in each category is positive is considered.  The best fit to the data
    (open cross) and the 68\% CL (solid line) and 95\% CL (dashed line) contours are also indicated,
    as well as the SM expectation (solid red $+$).  \subref{fig:mu_vbfoverggf} Results of a
    likelihood scan for
    $\mu_{\mathrm{VBF}+\mathrm{VH}}/\mu_{\mathrm{ggF}+b\bar{b}H+t\bar{t}H}$. \label{fig:mu_scan_2d}}
\end{figure*}

The production mechanisms are grouped into the ``fermionic'' and the ``bosonic'' ones.  The former
consists of ggF, $b\bar{b}H$ and $t\bar{t}H$, while the latter includes the VBF and VH modes.  In
\figref{fig:mu_vbfvsggf} the best fit value for $\mu_{\mathrm{ggF}+b\bar{b}H+t\bar{t}H}\times
B/B_{\mathrm{SM}}$ versus $\mu_{\mathrm{VBF}+\mathrm{VH}}\times B/B_{\mathrm{SM}}$ is presented.
The factor $B/B_{\mathrm{SM}}$, the scale factor of the branching ratio with respect to the SM
value, is included since with a single channel analysis the source of potential deviations from the
SM expectation cannot be resolved between production and decay.  The profile likelihood ratio
contours that correspond to the 68\% and 95\% confidence levels are also shown.  The measured values
for $\mu_{\mathrm{ggF}+b\bar{b}H+t\bar{t}H}\times B/B_{\mathrm{SM}}$ and
$\mu_{\mathrm{VBF}+\mathrm{VH}}\times B/B_{\mathrm{SM}}$ are respectively: \iftoggle{isPRD} {
  \begin{widetext}
    \begin{equation}
      \begin{aligned}
        \mu_{\mathrm{ggF}+b\bar{b}H+t\bar{t}H}\times B/B_{\mathrm{SM}} &= \hllllmuggFform \\
        \mu_{\mathrm{VBF}+\mathrm{VH}}\times B/B_{\mathrm{SM}}        &=\hllllmuVBFform.
      \end{aligned}
    \end{equation}
  \end{widetext}
}{ 
    \begin{equation}
      \begin{aligned}
        \mu_{\mathrm{ggF}+b\bar{b}H+t\bar{t}H}\times B/B_{\mathrm{SM}} &= \hllllmuggFform \\
        \mu_{\mathrm{VBF}+\mathrm{VH}}\times B/B_{\mathrm{SM}}        &=\hllllmuVBFform.
      \end{aligned}
    \end{equation}
} 
The rounded results, with statistical and systematic uncertainties combined, are:
$\mu_{\mathrm{ggF}+b\bar{b}H+t\bar{t}H}\times B/B_{\mathrm{SM}} = $\hllllmuggFForAbstract\ and
$\mu_{\mathrm{VBF}+\mathrm{VH}}\times B/B_{\mathrm{SM}} = $\hllllmuVBFForAbstract.

The fit to the categories can be constrained to extract a single overall signal strength for the
\htollllbrief\ final state.  This combined $\mu \times B/B_{\mathrm{SM}}$ is
\hllllmuCombmassCombCatValue.
The ambiguity between production and decay is removed in \figref{fig:mu_vbfoverggf}, where the ratio
$\mu_{\mathrm{VBF}+\mathrm{VH}}/\mu_{\mathrm{ggF}+b\bar{b}H+t\bar{t}H}$ is presented. The measured
value of this ratio is \hllllmuVBFOvermuggF.

Following the approach and benchmarks recommended by the LHC Higgs Cross Section Working
Group~\cite{Heinemeyer:2013tqa}, measurements of couplings are implemented using a leading-order
tree-level-motivated framework. This framework is based on the following assumptions:
\begin{inparaenum}[\itshape a\upshape)]
\item the central value of the ATLAS combined mass measurement of \atlasmassveryshort\ is assumed;
\item the width of the Higgs boson is narrow, justifying the use of the zero-width approximation; and
\item only modifications of coupling strengths are considered, while the SM tensor structure is
  assumed, implying that the observed state is a \cal{CP}-even scalar.
\end{inparaenum}
The zero-width approximation allows the signal cross section to be decomposed in the following way:
$\sigma\cdot B\ (i\to H \to f) = \sigma_{i}\cdot\Gamma_{f} / \Gamma_{H}$ where $\sigma_{i}$ is the
production cross section through the initial state $i$, $B$ and $\Gamma_{f}$ are the branching ratio
and partial decay width into the final state $f$, respectively, and $\Gamma_{H}$ the total width of
the Higgs boson.  This approach introduces scale factors applied to the Higgs boson coupling,
$\Cc_{j}$, for particle $j$, which correspond to deviations from the SM Higgs coupling.  For
example, ggF production of the $ZZ^*$ final state can be represented as $\sigma\cdot B\ (gg\to H\to
ZZ^{*}) = \sigma_{\rm SM}(gg\to H)\cdot B_{\rm SM}(H\to ZZ^*)\cdot (\Cc^2_g \cdot \Cc^2_Z) /
\Cc^2_H$, where $\Cc_g$, $\Cc_Z$, and $\Cc_H$ are the scale factors for the Higgs couplings to $g$
and $Z$, and a scale factor for the total Higgs width, respectively.  Results are extracted from
fits to the data using the profile likelihood ratio $\Lambda(\vec\Cc)$.  In the fit, the $\Cc_{j}$
are treated either as parameters of interest or as nuisance parameters, depending on the
measurement.

\begin{figure*}
  \centering
  \subfigure[\label{fig:kvkf}]{\includegraphics[width=\doublePlotSize]{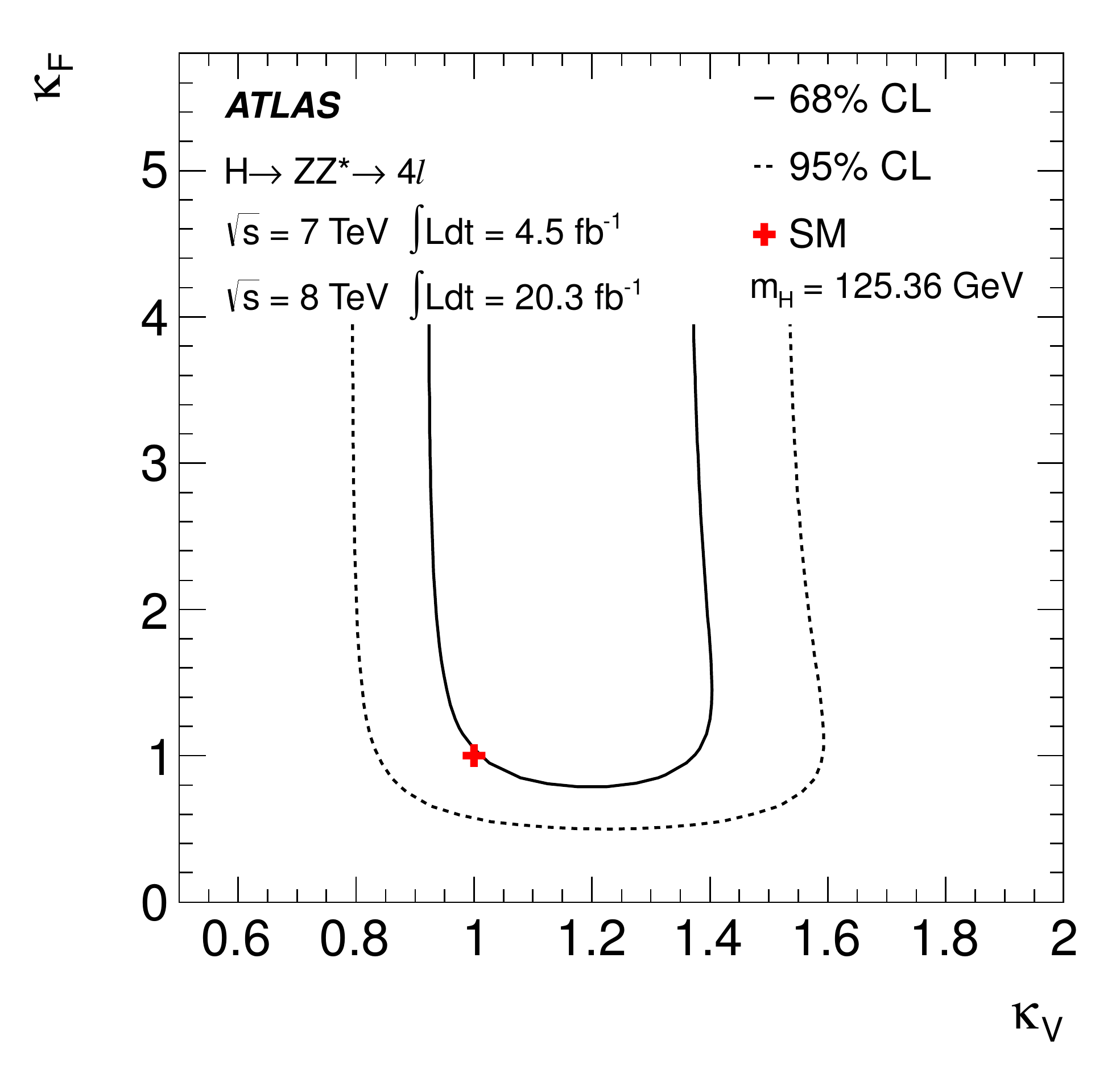}}
  \subfigure[\label{fig:lambda_FV}]{\includegraphics[width=\doublePlotSize]{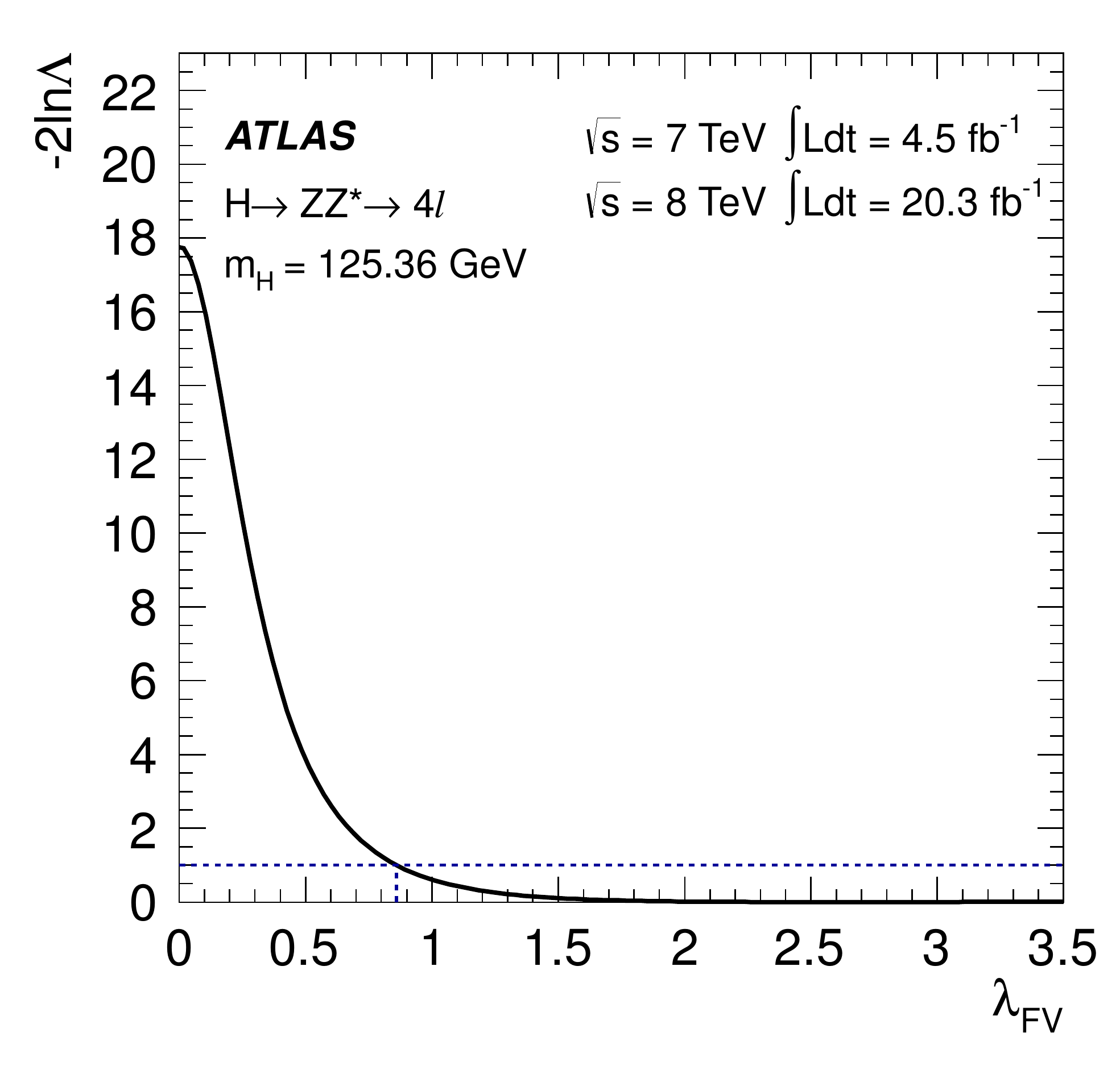}} 
  \caption{\subref{fig:kvkf} Likelihood contours at 68\% CL (solid line) and 95\% CL (dashed line)
    in the $\kappa_{\mathrm{V}}$ -- $\kappa_{\mathrm{F}}$ plane; the SM expectation (solid red
    cross) is also indicated.  \subref{fig:lambda_FV} Likelihood scan as a function of the ratio
    $\lambda_{\rm FV} = \kappa_{\mathrm{F}} / \kappa_{\mathrm{V}}$.  The Higgs boson mass is assumed
    to be the ATLAS combined value of \atlasmassveryshort. \label{fig:couplings}}
\end{figure*}

One benchmark model, which simplifies the measurement of possible deviations, groups the $\Cc_{j}$
for the electroweak vector bosons into a single scale factor, $\kappa_{\mathrm{V}}$, and defines
another coupling scale factor for all fermions, $\kappa_{\mathrm{F}}$.  The photon- and gluon-loop
couplings are derived from the tree-level couplings to the massive gauge bosons and fermions, and it
is assumed there is no non-SM contribution to the total decay width.  The likelihood contours in the
$\kappa_{\mathrm{V}}$--$\kappa_{\mathrm{F}}$ plane are shown in \figref{fig:kvkf}.  Since
$\kappa_{\mathrm{V}}$ and $\kappa_{\mathrm{F}}$ are related as $\kappa_{\mathrm{F}} =
\kappa_{\mathrm{V}} \times \mu_{\mathrm{ggF}+b\bar{b}H+t\bar{t}H}/\mu_{\mathrm{VBF}+\mathrm{VH}}$,
$\kappa_{\mathrm{F}}$ remains unbounded in \figref{fig:kvkf} because the present measurement of
$\mu_{\mathrm{VBF}+\mathrm{VH}}/\mu_{\mathrm{ggF}+b\bar{b}H+t\bar{t}H}$ cannot exclude the value of
zero, as can be seen in \figref{fig:mu_vbfoverggf}. The compatibility with the SM expectation is
30\%.  In \figref{fig:lambda_FV} the likelihood scan as a function of the ratio of fermion to
vector-boson coupling scale factors, $\lambda_{\rm FV} = \kappa_{\mathrm{F}} / \kappa_{\mathrm{V}}$,
is presented in the same benchmark model but where no assumption on the total decay width is made;
the branching ratio of the Higgs boson to a pair of $Z$ bosons cancels in the ratio.  The value
$\lambda_{\rm FV}=0$ is disfavored at the $4\sigma$ level.

\clearpage
\section{Summary}

The final Run I measurements of the Higgs boson production and couplings in the decay channel
\htollllp{} are presented. These measurements were performed using $pp$ collision data corresponding
to integrated luminosities of \lumia~and \lumib~at $\sqrt{s}=7$ TeV and $\sqrt{s}=8$ TeV,
respectively, recorded with the ATLAS detector at the LHC.  The signal and background simulation,
the electron and muon reconstruction and identification, the event selection and the reducible
background estimations are discussed in detail.  The analysis was performed both inclusively and
with events separated into categories for VBF, VH and ggF production modes.  Three multivariate
discriminants are employed to improve the separation of the Higgs signal from the
\zzstar\ background, to separate VBF from ggF Higgs boson production using jet kinematics, and to
distinguish hadronic decays of $W$ and $Z$ produced in association with a Higgs from ggF production.

For the inclusive analysis, in the $m_H$ range 120--130 \gev, 37 events are observed while 26.5
$\pm$ 1.7 events are expected, decomposed as 16.2 $\pm$ 1.6 events for a SM Higgs signal with
$m_H=125$ \gev, 7.4 $\pm$ 0.4 \zzstar\ background events and 2.9 $\pm$ 0.3 reducible background
events.  This excess corresponds to a \htollllbrief{} signal observed (expected) with a significance
of 8.1 (6.2) standard deviations at the combined ATLAS measurement of the Higgs boson
mass, \atlasmassveryshort~\cite{combmasspaper}.

For the VBF category, one event is seen with a high multivariate discriminant value and a mass of
123.4 \gev.  No VH candidate is found in the $m_H$ range 120--130 \gev\ with the $W$ or $Z$ decaying
either hadronically or leptonically.  The gluon fusion signal strength is found to be
\hllllmuggF\ and the signal strength for vector-boson fusion is found to be \hllllmuVBF.  At the
combined ATLAS measurement of the Higgs boson mass, \atlasmassveryshort, the measured combined
production rate relative to the SM expectation is \hllllmuCombmassCombCat. This measurement is based
on a fit to the categories assuming a single overall signal strength.  The ratio
$\mu_{\mathrm{VBF}+\mathrm{VH}}/\mu_{\mathrm{ggF}+b\bar{b}H+t\bar{t}H}$, which is independent of the
\htollllbrief{} branching ratio, is found to be \hllllmuVBFOvermuggF.  Finally, the observed event
yields in the categories are used to quantify the compatibility with the SM predictions in terms of
the Higgs coupling scale factor for weak vector bosons ($\kappa_{\mathrm{V}}$) and fermions
($\kappa_{\mathrm{F}}$); they are found to agree with the SM expectations.

The coupling measurements presented here for the Higgs boson decay to four leptons supersede those
of the previous ATLAS study~\cite{Aad:2013wqa} and are improved with respect to the earlier results.





\section{Acknowledgements}

We thank CERN for the very successful operation of the LHC, as well as the
support staff from our institutions without whom ATLAS could not be
operated efficiently.

We acknowledge the support of ANPCyT, Argentina; YerPhI, Armenia; ARC,
Australia; BMWFW and FWF, Austria; ANAS, Azerbaijan; SSTC, Belarus; CNPq and FAPESP,
Brazil; NSERC, NRC and CFI, Canada; CERN; CONICYT, Chile; CAS, MOST and NSFC,
China; COLCIENCIAS, Colombia; MSMT CR, MPO CR and VSC CR, Czech Republic;
DNRF, DNSRC and Lundbeck Foundation, Denmark; EPLANET, ERC and NSRF, European Union;
IN2P3-CNRS, CEA-DSM/IRFU, France; GNSF, Georgia; BMBF, DFG, HGF, MPG and AvH
Foundation, Germany; GSRT and NSRF, Greece; ISF, MINERVA, GIF, I-CORE and Benoziyo Center,
Israel; INFN, Italy; MEXT and JSPS, Japan; CNRST, Morocco; FOM and NWO,
Netherlands; BRF and RCN, Norway; MNiSW and NCN, Poland; GRICES and FCT, Portugal; MNE/IFA, Romania; MES of Russia and ROSATOM, Russian Federation; JINR; MSTD,
Serbia; MSSR, Slovakia; ARRS and MIZ\v{S}, Slovenia; DST/NRF, South Africa;
MINECO, Spain; SRC and Wallenberg Foundation, Sweden; SER, SNSF and Cantons of
Bern and Geneva, Switzerland; NSC, Taiwan; TAEK, Turkey; STFC, the Royal
Society and Leverhulme Trust, United Kingdom; DOE and NSF, United States of
America.

The crucial computing support from all WLCG partners is acknowledged
gratefully, in particular from CERN and the ATLAS Tier-1 facilities at
TRIUMF (Canada), NDGF (Denmark, Norway, Sweden), CC-IN2P3 (France),
KIT/GridKA (Germany), INFN-CNAF (Italy), NL-T1 (Netherlands), PIC (Spain),
ASGC (Taiwan), RAL (UK) and BNL (USA) and in the Tier-2 facilities
worldwide.




\iftoggle{isPRD} {
}{ 
    \bibliographystyle{atlasBibStyleWoTitle}
}
\bibliography{H4lpaper}


\onecolumn
\clearpage
\begin{flushleft}
{\Large The ATLAS Collaboration}

\bigskip

G.~Aad$^{\rm 85}$,
B.~Abbott$^{\rm 113}$,
J.~Abdallah$^{\rm 153}$,
S.~Abdel~Khalek$^{\rm 117}$,
O.~Abdinov$^{\rm 11}$,
R.~Aben$^{\rm 107}$,
B.~Abi$^{\rm 114}$,
M.~Abolins$^{\rm 90}$,
O.S.~AbouZeid$^{\rm 160}$,
H.~Abramowicz$^{\rm 155}$
% ...
\bigskip
\\
% ...
$^{11}$ Institute of Physics, Azerbaijan Academy of Sciences, Baku, Azerbaijan\\
% ...
$^{85}$ CPPM, Aix-Marseille Universit{\'e} and CNRS/IN2P3, Marseille, France\\
% ...
$^{90}$ Department of Physics and Astronomy, Michigan State University, East Lansing MI, United States of America\\
% ...
$^{107}$ Nikhef National Institute for Subatomic Physics and University of Amsterdam, Amsterdam, Netherlands\\
% ...
$^{113}$ Homer L. Dodge Department of Physics and Astronomy, University of Oklahoma, Norman OK, United States of America\\
$^{114}$ Department of Physics, Oklahoma State University, Stillwater OK, United States of America\\
% ...
$^{117}$ LAL, Universit{\'e} Paris-Sud and CNRS/IN2P3, Orsay, France\\
% ...
$^{153}$ Institute of Physics, Academia Sinica, Taipei, Taiwan\\
% ...
$^{155}$ Raymond and Beverly Sackler School of Physics and Astronomy, Tel Aviv University, Tel Aviv, Israel\\
% ...
$^{160}$ Department of Physics, University of Toronto, Toronto ON, Canada\\
% ...
\end{flushleft}


\end{document}
%